\documentclass{article}%
\usepackage{amsmath}
\usepackage{amsfonts}
\usepackage{amssymb}
\usepackage{graphicx}
\usepackage{subcaption}
\usepackage[bookmarks=true,colorlinks=true]{hyperref}
\hypersetup{linkcolor=blue}
\hypersetup{citecolor=red}
\newcommand{\bpartial}{\mathop{\partial\kern -4pt\raisebox{.8pt}{$|$}}}
\newcommand{\bra}{\mathopen{[\kern-1.6pt[}}
\newcommand{\ket}{\mathclose{]\kern-1.5pt]}}
\newcommand{\bbra}{\mathopen{[\kern-2.2pt[\kern-2.3pt[}}
\newcommand{\bket}{\mathclose{]\kern-2.1pt]\kern-2.3pt]}}

\makeindex
\begin{document}
	\title {\large{ \bf Anisotropic homogeneous  string
			cosmology with two-loop corrections}}
	
	\vspace{3mm}
	
	\author {  \small{ \bf  F. Naderi}\hspace{-1mm}{ \footnote{ e-mail: f.naderi@azaruniv.edu}} ,{ \small
		} \small{ \bf  A. Rezaei-Aghdam}\hspace{-1mm}{
			\footnote{Corresponding author. e-mail:
				rezaei-a@azaruniv.edu}} \\
		{\small{\em Department of Physics, Faculty of Basic Sciences, Azarbaijan Shahid Madani University}}\\
		{\small{\em   53714-161, Tabriz, Iran  }}}
	
	\maketitle
	
	\begin{abstract}
		The two-loop (order $\alpha'$) $\beta$-function equations, which are equivalent to the equations of motion of $\alpha'$-corrected string effective action, are considered for anisotropic homogeneous space-times. These equations are solved for all Bianchi-type models in two schemes of effective action, namely  $R^2$ and Gauss-Bonnet schemes with  zero cosmological constant and then 
		the  metric, dilaton and $B$-field   are found at $\alpha'$ perturbative corrections.

	\end{abstract}
	%%%%%%%%%%%%%%%%%%%%%%%%%%%%%%%%%%%%%%%%%%%%%%%%%%%%%%%%%%%%%%%%%%%%%%%%%%%%%%%%%%%%%%%%%%%%%
	%%%%%%%%%%%%%%%%%%%%%%%%%%%%%%%%%%%%%%%%%%%%%%%%%%%%%%%%%%%%%%%%%%%%%%%%%
	\section{Introduction}
	
	The application of  low energy, tree-level  string effective action for describing the evolution of  early  universe with a very low coupling, $g= e^{-\phi}$, and curvature is 
	well accepted
	\cite{oneloop1, oneloop2, oneloop3, Veneziano}.
	Solutions of  the low energy string effective action have been found in several cosmological backgrounds.
	For example, the case of Friedmann-Robertson-Walker (FRW) backgrounds have been  discussed in \cite{fr}, homogeneous anisotropic space-times with contributing of the antisymmetric tensor field  have been investigated   in \cite{batakis1,batakis2,batakis3}, $(4+1)$ dimensional homogeneous anisotropic models have been studied in \cite{mr} and  inhomogeneous models have been discussed in \cite{Inhomogeneous1,Inhomogeneous2}. However, evolving the universe  
	toward a strong coupling, dilaton and curvature will gain an inevitable accelerated growth \cite{7,8}. 
	But the singularity of curvature and coupling is widely believed to be regularized by the expansion of the leading order effective action and including higher-order  corrections \cite{9,1i,1ii}.   Generally, there are two types of expansions consisting of a quantum nature expansion in string coupling named  string loop expansion and  a stingy type $\alpha'$ expansion, where the string-length is $\lambda_s=\sqrt{\alpha'}$ \cite{Veneziano}. The $\alpha'$-corrections are commonly considered for regularizing the curvature.
	Consideration of the $\alpha'$-corrections to the low energy effective action  is important in the case of high curvature, where  the loop-corrections are significant when  the string coupling is  strong enough. So,  until the coupling is  sufficiently  small in the high curvature regime the  loop corrections can be neglected and  the $\alpha'$-corrections are included only \cite{Veneziano}.

	Vanishing of the two-loop $\beta$-function equations of the corresponding $\sigma$-model which are the requirement of the conformal invariance of the theory up to two-loop, are equivalent to the background field equations of motion of the   $\alpha'$-corrected string effective action in the string frame.
	The two-loop $\beta$-functions and the possible higher order $\alpha'$-correction terms to the string effective actions are investigated  in \cite{ts, hs}. These corrections are  generically quadratic type terms and give  higher order time
	derivative field equations.
	There is a renormalization scheme (RS) dependence in the two-loop order of $\beta$-functions and their   $\alpha'$-corrected effective action, where the two  Gauss-Bonnet and $R^2$ schemes are distinguished \cite{ts}. The effective action of the former case has  nice properties like being unitary, physical ghost free and having $O(d,d)$ symmetry which is related to the T-duality \cite{Zwiebach,gf1,gf2}.

	The   $\alpha'$-corrected field equations for massless string modes have been solved for 
	several  classes of backgrounds such as black hole \cite{bh, bh0, bh1, bh2} and cosmology
	\cite{cm1,cm2,cm3,cm4,cm5,cm6,cm7,cm8,cm9,cm1aaR. Lazkoz, Exirifard}.
	The dynamics of the Bianchi $IX$ and $I$ universe in Gauss-Bonnet gravity  from the 1-loop superstring effective action have been investigated in \cite{KawaiI}, \cite{KawaiIx}.
	There are some works on solving two-loop $\beta$-function equations,   for instance, in the \cite{Exirifard}, where the antisymmetric $B$-field is set to zero and the metric and dilaton fields are considered only. 
	
	In fact, in the presence of the  quadratic corrections  the equations are not easy to solve. Furthermore, if the $B$-field is considered too, the equations become more nontrivial. There are some few classes of such solutions, for example \cite{bf1, bf2}, where the equations of
	motion of the effective field theory are considered. Here our goal is to investigate the solutions of two-loop $\beta$-function equations with a dilaton and a $B$-field contributions on  homogeneous anisotropic space-times. Indeed, the RS dependence  of  the two-loop $\beta$-functions appears in  the presence of an antisymmetric $B$-field. Therefore, including the $B$-field requires  considering the  RS dependence and  the two useful schemes of $R^2$ and Gauss-Bonnet  are our  interests.

	The paper is organized as follows. In section $2$ we recall the general form of two-loop $\beta$-functions of the two RS. In section $3$ we derive the explicit forms of the two-loop $\beta$-function equations on the anisotropic homogeneous space-times in terms of Hubble coefficients of the string frame, dilaotn and $B$-field in the $R^2$ and Gauss-Bonnet schemes. In Section $4$ we present  the solutions of the  $\beta$-function equations in all Bianchi-type models. Finally, the
	main results of this paper are finally summarized in Section $5$ .

	%%%%%%%%%%%%%%%%%%%%%%%%%%%%%%%%%%%%%%%%%%%%%%%%%%%%%%%%%%%%%%%%%%%%%%%%%%%%%%%%%%%%%%%%%%%%%%%%%%%%%
	\section{Two-loop (order $\alpha'$) $\beta$-functions and $\alpha'$-corrected string effective action}
	Vanishing of  the $\beta$-functions of 
	$\sigma$-model is the requirement of conformal symmetry of the theory \cite{ts1}. On the other hand, the  $\beta$-functions are equivalent to   the equations of motion
	of the effective action \cite{ts}. 
	The two-loop   $\beta$-functions, the $\alpha'$-corrected effective actions and the  
	on-shell equivalence of  the equations of motion  of the  effective action and the conformal invariance conditions in $\alpha'$ order have been considered in \cite{ts}.
	Generally, the $\beta$-functions of the $\sigma$-model with the backgrounds of dilaton $\phi$, antisymmetric $B$-field and
	metric $G$ are given by \cite{ts}
	\begin{eqnarray}
	\bar{\beta}^{G}_{\mu\nu}&=&\beta^{G}_{\mu\nu}+2\alpha'\nabla_{\mu}\nabla_{\nu}\phi+\nabla_{(\mu}W_{\nu)},\\
	\bar{\beta}^{B}_{\mu\nu}&=&\beta^{B}_{\mu\nu}+\alpha'H_{\mu\nu}^{~~\lambda}\nabla_{\lambda}\phi+\frac{1}{2}H_{\mu\nu}^{~~\lambda}W_{\lambda}+\partial_{(\mu}L_{\nu)},\\
	\bar{\beta}^{{\phi}}&=&{\beta}^{{\phi}}+W_{2}^{\mu}\partial_{\mu}\phi,
	\end{eqnarray}
	where the $\beta^{G}_{\mu\nu}$, $\beta^{B}_{\mu\nu}$ and ${\beta}^{{\phi}}$ are the standard RG $\beta$-functions. The field strength $H$ is defined by $H_{\mu\nu\rho}=3\partial_{[\mu}B_{\nu\rho]}$ and the $W_{\mu}$ and $L_{\mu}$ are given by the renormalization matrix as follows \cite{ts1}
	\begin{eqnarray}
	W_{\mu}&=&\alpha'^2(l_1\nabla_{\mu}H^2+l_2H_{\mu\lambda\rho}\nabla_{\sigma}H^{\sigma\lambda\rho}),\\
	L_{\mu}&=&\alpha'^2l_3H_{\mu\lambda\rho}\nabla_{\sigma}H^{\sigma\lambda\rho}.
	\end{eqnarray}

	There is a RS dependence in $\beta$-function equations at two-loop order in the presence of $B$-field. These equations are calculated in \cite{ts} where the 
	$R^2$ and {Gauss-Bonnet} schemes are pointed especially.
	By choosing any of
	RS's, the resulting $\beta$-functions will
	be different, but transformation between the  RS's can be easily done by using the leading order of equations of motion and appropriate field redefinitions \cite{ts, hs}. Solutions of the  $\beta$-function equations of  any RS will be different, but still equivalent because  they belong to  different definitions of physical metric, dilaton and $B$-field.  
	The two-loop $\beta$-functions of the metric, $B$-field  are as follows \cite{ts} 
	\begin{equation}\label{betaGR}
	\begin{split}
	\frac{1}{\alpha'}\bar{\beta}^{G}_{\mu\nu}&={R}_{{\mu}{\nu}}-\frac{1}{4}{H}^{2}_{{\mu}{\nu}}-\nabla_{{\mu}}\nabla_{{\nu}}{\phi}+\frac{\alpha'}{2}\big[
	R_{\mu\alpha\beta\gamma}R_{\nu}^{~\alpha\beta\gamma}
	-\frac{3}{2}R_{(\mu}^{~~~\alpha\beta\gamma}H_{\nu)\alpha\lambda}H_{\beta\gamma}^{~~\lambda}
	-\frac{1}{2}R^{\alpha\beta\rho\sigma}H_{\mu\alpha\beta}H^{\nu\rho\sigma}	+\frac{1}{8}(H^4)_{\mu\nu}\\
	&
	+\frac{1}{4}\nabla_{\lambda}H_{\mu\alpha\beta}\nabla^{\lambda}H_{\nu}^{~\alpha\beta}
	+\frac{1}{12}\nabla_{\mu}H_{\alpha\beta\gamma}\nabla_{\nu}H^{\alpha\beta\gamma}
	+\frac{1}{8}H_{\mu\alpha\lambda}H_{\nu\beta}^{~~\lambda}(H^2)^{\alpha\beta} \\
	&
	-\frac{f}{2}\big(R_{\mu\alpha\beta\nu}(H^2)^{\alpha\beta}
	+2\,R_{(\mu}^{~~\alpha\beta\gamma}H_{\nu)\alpha\lambda}H_{\beta\gamma}^{~~\lambda}
	+R^{\alpha\beta\rho\sigma}H_{\mu\alpha\beta}H^{\nu\rho\sigma}
	-\nabla_{\lambda}H_{\mu\alpha\beta}\nabla^{\lambda}H_{\nu}^{~\alpha\beta}
	\big)	-\frac{1}{12}\nabla_{\mu}\nabla_{\nu}H^2\big],
	\end{split}
	\end{equation}
	\begin{eqnarray}\label{betaBR}
	\begin{split}
	\frac{1}{\alpha'}{\beta}_{{\mu}{\nu}}^{~{B}}(f=1)=-\frac{1}{2}\nabla^{{\mu}}{H}_{{\mu}{\nu}{\rho}}+\frac{\alpha'}{4}(&2R_{[\mu\gamma\alpha\beta}\nabla^{\gamma}H^{\alpha\beta}_{~~~\nu]}+\nabla_{\gamma}H_{\alpha\beta[\mu}H_{\nu]\rho}^{~~~\alpha}H^{\beta\gamma\rho}\\
	&+2\nabla_{\beta}H^2_{\alpha[\nu}H_{\mu]}^{~~\alpha\beta}+\frac{1}{2}H^2_{\alpha\beta}\nabla^{\alpha}H^{\beta}_{~\mu\nu}),
	\end{split}
	\end{eqnarray}
	\begin{equation}\label{betaBG}
	\begin{split}
	\frac{1}{\alpha'}{\beta}^{B}_{\mu\nu}(f=-1)&=\hat{R}_{[{\mu}{\nu}]}+\frac{\alpha'}{2}\bigg(\hat{R}^{\alpha\beta\gamma}_{~~~~[\nu}\hat{R}_{\mu]\alpha\beta\gamma}-\frac{1}{2}\hat{R}^{\beta\gamma\alpha}_{~~~~[\nu}\hat{R}_{\mu]\alpha\beta\gamma}+\frac{1}{2}\hat{R}_{\alpha[\mu\nu]\beta}H^{\alpha\beta}\bigg)
	\end{split},
	\end{equation}
	where $H^4=H_{\mu\nu\lambda}H^{\nu\rho\kappa}H_{\rho\sigma}^{~~\lambda}H^{\sigma\mu}_{~~\kappa}$,  $H_{\mu\nu}^2=H_{\mu\rho\sigma}H_{\nu}^{\rho\sigma}$ and	 the $\hat{R}^{\rho}_{\mu\nu\sigma}$ is the Riemann tensor associated with the  generalized connection for the    $\sigma$-model torsion ${\hat{\Gamma}}^{\rho}_{\mu\nu}={{\Gamma}}^{\rho}_{\mu\nu}-\frac{1}{2}H^{\rho}_{\mu\nu}$.\footnote{
		${\hat{R}}^{\kappa}_{~\lambda\mu\nu}= R^{\kappa}_{~\lambda\mu\nu}-\frac{1}{2}\nabla_{\mu}H^{\kappa}_{\nu\lambda}-\frac{1}{2}\nabla_{\nu}H^{\kappa}_{\mu\lambda}+\frac{1}{4}H_{~\nu\lambda}^{\gamma}H_{~\mu\gamma}^{\kappa}-\frac{1}{4}H_{~\mu\lambda}^{\gamma}H^{\kappa}_{~\nu\gamma}.
		$} 
	The $f$ parameter indicates the RS dependence and especially the $f=1$ and $f=-1$  are correspond to the $R^2$ and  Gauss-Bonnet schemes, respectively.
	
	For the effective action,	there are two parameterizations   namely  the $\sigma$-parameterization and the  $s$-parameterization where their metrics  are related with each other in 4 dimensional space-time by
	\begin{equation}\label{3}
	\bar{	G}_{\mu\nu}^{(\sigma)}=e^{\phi}G_{\mu\nu}^{(s)}.
	\end{equation}
	Alternatively, these two parameterizations  indicate two frames called respectively  {string frame}, ${G}_{\mu\nu}^{(\sigma)}$, and {Einstein frame}, $\bar{G}_{\mu\nu}^{(s)}$. 
	The $\alpha '$-corrected effective action  in the $\sigma$-parameterization   is given by \cite{ts}
	\begin{eqnarray}\label{action}
	S=\int d^4 x\sqrt{G}e^{\phi}(R-\frac{1}{12}H^2+(\nabla\phi)^2-\Lambda+\frac{\alpha'}{4}(R^2_{\mu\nu\rho\lambda}-\frac{1}{2}R^{\alpha\beta\rho\sigma}H_{\alpha\beta\lambda}H_{\rho\sigma}^{~~\lambda}
	+\frac{1}{24}H^4-\frac{1}{8}(H_{\mu\nu}^{~~2})^2).
	\end{eqnarray}
	Its variation with respect to the dilaton gives  the averaged 
	$\beta$-function of dilaton, $	\tilde{\beta}^{~({\phi})}$, which can be written in terms of metric and dilaton $\beta$-functions as follows
	\begin{eqnarray}
	\tilde{\beta}^{~{\phi}}=\bar{\beta}^{~{\phi}}-\frac{1}{4}\bar{\beta}^{G}_{\mu\nu}G^{\mu\nu},
	\end{eqnarray}
	and is given by
	\begin{eqnarray}\label{betafR}
	\frac{1}{\alpha'}\tilde{\beta}^{~{\phi}}=-{R}+\frac{1}{12}{H}^{2}+2\nabla_{{\mu}}\nabla^{{\mu}}{\phi}+(\partial_{{\mu}}{\phi})^{2}+\Lambda-\frac{\alpha'}{4}(R^2_{\mu\nu\rho\lambda}-\frac{1}{2}R^{\alpha\beta\rho\sigma}H_{\alpha\beta\lambda}H_{\rho\sigma}^{~~\lambda}
	+\frac{1}{24}H^4-\frac{1}{8}(H_{\mu\nu}^{~~2})^2).
	\end{eqnarray}
	Furthermore, the  Gauss-Bonnet effective action  $s$-parameterization in 4-dimensions for bosonic string is given by \cite{ts}
	\begin{eqnarray}
	\begin{split}
	S=\int d^4 x\sqrt{G}\bigg(R-\frac{1}{12}\,e^{2\phi}H^2-\frac{1}{2}(\nabla\phi)^2&-\Lambda e^{\phi}+\frac{\alpha'e^{\phi}}{4}\big[R^2_{\mu\nu\rho\lambda}-4\,R_{\mu\nu}^2+R^2\\
	&+e^{2\phi}(-\frac{1}{2}R^{\alpha\beta\rho\sigma}H_{\alpha\beta\lambda}H_{\rho\sigma}^{~~\lambda}
	+\frac{1}{2}H^2_{\mu\nu}\nabla^{\mu}\phi\nabla^{\nu}\phi-\frac{1}{6}H^2(\nabla\phi)^2)\\
	&+e^{2\phi}(\frac{1}{24}H^4+\frac{1}{8}(H_{\mu\nu}^{~~2})^2-\frac{5}{144}(H^2)^2\big)]\bigg).
	\end{split}
	\end{eqnarray}
	
	In fact,    the  $\alpha'$-corrected  effective action in its most general form is parametrized by $20$ constants \cite{Metsaev}.  Since $S$-matrix is invariant under the field redefinitions of type \cite{Metsaev}
	\begin{eqnarray}\label{mm}
	\begin{split}
	\delta G_{\mu\nu}=\alpha'(b_1R_{\mu\nu}+b_2\partial_{\mu}\phi\partial_{\nu}\phi+b_3H^2_{\mu\nu}+G_{\mu\nu}(b_4R+b_5(\partial \phi)^2+b_6\nabla^2\phi+b_7H^2)),
	\end{split}
	\end{eqnarray}
	\begin{eqnarray}
	\begin{split}
	\delta B_{\mu\nu}=\alpha'(b_8\nabla^{\lambda}H_{\lambda\mu\nu}+b_9H_{\mu\nu}^{~~\lambda}\partial_{\lambda}\phi),
	\end{split}
	\end{eqnarray}
	\begin{eqnarray}\label{nn}
	\begin{split}
	\delta \phi=\alpha'(b_{10}R+b_{11}(\partial\phi)^2+b_{12}\nabla^2\phi+b_{13}H^2),
	\end{split}
	\end{eqnarray}
	there is  a field redefinition ambiguity and we have a class of physically equivalent effective actions  \cite{Tseytlin2}.
	The \eqref{mm}-\eqref{nn}  can  leave $3$  constants together with  $5$ combinations of the $20$ constants invariant. Thus, the $\alpha'$-corrected effective action is parametrized by $8$ essential coefficients and it is easy to relate it to the Gauss-Bonnet  effective action   by an appropriate  field redefinition and using the leading order equations of motion \cite{ts}. Hence, in the $\sigma$-model context, if one works with the higher order corrections in effective field theory and the equations of motions, it is convenience to transform  by the field redefinitions to the Gauss-Bonnet  effective action which has no higher than second derivative in its field equations and is free of ghost.
	The $8$ essential coefficients that parametrize the $\alpha'$-corrected effective action  must be fixed by comparison with the $S$-matrix \cite{ts}.
	Also, the two-loop $\beta$-functions  and the equations of motion of the effective action   are identical if a field redefinition is applied \cite{ts}. Working with the field equation in the effective field theory, the  field redefinitions may be considered, for instance as it has been done in \cite{bf1}.
	In this work, we will not redefine the fields and maintain in the convention provided by the $\beta$-functions.

	Moreover, it is worth mentioning that in a non-critical $D$ dimensional bosonic theory, the constant $\Lambda$ equals to   $\frac{2\,(D-26)}{3\alpha'}$    and is related to the central charge deficit of the original theory  \cite{lambda2, landa1}.  From a cosmological point of view,  the $\Lambda$  is analogous to the non-vanishing cosmological constant in  standard
	theory of gravity  \cite{lambda1}. 
	We will set $\Lambda=0$ in our analysis. This may  be  a  good approximation at early times
	if compared with the  $\Lambda $ the curvature and/or kinetic energy  are large, $\Lambda \ll R, (\nabla \phi)^2, H^2$ \cite{landa2}. However,  $\Lambda $ is required to be considered for expanding universe at late times.

	%%%%%%%%%%%%%%%%%%%%%%%%%%%%%%%%%%%%%%%%%%%%%%%%%%%%%%%%%%%%%%%%%%%%%%%%%%%%%%%%%%%%%%%%%%%%%%%%%555
	\section{Anisotropic homogeneous two-loop string cosmology}
	Anisotropic homogeneous space-times are achieved by relaxing the isotropy requirement of FRW-type models. These type of space-times have been  revealed in string cosmology models when approaches to find special backgrounds were being investigated to describe the early evolution of the universe. Their   general metric  in  string frame  is written as the following form \cite{landau}
	\begin{eqnarray}\label{metric}
	ds^2=G_{\mu\nu}\,dx^{\mu}\,dx^{\nu}=-g_{00}(t)\,dt^{2}+e_{\alpha}^{~~i}(x) \,g_{ij}(t)\,e_{\beta}^{~~j}(x)\,dx^{\alpha}\,dx^{\beta},
	\end{eqnarray}
	where $\alpha$ and $\beta$  are indices of spatial the submanifold     and the $i$,$ j$ are indices
	of the bases of an isometry group $G$. The $ e_{\alpha}^{~~i}(x)$ denote vielbeins which for any Bianchi model are given in appendix $A$.
	Non-diagonal components of the algebraic metric $g_{ij}$ which give constraint in the $\beta$-function equations can be taken  zero. Hence, as
	a choice one can set \cite{batakis2}
	\begin{eqnarray}\label{d}
	g_{ij}(t)=a_i^2(t)\delta_{ij},
	\end{eqnarray}
	where $a_i^2$ are the string frame scale factors. Hence, according to the \eqref{3}, the Einstein frame scale factors will be $\bar{a}_{i}^2=e^{\phi}a_i^2$.  If $\{\sigma^{i},~i=1,2,3\}$ indicate the  left invariant basis of $1$-forms, the 3-form $H$ may be chosen as \cite{ batakis2}
	\begin{eqnarray}\label{H0}
	H=A\,\sigma^1\wedge \sigma^2\wedge \sigma^3.
	\end{eqnarray}
	The requirement of $\,d\,H=0$ implies  $A$ to be a constant.

	Substituting the \eqref{metric} (along with \eqref{d}) and \eqref{H0} in  \eqref{betaGR} and using the relations \eqref{rnon}-\eqref{gnon},
	the $(i,i)$ and time-time components of $\beta$-function  of metric   \eqref{betaGR} cast the following  general forms, respectively\footnote{Because  the results may be complicated, we present the following formalism  with more details for a particular isotropic case on Bianchi-type $V$ in appendix $B$.}
	\begin{eqnarray}\label{ii}
	\frac{1}{\alpha'}\bar{\beta}^{G~i}_{~i}=	\dot{H_{i}}+ H_i{\sum}H_{k} + H_{i} \dot{\phi}-H_{i}\,\dot{(\ln{g_{00}})}+V_{i}^{(1)}-\frac{1}{2}A^2{\bar{a}}^{-6}e^{3\phi}+\alpha'(K_{i}+V_{i}^{(2)})
	,
	\end{eqnarray}
	\begin{eqnarray}\label{00}
	\frac{1}{\alpha'}\bar{\beta}^{G~0}_{~0}=	{\sum}(\dot{H_i}+ H_i^2)+\ddot{\phi}-\dot{(\ln{g_{00}})}(\dot{\phi}+{\sum}{H_i})-\alpha'(K_{0}+V_{0}^{(2)}),
	\end{eqnarray}
	where $\bar{a}^3= \bar{a}_{1} \bar{a}_{2} \bar{a}_{3}$ and the dot stands for ${d}/{d\,t}$.  The   Hubble coefficients of string frame are defined with
	$H_{i}=\dot{(\ln a_{i}(t))}$.
	The above equations have been written in terms of  auxiliary functions of $V$ and $K$ in order to get a general and shorthanded form of $\beta$-equations in  all Bianchi-type models. All the terms with dependence on structure constants of lie-group of models are collected in the $V$ terms and the others with no dependence on structure constants are gathered in $K$ part.  So, throughout this paper $V$ and  $K$   denote  Bianchi-type dependent and independent terms, respectively.
	The $V^{(1)}$ and $V^{(2)}$  are originated from  one-loop  and two-loop orders of $\beta$-equations, receptively.  This terms  will be declared  for each Bianchi model in the next section.
	The $K$ terms of \eqref{ii} and \eqref{00} that are common for all types  are given   as following
	\begin{eqnarray}\label{kii}
	\begin{split}
	K_{i}=&\dot{H_i}^2+2H^2_{i}\dot{H_{i}}+H^2_{i}\sum H^2_{k}+\frac{3}{8}A^4 \,{\bar{a}}^{-12}e^{6\phi}\\
	&-\frac{1}{4}A^2\,\bar{a}^{-6}e^{3\phi}\big(2({f+4})H^2_{i}+\sum_{k,j\neq i}\left(4\left(2f+1\right)H_k^2+2\left(4f+{11}\right)H_{i}H_{k}+{(5f+3)}H_{j}H_{k}\right)\big),
	\end{split}
	\end{eqnarray}
	\begin{eqnarray}\label{koo}
	K_{0}=-\sum(\dot{H^2_{i}}+2\dot{H_{i}}H_{i}^2+H^4_{i})-\frac{1}{2}A^2\,\bar{a}^{-6}e^{3\phi}(\sum(\dot{H_{i}}+2\,fH_{i}^2-2\sum_{k< i }H_{k}H_{i})).
	\end{eqnarray}
	Obviously, there is a RS dependence in the $K_i$ terms expressed by $f$ parameter which equals to $1$ for $R^2$ scheme and $-1$ for Gauss-Bonnet scheme.	
	Furthermore, with the considered metric \eqref{metric}, in  the Bianchi-type of class $\cal{B}$\footnote{The classification of Bianchi-type models and their structure constants are given on the appendix $A$.} the $(0,i)$ components of 
	$\beta$-function equations of metric \eqref{betaGR} give constraint equations  that will be provided in the following section.

	Furthermore, in the same way the dilaton $\beta$-function   \eqref{betafR}  casts  the following general form
	\begin{eqnarray}\label{fi}
	\begin{split}
	\frac{1}{\alpha'}	\tilde{\beta}^{\phi}=-2\ddot{\phi}-\dot{\phi}^2&-{\sum}(2\,\dot{\phi} H_i+V_{i}^{(1)}+{H_i}^2+2\,\dot{H_i})
	- ({\sum} H_i)^2\\
	&+2(\phi'+\sum H_i)\dot{(\ln{g_{00}})}+\frac{1}{2}A^2\bar{a}^{-6}e^{3\phi}
	+\alpha'[K_{\phi}+V_{\phi}^{(2)}],
	\end{split}
	\end{eqnarray}
	where its $K$ term is given by
	\begin{eqnarray}\label{kfii}
	K_{\phi}=
	-\sum(\dot{H_i}^2+2H_{i}^2\dot{H_i}+{H^4_{i}})-\sum_{i<j}H_i^2H_j^2-\frac{1}{4}A^2\bar{a}^{-6}e^{-3\phi}(\sum_{k< i }H_{k}H_{i}-5A^2\bar{a}^{-6}e^{-3\phi}).
	\end{eqnarray}
	The Bianchi-type dependent term $V_{\phi}^{(2)}$ term which is produced by the two-loop order of dilaton $\beta$-function will be given in the next section.
	
	The solutions of  the one-loop
	$\beta$-function equations on anisotropic homogeneous space-time with the   field strength of \eqref{H0} have been investigated in \cite{ batakis2}. Here, our 	
	purpose is to solve the  two-loop $\beta$-functions on  both RS.
	For performing this, we  implement a perturbative series expansion in $\alpha'$ up to first order on the background fields
	as following
	\begin{eqnarray}
	\phi=\phi_{0}+\alpha'\phi_{1},\label{18}\\
	\bar{a}_i^2= a_{i}^2 e^{\phi_{0}}(1+2\,\alpha'X_i),\label{sf}\\
	g_{00}=1+2\alpha'\xi_4.\label{17}
	\end{eqnarray}
	In \eqref{sf}   a series expansion for the Einstein frame scale factors $a_{i}^2 e^{\phi}$ has been introduced. In fact, all of the $\beta$-function equations are in the string frame. However, we will briefly write them in terms of Einstein frame scale factors.

	Now, it is worth to  introduce a new time coordinate $\tau$ \cite{ batakis2}
	\begin{eqnarray}\label{tau}
	d\tau= a^{-3}e^{-\phi}dt.
	\end{eqnarray}
	where $a^3=a_1a_2a_3$. In the new coordinate $\tau$, putting the $\beta$-functions \eqref{ii}, \eqref{00} and \eqref{fi} zero, we come to solve the equations. 
	The $\tilde{\beta}^{\phi}$ equation \eqref{fi} with using  \eqref{ii} and \eqref{00}  leads to the following  two equations in the zeroth and first order of $\alpha'$, respectively
	\begin{eqnarray}\label{fio}
	\phi_0''+A^2 e^{2\phi_0}=0,
	\end{eqnarray}
	\begin{eqnarray}\label{fi1}
	\phi_1''+2A^2 e^{2\phi_0}\phi_1+2A^2e^{2\phi_0}\xi_4-\phi_0'\xi_4'+\rho=0,
	\end{eqnarray}
	where the prime stands for the  derivative with respect to the $\tau$. The $\rho$ term denotes the  inhomogeneous part of the nonlinear equation and is     given  as following 
	\begin{eqnarray}\label{ro}
	\begin{split}
	\rho=-\hat{K}_{\phi}-\hat{V}_{\phi}^{(2)}-\sum(\hat{K}_{i}+\hat{V}_{i}^{(2)})+\hat{K}_{0}+\hat{V}_{0}^{(2)}.
	\end{split}
	\end{eqnarray}
	The hatted $K$ and $V$  stand for   the corresponding terms in the $\tau$ coordinate multiplied with a ${a}^6e^{2\phi_0}$ factor. The  $\hat{V}$ terms of every Bianchi model will be given in the next section. The $\hat{K}$ terms based on \eqref{kii}, \eqref{koo} and \eqref{kfii} with using \eqref{tau} are given as follows
	\begin{eqnarray}\label{ki}
	\begin{split}
	\hat{K}_{i}=&\big[((\ln a_i)''-(\ln a_i)'(\phi_0'+\sum (\ln a_j)'))^2+2(\ln a_i)'^2((\ln a_i)''-(\ln a_i)'(\phi_0'+\sum (\ln a_j)')\\
	&+(\ln a_i)'^2\sum (\ln a_k)'^2)\big]{a}^{-6}e^{-2\phi_0}+\frac{3}{8}A^4 \,{a}^{-6}e^{2\phi}
	-\frac{1}{4}A^2\big[2({f+4})(\ln a_i)'^2+\sum_{k,j\neq i}(4\left(2f+1\right)(\ln a_k)'^2\\
	&+2\left(4f+{11}\right)(\ln a_i)'(\ln a_k)'+{(5f+3)}(\ln a_j)'(\ln a_k)')\big],
	\end{split}
	\end{eqnarray}
	\begin{eqnarray}\label{k0}
	\begin{split}
	\hat{K}_{0}=&\sum\big[((\ln a_i)''-(\ln a_i)'(\phi_0'+\sum (\ln a_j)'))^2+2((\ln a_i)''-(\ln a_i)'(\phi_0'+\sum (\ln a_j)'))(\ln a_i)'^2\\
	&+(\ln a_i)'^4\big]{a}^{-6}e^{-2\phi_0}+\frac{1}{2}A^2\,{a}^{-6}[\sum(((\ln a_i)''-(\ln a_i)'(\phi_0'+\sum (\ln a_j)'))\\
	&+2\,f(\ln a_i)'^2-2\sum_{k< i }(\ln a_k)'(\ln a_i)')],
	\end{split}
	\end{eqnarray}
	\begin{eqnarray}
	\begin{split}
	\hat{K}_{\phi}=&
	(-\sum\big[((\ln a_i)''-(\ln a_i)'(\phi_0'+\sum (\ln a_j)'))^2+2H_{i}^2((\ln a_i)''-(\ln a_i)'(\phi_0'+\sum (\ln a_j)'))\\
	&+{(\ln a_i)'^4}-\sum_{i<j}(\ln a_i)'^2(\ln a_j)'^2\big]){a}^{-6}e^{-2\phi_0}-\frac{1}{4}[A^2{a}^{-6}\sum_{k< i }(\ln a_k)'(\ln a_i)'-5A^4\,{a}^{-6}e^{2\phi}].
	\end{split}
	\end{eqnarray}
	Now, the solution of \eqref{fio} gives the zeroth order of dilaton as following 
	\begin{eqnarray}\label{fioa}
	\phi_0=-\ln(\frac{A}{n}\,{\cosh{(n\tau)}}),
	\end{eqnarray}
	and the \eqref{fi1} equation has the following general solutions for the first order of dilaton 
	\footnote{An another solution for \eqref{fi1}  can be of the following form
		$$
		\phi_1= Q_1\,\tau\tanh(n\tau)-\frac{Q_1}{n}+Q_2\,\tanh(n\tau)+Q_3\ln(\sinh(n\tau))\tanh(n\tau)-\tanh(n\tau)\int  \coth(n\tau)\int \rho\tanh(n\tau)\,d\tau\,d\tau-n\tanh(n\tau)\int \xi_4 \,d\tau.
		$$}
	\begin{eqnarray}\label{fi1s}
	\begin{split}
	\phi_1=& Q_1\,\tanh(n\tau)+Q_2\,(n\tau \tanh(n\tau)-1)-n\tanh(n\tau)\int \xi_4 \,d\tau\\
	&+\frac{1}{n}[\tanh(n\tau)\int (n\tau \tanh(n\tau)-1)\,\rho\,d\tau-(n\tau \tanh(n\tau)-1)\int \tanh(n\tau)\,\rho\,d\tau],
	\end{split}
	\end{eqnarray}
	where the $n$, $Q_1$ and $Q_2$ are arbitrary constants.  The $\rho$ varies in Bianchi models and hence the $\phi_1$  will be different for any type after integrating. 
	The $\rho$ term of any Bianchi-type will be expressed in the next section. 
	
	Furthermore, in the new time coordinate $\tau$,  the  $(i,i)$ components of $\beta$-function equations of metric  \eqref{ii} can be rewritten  with using the \eqref{fio} and \eqref{fi1}  to get the following general form
	\begin{eqnarray}\label{iifinal}
	\frac{1}{2}	{(\ln a_{i}^2 e^{\phi_{0}})}''+\hat{V}_{i,0}^{(1)}+\alpha'\big(X_i''+\hat{V}_{i,1}^{(1)}-\frac{1}{2}\,	{(\ln a_{i}^2 e^{\phi_{0}})}'\xi_4'+g_i\big)=0,
	\end{eqnarray}
	where  the   $g_i$ terms are defined  as  follows 
	\begin{eqnarray}\label{gi}
	\begin{split}
	g_i=&\frac{{\rho}}{2}+\hat{K}_{i}+V_i^{(2)},~~~~~i=1,\,2,\,3.
	\end{split}
	\end{eqnarray}
	where the $\rho$ and $\hat{K}_{i}$ are given by \eqref{ro} and \eqref{ki}.  The $\hat{V}_{i,0}^{(1)}$ and $\hat{V}_{i,1}^{(1)}$ are taken to indicate the zeroth and first order of the term $\hat{V}_{i}^{(1)}$, i.e.  $\hat{V}_{i}^{(1)}=\hat{V}_{i,0}^{(1)}+\alpha' \hat{V}_{i,1}^{(1)}$, and will be obtained expansion of $\hat{V}_{i}^{(1)}$ by using the \eqref{18} and \eqref{sf}.
	The $\hat{V}_{i}^{(1)}$ terms  depend on the three radii $a_i$ and briefly are called one-loop potentials. The $\hat{V}_i^{(2)}$  terms  depend on $a_{i}$,  $\phi_0$ and their derivatives. However, the zeroth order of scale factors and dilaton are totally characterized  by the solutions of the one-loop order of $\beta$-function equations and from the two-loop equations point of view $V_i^{(2)}$   can be regarded as two-loop potentials.
	As it will be seen in the next section, in the first order of $\alpha'$ the $\hat{V}_{i,1}^{(1)}$ terms will determine the nonlinear part  of the equations while the $g_i$ will be the inhomogeneous part of differential equations. Depending on the $\hat{V}$ terms, the solutions of \eqref{iifinal}   differ in each Bianchi-type and will be discussed in the next section.

	Moreover, by rewriting the time-time component of metric $\beta$-function equation \eqref{00} in the new coordinate and using the \eqref{fio}, \eqref{fi1} and \eqref{iifinal}, the following initial value equation is obtained
	\begin{eqnarray}\label{initial}
	\begin{split}
\frac{1}{2}\big[&\sum_{i< j}(\ln a_{i}^2e^{\phi_0})'(\ln a_{j}^2e^{\phi_0})'+2\sum \hat{V}_{i,0}^{(1)}-\phi_0'^2-A^2e^{2\phi_0}\big](1+2\alpha'\xi_4)\\
	&+\alpha'\big[(\sum(\ln a_{i}^2e^{2\phi_0})'-\phi_0')\sum X_j'-\phi_0'\phi_1'+A^2e^{2\phi_0}\phi_1+\sum (\hat{V}_{i,1}^{(1)}+g_i)+\frac{\rho}{2}+\hat{K}_{0}+\hat{V}_{0}^{(2)}\big]=0.
	\end{split}
	\end{eqnarray}
	The first brackets  in this equation (the coefficient of the  $(1+\alpha'\xi_4)$) contain terms which are originated from the leading order of $\beta$-function equations.  It has been shown in \cite{batakis2} that   this is just a constraint  between some integrating constants which  appears in the solutions of dilaton and scale factors. Here, we have a generalization of this equation which its common feature between all Bianchi-type is that the first brackets and the last brackets  of it could not  vanish simultaneously. 	Thus,  we regard this equation as an initial value equation which determines the  correction of the lapse function, $\xi_4$ in \eqref{17}. In this equation the $\alpha'(-\phi_0'\phi_1'+A^2e^{2\phi_0}\phi_1)$ term in last brackets brings about a $\alpha' Q_2$ term.
	For consistency in the order of $\alpha'$, we may choose the $Q_2$  as  a constant in order $\alpha'^{-1}$ with
	$
	Q_2=\alpha'^{-1}q_2
	$
	and then, solve the equation \eqref{initial} to get
	\begin{eqnarray}\label{q2}
	q_2=-\big(\frac{1}{2}\sum_{i< j}(\ln a_{i}^2e^{\phi_0})'(\ln a_{j}^2e^{\phi_0})'+\sum V_{i,0}^{(1)}-\phi_0'^2-A^2e^{2\phi_0}\big).
	\end{eqnarray}	
	and 
	\begin{eqnarray}\label{xi4}
	\begin{split}
	\xi_4=&\frac{1}{2q_2}\,\big[(\sum(\ln a_{i}^2e^{2\phi_0})'-\phi_0')\sum X_j'+\sum (V_{i,1}^{(1)}+g_i+\frac{\rho}{2}+\hat{K}_{0}+\hat{V}_{0}^{(2)})\\
	&-\frac{\phi_0'}{n}\big((\tanh(n\tau))'\int (n\tau \tanh(n\tau)-1)\,\rho\,d\tau-(n\tau \tanh(n\tau))'\int \tanh(n\tau)\,\rho\,d\tau\big)\big].
	\end{split}
	\end{eqnarray}
	with demanding $q_2\neq 0$.
	Here, the general form of $\xi_4$ is found in terms of $\rho$ and $\hat{V}$ and the explicit answers for any case of Bianchi-types will be calculated in the next section.

	The solution of two-loop $\beta$-function equations \eqref{betaGR}-\eqref{betaBG} and \eqref{betafR} provides a conformal invariance only to the first order in $\alpha'$. In this order, the $\alpha'$-corrections of  quadratic curvature corrections type $\alpha' R^2$ are introduced by the  $\beta$-functions  to the effective action which are significant  when curvature grows. Principally, recommended by the conformal invariance condition, when $R\alpha'\gtrsim 1$, $\alpha' H^2\gtrsim 1$ and $\alpha' (\nabla \phi)^2\gtrsim 1$, all the orders of $\alpha'$-corrections series should be considered \cite{Witten}. However, in this paper, we consider only  the first order $\alpha'$-corrections. It is worth mentioning, even to this level the higher-derivative corrections are significantly applicable to remove curvature singularities \cite{cm2, Buonanno}.

	According to \eqref{H0} and dilaton solution \eqref{fioa}, 	the kinetic terms of the $B$-field and dilaton in \eqref{action} and   the string coupling $g_s=e^{-\phi}$ are given by
	\begin{eqnarray}
	H^2=A^2 \,\bar{a}^6e^{-3\phi}=\frac{\cosh(n\tau)^3}{n^3A}\bar{a}^6,\label{H}\\
	(\nabla \phi)^2=\frac{n^2(\tanh(n\tau))^2}{A(\cosh(n\tau))^2}\bar{a}^{-6},\label{fii}
	\end{eqnarray}
	\begin{eqnarray}\label{co}
	g_s=\frac{A \cosh(n\tau)}{n}.
	\end{eqnarray}
	It will be seen in the next section that  the zeroth order scale factors $\bar{a}_i$ have no dependence on the  $A$ constant. Also,  string frame curvatures of the Bianchi model with the considered metric here, are usually  proportional to  $A^{-1}$.
	Hence, if the field strength magnitude $A$   be small enough, then the curvature and kinetics of models will be in high limit so the $\alpha'$-corrections become important.
	For instance, if the $A$ is supposed to be of order $\alpha'$, the models will be in $R\alpha'\gtrsim 1$, $\alpha' H^2\gtrsim 1$ and $\alpha' (\nabla \phi)^2\gtrsim 1$ limit. 
	Furthermore, with a small $A$ constant  the string coupling $g_s$ start in the weak limit where the loop-corrections in the effective action are negligible.
	However, the $g_s$  is an increasing function of time and it may creep into the strong coupling limits at late times.

	\section{The solutions}
	In the previous section the general form of $\beta$-function  equations  in terms of dilaton and Hubble coefficients      has been presented in \eqref{fio}, \eqref{fi1}, \eqref{iifinal} and \eqref{initial}.
	The general solutions of dilaton in zeroth and first order of $\alpha'$  and the correction of lapse function have been found in \eqref{fioa}, \eqref{fi1s} and \eqref{xi4}. In this section, by finding the   Bianchi-type dependent $\hat{V}$ terms  of any Bianchi-type group examples, the solution of the \eqref{iifinal} equations   is investigated.
	\subsection{Bianchi-type $I$}
	
	The structure constants of $I$ type are zero and 	all $\hat{V}$ terms this model vanishes. So, with  $\hat{V}_i=\hat{V}_{0}^{(2)}=\hat{V}_{\phi}^{(2)}=0$ the equations of \eqref{iifinal}  give the following equations respectively in the zeroth and first order of $\alpha'$ 
	\begin{eqnarray}\label{I1}
	{(\ln a_{i}^2 e^{\phi_{0}})}''=0,
	\end{eqnarray}
	\begin{eqnarray}\label{I2}
	X_i''-\frac{1}{2}\,	{(\ln a_{i}^2 e^{\phi_{0}})}'\xi_4'+g_i=0.
	\end{eqnarray}
	The $g_i$, which have been defined in \eqref{gi},  are reduced to  $\frac{\rho}{2}+\hat{K}_{i}$ in this Bianchi-type. The solution of \eqref{I1} gives the zeroth order of the Einstein frame scale factors of \eqref{sf} as following \cite{batakis2}
	\begin{eqnarray}\label{45}
	\bar{a}_i^2=	a_{i}^2 e^{\phi_{0}}=L_ie^{p_i\tau},
	\end{eqnarray}
	where the $p_i$ and $L_i$ are arbitrary constants. Then, with using \eqref{45}, the solution of \eqref{I2} determines  general  form of the first order of the Einstein frame scale factors $X_i$, as  following
	\begin{eqnarray}\label{46}
	X_i=-\iint (\frac{\rho}{2}+\hat{K}_{i})\,d\tau\,d\tau+\frac{p_i}{2}
	\int\xi_4 \,d\tau.
	\end{eqnarray}
	Now, substituting the dilaton \eqref{fioa} and scale factors \eqref{45} in  the $\rho$ \eqref{ro}, $\hat{K}_{i}$   \eqref{ki} and $ \hat{K}_{0}$  \eqref{k0} we obtain
	\begin{eqnarray}
	\begin{split}
	\hat{K}_{i}=-&\frac {3\,n  {{\rm e}^{-\sum p_i\tau    }
	}}{2{A}\,{L}^{3} \left( \cosh \left( n\tau \right)  \right) ^{5}}
	\big( -\frac{a}{12} \left( \cosh \left( n\tau \right)  \right) ^{4
	}-\frac{b}{6}\sinh \left( n\tau \right)  \left( \cosh \left( n\tau \right) 
	\right) ^{3}+c \left( \cosh \left( n\tau \right)  \right) ^{2}\\
	&+n^3((f+\frac{5}{3})p_i+(\frac{17}{24}f+\frac{4}{6})\sum_{j\neq i}p_j )\sinh
	\left( n\tau \right) \cosh \left( n\tau \right) -\frac{n^4}{12}(23f+28)
	\big)
	\end{split}
	\end{eqnarray}
	\begin{eqnarray}
	\begin{split}
	\hat{K}_{0}=&\frac {n  {{\rm e}^{-\sum p_i\tau }}}{2{A}\,{L}^{3} \left( \cosh \left( n\tau \right) 
		\right) ^{5}}\big( d \left( \cosh \left( n\tau \right)  \right) ^{4
	}-(\frac{n}{4}  (\sum_{i\neq j}p_i^2 p_j+6 p_1p_2p_3))\sinh \left( n\tau \right)  \left( \cosh \left( n\tau \right) 
	\right) ^{3}\\
	&+e \left( \cosh \left( n\tau \right)  \right) ^{2}+{n}^{3
	} \left( { f}-1 \right) \sum p_i\sinh \left( n
	\tau \right) \cosh \left( n\tau \right) -{n}^{4} \left(3/2\, { f}-
	1 \right)  \big)
	\end{split}
	\end{eqnarray}			
	\begin{eqnarray}
	\begin{split}
	\rho=\frac{25}{4}&\frac {n{{\rm e}^{-\sum p_i\tau }}}{{A}\,{L}^{3} \left( \cosh \left( n\tau \right) 
		\right) ^{5}} \big( h \left( \cosh \left( n\tau \right)  \right) ^{4
	}-\frac{3n}{25} \sum_{j<k}(p_jp_k+n^2) \sum p_j
	\sinh \left( n\tau \right)  \left( \cosh \left( n\tau \right) 
	\right) ^{3}
	\\
	&	+m\sinh \left( n\tau \right)  \left( \cosh \left( n\tau \right) 
	\right) ^{2}+{n}^{3} \sum p_i  \left( { f}+{ {11}/{10}} \right) \sinh \left( n\tau \right) \cosh
	\left( n\tau \right) -\frac{3}{2} \left( { {47}/{25}}+{ 3f}/2 \right) {
		n}^{4} \big),
	\end{split}
	\end{eqnarray}	
	where $L^3=L_1L_2L_3$ and
	\begin{eqnarray}
	a&=&n^4+n^2( \sum p_j)^2 +\sum_{k,j\neq i}  (p_j^2+\frac{1}{2}(p_i^2-n^2)p_j p_k)\\
	b&=&2  n (n^2\sum p_j +\frac{1}{2} (n^2+\sum_{j,k\neq i} (p_j p_i+2 p_j^2+2 p_j p_k)) p_i),\\
	c&=&\frac{1}{12} n^2 ((23 f+24) n^2+(2 f+4) p_i^2+\sum_{j,k\neq i} ((4 f+1) p_j p_i+(4 f+3) p_j^2+ (5 f+4) p_j p_k/2))\\
	d&=&(\frac{1}{2}((\sum p_i)^2+\sum p_i^2)) n^2+\sum_{j<i}p_i^2 p_j^2 - p_2 p_3  p_1\sum p_i,\\
	e&=& ((6 f-9) n^2+3 \sum  p_i^2 f+\sum_{i<j}p_ip_j) n^2/4\\
	h&=&\frac{-9}{10}( N^4+\frac{2n^2}{3}(\frac{1}{2}(3(\sum p_i)^2+\sum p_i^2)) +\frac{2}{3}( p_2 p_3 p_1^2+ p_2 p_3 p_1(p_2+p_3) )),\\
	m&=&(((f+\frac{9}{10})n^2+(\frac{4}{25}(f+1))\sum p_i^2+\frac{13}{75}(f+\frac{5}{2}) \sum_{j < i}p_ip_j))3n^2/2.
	\end{eqnarray}
	The $q_2$ constant in \eqref{q2} in this Bianchi-type is given by   
	\begin{eqnarray}
	\begin{split}
	q_2=	(n^2-\sum_{j<k}p_jp_k)/2.
	\end{split}
	\end{eqnarray}
	Now, the appeared integrals in the $\phi_1$ \eqref{fi1s},  $\xi_4$ \eqref{xi4} and $X_i$ \eqref{46} can be calculated for finding  the metric and dilaton corrections.
	For choosing any value for $p_i$s, the $q_2\neq 0$ condition should be considered. 
	\subsubsection{Some Physical Properties of an isotropic example in the Bianchi-type $I$}	
	
	In Appendix C, a set of answers for metric and dilaton corrections    is given with an isotropic parametrization $p_1=p_2=p_3=1/3$  and $n=1$ through  \eqref{ix1}-\eqref{53}.
	Regarding  the kinetic of $B$-field and dilaton in \eqref{H} and \eqref{fii} and the string Ricci scalar of this example which is given by
	\begin{eqnarray}
	R={\frac {{{\rm e}^{-\tau}} \left( 8\, \left( \cosh \left( \tau
			\right)  \right) ^{2}+9 \right) }{6A\,{L}^{3} \left( \cosh \left( 
			\tau \right)  \right) ^{3}}},
	\end{eqnarray}
	it is clear that if the $A$ supposed to be very small (of order $\alpha'$), the model will be in the high curvature and high kinetics limits and  the inclusion of the $\alpha'$-correction becomes necessary. The curvature of this model is  decreasing  so the corrections become significant in the early times near $\tau=0$.
	Here, we come to investigate some properties of this specific example and consider some physical quantities such as  
	Einstein frame mean Hubble parameter  and deceleration parameter which are given by
	\begin{eqnarray}\label{hbar}
	\bar{H}= \frac{\dot{\bar{a}}}{\bar{a}},~~~q=-\frac{\bar{a}\ddot{\bar{a}}}{\dot{\bar{a}}^2}.
	\end{eqnarray}
	Here the dot stands for derivation with respect to  Einstein frame time, which is called cosmic time, such that $\dot{\bar{a}}={d \,\bar{a}}/{d\,\bar{t}}$ and $\ddot{\bar{a}}={d^2 \,\bar{a}}/{d\,\bar{t}^2}$.
	The answers have been given in terms of $\tau$. Now, there are two equivalent ways for investigating the behavior of $\dot{\bar{a}}$ and $\ddot{\bar{a}}$; first, by rewriting the solutions in $\bar{t}$ terms by finding the relation between $\bar{t}$ and $\tau$  using \eqref{tau} and \eqref{3}; second, just  rewriting the derivatives of $\dot{\bar{a}}$ and $\ddot{\bar{a}}$ in terms of $\tau$ derivatives.
	The former way  is applicable here but  it is not generally suitable  for other models, where the latter one  can be used in every model. Here,  following the first procedure, we consider the  Einstein frame metric which according to \eqref{3}, \eqref{metric} and  \eqref{d}  takes the following form  in this isotropic example
	\begin{eqnarray}	
	ds^2=-N(\bar{t})\,d\bar{t}^2+\bar{a}_1^2 (1+2\,X_i) \delta_{\alpha\beta}\,dx^{\alpha}\,dx^{\beta}
	\end{eqnarray}
	where $\bar{a}_1^2=Le^{\tau/3}$.	Accordingly, with $N=(1+\alpha'(\sum X_1+\xi_4))$, the cosmic time $\bar{t}$ and the $\tau$, which has been introduced by \eqref{tau}, would have the following relation
	\begin{eqnarray}	\label{tautot}
	\bar{t}+\bar{t}_0=\int \bar{a}_{1}^3\,d\tau,
	\end{eqnarray}
	where the $\bar{t}_0$ is  used for coinciding the beginning of the two time coordinates.
	Fortunately, in the Bianchi-type $I$ model with the scale factors given by  \eqref{45} it is easy to find  $\tau$ in terms of $\bar{t}$   as following		
	\begin{eqnarray}\label{ti}	
	\tau=\ln  \left( (1+{L^{-3/2}}\bar{t}) \right) ,
	\end{eqnarray}
	where $\bar{t}_0$ is fixed as $\frac {2\,L^{3/2}}{\sum p_i}$.
	Now,  substituting \eqref{ti} \eqref{45}, \eqref{fioa}, \eqref{ix1}, \eqref{48} and \eqref{53}, the $\alpha'$-corrected dilaton and  Einstein frame scale factors  are  found in terms of cosmic time and their behavior could be investigated.
	
	In this example, if the zeroth order is considered only, the solution \eqref{45} will have $\dot{\bar{a}}>0 $ and  $\ddot{\bar{a}}>0$ and consequently describe a decelerated  expanding universe with no inflation. Furthermore, in this order the strong energy condition is satisfied and there is an initial singularity \cite{batakis2}.
	
	The behavior of $\dot{\bar{a}}$ and $\ddot{\bar{a}}$ of the corrected scale factors are being investigated, with choosing
	$A=\alpha', L=1.1$, $l_1=1/11, c_1=-1, \varphi_0=1$ and $ Q_1=11$ in the Gauss-bonnet scheme and $A=\alpha', L=1.1$, $l_1=0.5, c_1=1.5, \varphi_0=2.2$ and $ Q_1=11$ for the $R^2$ scheme.\footnote{
		Regarding the signature of metric to be adopted $(-1,1,1,1)$ after adding corrections, the constants have been chosen and these set of constants are not unique.}
	In the high curvature limit (which is until about $\bar{t}=0.5$ in these parameterizations) we have $\dot{\bar{a}}>0$, but the sign of $\ddot{\bar{a}}$ changes such that it is negative near $\bar{t}=0$ and then, before leaving the high curvature limit,  becomes positive. So, until the corrections are significant, the $\alpha'$-corrected answer describes an expanding universe which has a decelerated phase in the early times and then becomes accelerated. After leaving the high curvature limit, the behaviors of $\ddot{\bar{a}}$ and $\dot{\bar{a}}$ turn into the leading order behavior and become negative.

	Also, obeying the strong energy condition  in the case that the higher order correction terms contribute to the right hand of Einstein frame field equations, may be considered. For the Bianchi-type models the strong energy condition
	requires $\dot{k}+\frac{1}{3}k^2<0$, where $k$ is the  extrinsic curvature \cite{wald}. With the considered diagonal metric in this paper, we have $k=\sum \dot{\bar{a}}_i/\bar{a}_i$ and the strong energy condition leads to $\ddot{a}<0$.
	In the accelerated phase of our considered example, with  $\ddot{a}>0$, there is a violation of the strong energy condition which is a necessary (but not sufficient) condition of avoiding  singularity \cite{Hawking}. So, the singularity of the leading order  has been cured in some regions in the early time by introducing the higher order correction in the action. This period in the Bianchi-type $I$ model may be regarded as a cosmological inflationary phase \cite{inf}.

	Furthermore,  the expansion is rapid at first but then slow down without any singularity in $\bar{H}$.  Hence, if it is considered as an inflationary phase followed by a standard phase, it will be free of graceful exit problem. This kind of nonsingular behavior (no  graceful exit problem) of the Bianchi-type $I$ model with superstring-motivated  Gauss-Bonnet term with dilaton and modules fields  have already been  investigated  in \cite{KawaiI}.

	\subsection{Bianchi-type $II$}
	In Bianchi-type $II$ the   $\hat{V}$   terms are given by
	\begin{eqnarray}
	\begin{split}
	\hat{V}_{1}^{(1)}=-	\hat{V}_{2}^{(1)}=-	\hat{V}_{3}^{(1)}=\frac{a_1^4e^{2\phi_0}}{2}(1+2\alpha'(2X_1+\xi_4)),
	\end{split}
	\end{eqnarray}
	\begin{eqnarray}
	\hat{V}_1^{(2)}&=&\frac{a_{1}^2}{8a_{2}^2a_{3}^2}\big(a_{1}^4e^{2\phi_0}-4(\sum (\ln a_{i})'^2-2(\ln a_{1})'(\ln a_{2}^2a_{3}^2a_{1}^{-1})'+(\ln a_{2}a_{3})')+3(f-1)\,A^2e^{2\phi_0}\big),\\
	\hat{V}_2^{(2)}&=&\frac{a_{1}^2}{8a_{2}^2a_{3}^2}\big(5a_{1}^4e^{2\phi_0}-4(\sum (\ln a_{i})'^2+4(\ln a_{2})'(\ln\frac{a_{1}}{a_{3}})'+(\ln a_{1})'(\ln \frac{a_{1}^2}{a_{3}^3})')-(f-5)\,A^2e^{2\phi_0}\big),\\
	\hat{V}_3^{(2)}&=&\frac{a_{1}^2}{8a_{2}^2a_{3}^2}\big(5a_{1}^4e^{2\phi_0}-4(\sum (\ln a_{i})'^2+4(\ln a_{3})'(\ln \frac{a_{1}}{a_{2}})'+(\ln a_{1})'(\ln \frac{a_{1}^2}{a_{2}^3})')-(f-5)\,A^2e^{2\phi_0}\big),\\
	\hat{V}_{0}^{(2)}&=&\frac{a_{1}^2}{2a_{2}^2a_{3}^2}(\sum (\ln a_{i})'^2+(\ln a_{i})'(\ln\frac{a_{1}^2}{a_{2}^3a_{3}^3})'+(\ln a_{2})'(\ln a_{3})'),\\
	\hat{V}_{\phi}&=&\frac{a_{1}^2}{2a_{2}^2a_{3}^2}\big(-2\sum (\ln a_{i})'-4(\ln a_{1})'+7(\ln a_{1})'(\ln a_{2}a_{3})'-5(\ln a_{2})'(\ln a_{3})'+\frac{(11a_{1}^4+A^2)e^{2\phi_0}}{16}\big).
	\end{eqnarray}
	With these potentials, the $\beta$-function equations \eqref{iifinal} give the following nonlinear equations  in the zeroth order of $\alpha'$ \cite{batakis2}
	\begin{eqnarray}\label{II1}
	{(\ln a_{1}^2 e^{\phi_{0}})}''+a_{1}^4 e^{2\phi_{0}}=0,\label{62}\\
	{(\ln a_{2}^2 e^{\phi_{0}})}''-a_{1}^4 e^{2\phi_{0}}=0,\label{63}\\
	{(\ln a_{3}^2 e^{\phi_{0}})}''-a_{1}^4 e^{2\phi_{0}}=0,\label{64}
	\end{eqnarray}
	and in the first order of $\alpha'$ give
	\begin{eqnarray}
	X_1''+a_{1}^4 e^{2\phi_{0}}(2X_1+\xi_4)-\frac{1}{2}\,	{(\ln a_{1}^2 e^{\phi_{0}})}'\xi_4'+g_1=0,\label{65}\\
	X_2''-a_{1}^4 e^{2\phi_{0}}(2X_1+\xi_4)-\frac{1}{2}\,	{(\ln a_{2}^2 e^{\phi_{0}})}'\xi_4'+g_2=0,\label{66}\\
	X_3''-a_{1}^4 e^{2\phi_{0}}(2X_1+\xi_4)-\frac{1}{2}\,	{(\ln a_{3}^2 e^{\phi_{0}})}'\xi_4'+g_3=0.\label{67}
	\end{eqnarray}
	The general solutions of \eqref{62}-\eqref{64}  are given by \cite{batakis2}
	\begin{eqnarray}
	a_{1}^2 e^{\phi_{0}}=\frac{p_1}{\cosh(p_1\tau)},~~~~~\label{68}\\
	a_{2}^2 e^{\phi_{0}}=\cosh(p_1\tau)e^{p_2\tau},\label{69}\\
	a_{3}^2 e^{\phi_{0}}=\cosh(p_1\tau)e^{p_3\tau}.\label{70}
	\end{eqnarray}
	The general solutions of  \eqref{65}-\eqref{67} equations are found as follows
	\begin{eqnarray}\label{IIx1}
	\begin{split}
	X_1=& Q_3\,\tanh(p_1\tau)+Q_4\,(p_1\tau \tanh(p_1\tau)-1)-\frac{p_1}{2}\tanh(p_1\tau)\int \xi_4 \,d\tau\\
	&+\frac{1}{p_1}(\tanh(p_1\tau)\int (p_1\tau \tanh(p_1\tau)-1)\,g_1\,d\tau-(p_1\tau \tanh(p_1\tau)-1)\int \tanh(p_1\tau)\,g_1\,d\tau),
	\end{split}
	\end{eqnarray}
	\begin{eqnarray}\label{IIx2}
	X_2=-X_1-\iint(g_1+g_2)\,d\tau\,d\tau +\frac{p_2}{2}\int \xi_4 \,d\tau,
	\end{eqnarray}
	\begin{eqnarray}\label{IIx3}
	X_3=-X_1-\iint(g_1+g_3)\,d\tau\,d\tau +\frac{p_3}{2}\int \xi_4 \,d\tau,
	\end{eqnarray}
	where the $Q_3$ and $Q_4$ are integrating  constants and the  $g_i$ has been introduced in \eqref{gi}. 
	With the dilaton \eqref{fioa} and scale factors given by \eqref{68}-\eqref{70}, the explicit forms of $\rho$ introduced in \eqref{ro} and $g_i$  can be derived and then the integrals of  $X_i$ and $\phi_1$ \eqref{fi1s} can be computed. In this Bianchi-type the $q_2$ constant of \eqref{q2} is given by
	\begin{eqnarray}
	\begin{split}
	q_2=	(n^2+p_1^2-p_2p_3)/2.
	\end{split}
	\end{eqnarray}
	Keeping in mind the condition $q_2\neq 0$, one could choose any value for the $p_i$ and $n$. 
	For example, with the special choice of parameterization as  $p_1=p_2=p_3=n=1$ the  $g_i$ and $\rho$ are given as following\footnote{Here and hereafter, avoiding dense mathematical	formulae, we will choose a  set of values for the constants of leading order scale factors and present a form of $g_i$ and $\rho$ as an example. Final forms of metric and dilaton corrections, obtained after performing their integrals, of these example can be found in the appendix $C$.}
	\begin{eqnarray}
	\begin{split}
	g_1=&\frac {{{\rm e}^{-2\,\tau}} }{16{A}
		\left( \cosh \left( \tau \right)  \right) ^{6}}\big(-153\, \left( \cosh \left( \tau \right) 
	\right) ^{4}-144\, \left( \cosh \left( \tau \right)  \right) ^{3}
	\sinh \left( \tau \right) + \left( 110\,f+561 \right)  \left( 
	\cosh \left( \tau \right)  \right) ^{2}\\
	&+ \left( 88\,f+444 \right) 
	\sinh \left( \tau \right) \cosh \left( \tau \right) -78\,f-470
	\big),
	\end{split}
	\end{eqnarray}
	\begin{eqnarray}
	\begin{split}
	g_2=&g_3=\frac {{{\rm e}^{-2\,\tau}}  }{ {32A}\left( \cosh \left( \tau
		\right)  \right) ^{6}}\big( -61\, \left( \cosh \left( \tau \right)  \right) 
	^{4}-56\, \left( \cosh \left( \tau \right)  \right) ^{3}\sinh \left( 
	\tau \right) + \left( 170\,f+285 \right)  \left( \cosh \left( \tau
	\right)  \right) ^{2}\\
	&+ \left( 136\,f+236 \right) \sinh \left( 
	\tau \right) \cosh \left( \tau \right) -142\,f-246
	\big)
	,
	\end{split}
	\end{eqnarray}
	\begin{eqnarray}
	\begin{split}
	\rho=&\frac {{{\rm e}^{-2\,\tau}}}{16A
		\left( \cosh \left( \tau \right)  \right) ^{6}} \big( -153\, \left( \cosh \left( \tau \right) 
	\right) ^{4}-144\, \left( \cosh \left( \tau \right)  \right) ^{3}
	\sinh \left( \tau \right) + \left( 370\,f+761 \right)  \left( 
	\cosh \left( \tau \right)  \right) ^{2}\\
	&+ \left( 296\,f+604
	\right) \sinh \left( \tau \right) \cosh \left( \tau \right) -298\,f-634\big).
	\end{split}
	\end{eqnarray}
	Now, using the \eqref{fi1s}, \eqref{xi4} and \eqref{IIx1}-\eqref{IIx3}, the $\alpha'$-corrections of metric and dilaton  in both  scheme can be found in this special choice of parameters. The detiled answers are given in Appendix C through \eqref{iix1}-\eqref{53ii}.
	
	\subsection{Bianchi-type III}
	Here, as a general feature of the Bianchi-type of class $\cal{B}$, the $\beta$-function equations are subject to a constraint equation where in this case the $(0,3)$ component of $\beta$-function equations of metric imposes  it as following
	\begin{eqnarray}
	(\ln{\frac{a_{1}}{a_{3}}})'[1+\alpha'(a_0^{-6} e^{-2\phi}({(\ln a_{1} )}''- (\ln a_{1})'(\ln a_{2})'-\phi_0'(\ln a_{1})') -{a_{3}}^{-2})-\frac{3A^2a_0^{-6}}{4})]+\alpha'(X_1'-X_3')=0,
	\end{eqnarray}
	which essentially imposes the conditions of $a_{1}=a_{3}$ and $X_1=X_3$. The $\beta$-function equations of $B$-field \eqref{betaBR} and \eqref{betaBG} vanish with the choice \eqref{H0} in the zeroth order of $\alpha'$. But at first order of $\alpha'$,
	despite of the $\cal{A}$ class, in the $\cal{B}$ class models they do not vanish. However, they   will be vanished automatically by imposing the conditions that resulted form the constraints equations. 
	
	The Bianchi-type dependent terms  $\hat{V}$  of $III$ model are given as following
	\begin{eqnarray}
	\hat{V}_1^{(1)}=\hat{V}_3^{(1)}=a_{1}^2a_{2}^2 e^{2\phi_{0}}(1+2\alpha'(X_1+X_2+\xi_4)),~~~~~\hat{V}_2^{(1)}=0,
	\end{eqnarray}	
	\begin{eqnarray}
	\hat{V}_1^{(2)}&=&\hat{V}_3^{(2)}=a_{3}^{-2}\big(-2(( \ln a_{1} )'^2+(\ln a_{3}) '^2-{(\ln a_{1}) }'{(\ln a_{3}) }')+a_{1}^2a_{2}^2e^{2\phi}+{3}A^2e^{2\phi}/2\big),\\
	\hat{V}_2^{(2)}&=&{(f+1)}A^2e^{2\phi}/2,\\
	\hat{V}_{0}^{(2)}&=&a_{3}^{-2}((\ln a_{1} )'-(\ln a_{3} )')^2,\\
	\hat{V}_{\phi}~~&=&a_{3}^{-2}(-2((\ln a_{1} )'^2+(\ln a_{3} )'^2-(\ln a_{1}) '(\ln a_{3}) ')+a_{1}^2a_{2}^2e^{2\phi_0}+A^2e^{2\phi_0}/4).
	\end{eqnarray}			
	With this form of potentials, the $\beta$-function equations \eqref{iifinal} lead to the following equations  at the zeroth order \cite{ batakis2}
	\begin{eqnarray}
	{(\ln a_{1}^2 e^{\phi_{0}})}''-2a_{1}^2a_{2}^2 e^{2\phi_{0}}=0,\label{31}\\
	{(\ln a_{2}^2 e^{\phi_{0}})}''=0,\\
	{(\ln a_{3}^2 e^{\phi_{0}})}''-2a_{1}^2a_{2}^2 e^{2\phi_{0}}=0\label{33},
	\end{eqnarray}
	and at the first order of $\alpha'$ give
	\begin{eqnarray}
	X_1''-2a_{1}^2a_{2}^2 e^{2\phi_{0}}(X_1+X_2+\xi_4)-\frac{1}{2}\,	{(\ln a_{1}^2 e^{\phi_{0}})}'\xi_4'+g_1=0,\label{96}\\
	X_2''-\frac{1}{2}\,	{(\ln a_{2}^2 e^{\phi_{0}})}'\xi_4'+g_2=0,\label{97}\\
	X_3''-X_1''=0\label{98}.
	\end{eqnarray}
	Solutions of \eqref{31}-\eqref{33} are given by \cite{ batakis2}
	\begin{eqnarray}
	a_{1}^2 e^{\phi_{0}}=a_{3}^2 e^{\phi_{0}}&=&\frac{p_1e^{p_2\tau}}{\sinh(p_1\tau)},\\
	a_{2}^2 e^{\phi_{0}}&=&e^{-p_2\tau}.
	\end{eqnarray}
	The equations of \eqref{96}-\eqref{98} admit the following  solutions
	\begin{eqnarray}\label{94}
	\begin{split}
	X_1=X_3=&-X_2+{ Q_3}\,{\rm coth} (p_1\tau)+{Q_4}\, ( p_1\,\tau\,{\rm coth} (p_1\tau)-1 ) +(\frac{p_2}{2}-p_1{\rm coth} (p_1\tau))\int \xi_4\, d\tau\\
	&+\frac{1}{p_1}\bigg({\rm coth} (p_1\tau)\int ( p_1\,\tau\,{\rm coth} (p_1\tau)-1 ) (g_1+g_2)\,d\tau\\
	&~~~~~~~~~~~~~~~~~~~~~~~~~~~~~~~~~~~~~-( p_1\,\tau\,{\rm coth} (p_1\tau)-1 )\int {\rm coth} (p_1\tau) (g_1+g_2)\,d\tau\bigg),
	\end{split}
	\end{eqnarray}
	\begin{eqnarray}\label{95}
	X_2=-\frac{p_2}{2}\int \xi_4\, d\tau-\iint g_2\,d\tau\,d\tau,
	\end{eqnarray}
	where the $Q_3$ and $Q_4$ are constant. Here the $q_2$ constant \eqref{q2} is given by
	\begin{eqnarray}
	q_2=(p_2^2+n^2-4\,p_1^2)/2.
	\end{eqnarray}
	As the other Bianchi-types, the $g_i$ terms in the answer of metric corrections \eqref{94}, \eqref{95} and $\rho$ in dilaton \eqref{fi1s} can be derived by substituting the zeroth order solution of scale factors and dilaton. For example, with $p_1=p_2=n=1$ they are given by
	\begin{eqnarray}
	\begin{split}
	g_2=&
	\frac { \left( \sinh \left( \tau \right)  \right) ^{2}{{\rm e}^
			{-\tau}}}{16A \left( \cosh \left( \tau \right)  \right) ^{5}}
	\big( -144\, \left( \cosh \left( \tau \right)  \right) ^{6}+ \left( 
	26\,f+73 \right)  \left( \cosh \left( \tau \right)  \right) ^{4}+
	\left( 9\,f-7 \right)  \left( \cosh \left( \tau \right)  \right) 
	^{2}\\
	&+\sinh \left( \tau \right)  ( -26\, ( f+15/26) \cosh \left( \tau \right) +144\, \left( \cosh \left( \tau
	\right)  \right) ^{5}-26\, \left( f+1/26 \right)  \left( \cosh
	\left( \tau \right)  \right) ^{3} ) +29\,f+38 \big) 
	\end{split}
	\end{eqnarray}
	\begin{eqnarray}
	\begin{split}
	g_2=
	&\frac { \left( \sinh \left( \tau \right)  \right) ^{2}{{\rm e}^{-
				\tau}}}{8 \left( \cosh \left( \tau \right)  \right) ^{5}}
	\big( -{ {40\, \left( \cosh \left( \tau \right)  \right) ^{6}}}+ ( 9f+73/2 )  \left( \cosh \left( \tau
	\right)  \right) ^{4}+1/18\, \left( f-55 \right)  \left( \cosh
	\left( \tau \right)  \right) ^{2}\\
	&+\sinh \left( \tau \right) 
	( -(6 f+{31} ) \cosh \left( \tau
	\right) +{ {40 \left( \cosh \left( \tau \right)  \right) ^{5}
	}}- 3( 3\,f+{11}/{2} )  \left( \cosh \left( \tau \right) 
	\right) ^{3} ) +{ {29\,f}/{2}}+ {19}
	\big) 
	,
	\end{split}
	\end{eqnarray}	
	\begin{eqnarray}
	\begin{split}
	\rho=&\frac {{{\rm e}^{-\tau}}}{8A\, \left( \cosh \left( \tau
		\right)  \right) ^{5}} \big(  \left( \sinh \left( \tau \right) 
	\right) ^{3} ( -46\, ( f+{\frac {17}{46}} ) 
	\left( \cosh \left( \tau \right)  \right) ^{3}+168\, \left( \cosh
	\left( \tau \right)  \right) ^{5}-50\, ( f+{\frac {11}{10}}
	) \cosh \left( \tau \right)  ) \\
	&+ \left( \sinh \left( \tau
	\right)  \right) ^{2} \left( 75\,f+94+ \left( 23\,f-27
	\right)  \left( \cosh \left( \tau \right)  \right) ^{2}+ \left( 46\,f
	_{1}+101 \right)  \left( \cosh \left( \tau \right)  \right) ^{4}-168\,
	\left( \cosh \left( \tau \right)  \right) ^{6} \right)  \big) 
	,
	\end{split}
	\end{eqnarray}	
	The explicit answers  for $X_i$, $\xi_i$ and $\phi_1$ with this parameterization can be found in appendix $C$ through \eqref{iiix1}-\eqref{53iii}.	
	
	%%%%%%%%%%%%%%%%%%%%%%%%%%%%%%%%%%%%%%%%%%%%%%%%%%%%%%%%%%%%%%%%%%%%%%%5
	\subsection{Bianchi-type $IV$}
	For this type the constraint equation is
	\begin{eqnarray}
	\frac{a_{3}^2}{a_{1}^2}	+\alpha'(2\,(X_3-X_1)+\frac{4(\ln a_{1} )'^2-(\ln{\frac{a_{2}}{a_{3}^3}})'(\ln{\frac{a_{2}a_{3}}{a_{1}^2}})'-A^2}{a_{1}^4a_{2}^2}-\frac{2a_{3}^2}{a_{1}^4}-\frac{a_{3}^4}{2a_{1}^4a_{2}^2})=0,
	\end{eqnarray}
	where in the zeroth order of $\alpha'$ leads to the following condition
	\begin{eqnarray}
	\frac{a_{3}^2}{a_{1}^2}=0.
	\end{eqnarray}
	So,  all solution in zeroth order of $\alpha'$ is singular and because of this inconsistency in one-loop order, we abandon its two-loop calculations. 
	%%%%%%%%%%%%%%%%%%%%%%%%%%%%%%%%%%%%%%%%%%%%%%%%%%%%%%%%%%%%%%%%%%%%%%%%%%
	
	\subsection{Bianchi-type $V$}
	Bianchi-type $V$  belongs to the   $\cal{B}$ class. Here the $(0,1)$ component of $\beta$-function of the metric \eqref{betaGR} is a constraint equation and gives
	\begin{eqnarray}\label{vv}
	\begin{split}
	(\ln{\frac{a_{1}^2}{a_{2}a_{3}}})'&+\alpha'[2X_1'-X_1'-X_3'+(\ln{\frac{a_{1}^2}{a_{2}a_{3}}})'(-\frac{3A^2a_0^{-6}}{4}-\frac{1}{a_{1}^2}+a_0^{-6} e^{-2\phi}(\ln a_{2}a_{3}) ')\\
	&+a_0^{-6} e^{-2\phi}((\ln a_{2}) ''(\ln \frac{a_{3} }{a_{1} })'+(\ln a_{3}) ''(\ln \frac{a_{2} }{a_{1} })'+\phi'((\ln a_{2} )'(\ln \frac{a_{3} }{a_{1} })'+(\ln a_{3} )'(\ln \frac{a_{2} }{a_{1} })')]=0.
	\end{split}
	\end{eqnarray}
	This equation  in zeroth order of $\alpha'$ implies
	\begin{eqnarray}
	a_{1}^2=a_{2}a_{3},
	\end{eqnarray}
	and its first order of $\alpha'$  will be discuses afterward. 
	
	The potential terms in this Bianchi-type are given by
	\begin{eqnarray}
	\hat{V}_1^{(1)}=\hat{V}_2^{(1)}= \hat{V}_3^{(1)}=-2a_2^2a_3^2e^{2\phi_0}(1+2\alpha'(X_2+X_3+\xi_4)),
	\end{eqnarray}
	\begin{eqnarray}
	\hat{V}_1^{(2)}&=&a_{1}^{-2}\big(2\,(a_{2}a_{3}e^{\phi_0})^2-2(\ln a_{1})'^2-(\ln a_{2})'^2-(\ln a_{3})'^2+(f+7)A^2e^{2\phi_0}/2\big),\\
	\hat{V}_2^{(2)}&=&a_{1}^{-2}\big(2\,(a_{2}a_{3}e^{\phi_0})^2-2(\ln a_{1})'^2-(\ln a_{2})'^2+(\ln a_{2})'(\ln \frac{a_{1}}{a_{3}})'+(f+7)A^2e^{2\phi_0}/2\big),\\
	\hat{V}_3^{(2)}&=&a_{1}^{-2}\big(2\,(a_{2}a_{3}e^{\phi_0})^2-2(\ln a_{1})'^2-(\ln a_{3})'^2+(\ln a_{3})'(\ln \frac{a_{1}}{a_{2}})'+(f+7)A^2e^{2\phi_0}/2\big),\\
	\hat{V}_{0}^{(2)}&=&-a_{1}^{-2}{(2(\ln a_{1} )'^2+(\ln a_{0} )'^2+(\ln a_{3} )'^2-(\ln a_{1} )'(\ln a_{2}a_{3}) ')},\\
	\hat{V}_{\phi}&=&a_{1}^{-2}\big(-2(\ln a_{1} ')^2-(\ln a_{2} ')^2-(\ln a_{3} ')^2+(\ln a_{1}) '(\ln a_{2}a_{3}) '+(\ln a_{2} )'(\ln a_{3}) '\\
	&~&~~~~~~~~~~~~~~~~~~~~~~+3a_{3}^2a_{2}^2e^{2\phi_0}+3A^2e^{2\phi_0}/4\big).
	\end{eqnarray}				
	With this form of potentials, the $\beta$-function equations \eqref{iifinal} give the following equations  in the zeroth order of $\alpha'$\cite{batakis2}
	\begin{eqnarray}
	{(\ln a_{1}^2 e^{\phi_{0}})}''-4a_{2}^2a_{3}^2 e^{2\phi_{0}}=0,\label{119}\label{v1}\\
	{(\ln a_{2}^2 e^{\phi_{0}})}''-4a_{2}^2a_{3}^2 e^{2\phi_{0}}=0,\label{12}\\
	{(\ln a_{3}^2 e^{\phi_{0}})}''-4a_{2}^2a_{3}^2 e^{2\phi_{0}}=0,\label{121}
	\end{eqnarray}
	and in  the first order of $\alpha'$ give
	\begin{eqnarray}
	X_1''-4\,a_{2}^2a_{3}^2 e^{2\phi_{0}}(X_2+X_3+\xi_4)-\frac{1}{2}\,	{(\ln a_{1}^2 e^{\phi_{0}})}'\xi_4'+g_1=0,\label{122}\\
	X_2''-4\,a_{2}^2a_{3}^2 e^{2\phi_{0}}(X_2+X_3+\xi_4)-\frac{1}{2}\,	{(\ln a_{2}^2 e^{\phi_{0}})}'\xi_4'+g_2=0,\label{123}\\
	X_3''-4\,a_{2}^2a_{3}^2 e^{2\phi_{0}}(X_2+X_3+\xi_4)-\frac{1}{2}\,	{(\ln a_{3}^2 e^{\phi_{0}})}'\xi_4'+g_3=0.\label{124}
	\end{eqnarray}
	The general solutions of \eqref{119}-\eqref{121} are as following \cite{ batakis2}
	\begin{eqnarray}
	a_{1}^2 e^{\phi_{0}}&=&\frac{p}{2\,\sinh(p\tau)},\\
	a_{2}^2 e^{\phi_{0}}&=&\frac{p}{2\,\sinh(p\tau)}\,e^{q\tau},\\
	a_{3}^2 e^{\phi_{0}}&=&\frac{p}{2\,\sinh(p\tau)}\,e^{-q\tau}.
	\end{eqnarray}
	Now, the $g_i$ terms \eqref{gi} can be derived.  The general solutions admitted by\eqref{122}-\eqref{124} are as following
	\begin{eqnarray}\label{vx1}
	\begin{split}
	X_1&=X_2  -\iint(g_1-g_2)d\tau d\tau-\frac{q}{2}\int \xi_4\, d\tau,
	\end{split}
	\end{eqnarray}
	\begin{eqnarray}\label{vx2}
	\begin{split}
	X_2&={ Q_3}\,{\rm coth} (p\tau)+{Q_4}\, ( p\,\tau\,{\rm coth} (p\tau)-1 ) -\frac{1}{2}(p\,{\rm coth} (p\tau)+q)\int \xi_4\, d\tau\\
	&+\frac{1}{2p}\bigg({\rm coth} (p\tau)\int ( p\,\tau\,{\rm coth} (p\tau)-1 ) (g_2+g_3)\,d\tau\\
	&~~~~~~~~~~-( p\,\tau\,{\rm coth} (p\tau)-1 )\int {\rm coth} (p\tau) (g_2+g_3)\,d\tau\bigg)+\frac{1}{2}\iint(g_3-g_2)d\tau d\tau,
	\end{split}
	\end{eqnarray}
	\begin{eqnarray}\label{vx3}
	\begin{split}
	X_3&=X_2-\iint(g_3-g_2)d\tau d\tau-q\int \xi_4\, d\tau.
	\end{split}
	\end{eqnarray}
	Here, the $q_2$ \eqref{q2} is given by
	\begin{eqnarray}
	q_2=(n^2-3p^2+q^2)/2
	\end{eqnarray}
	With the above solutions,  the first order of $\alpha'$ of the constraint equation \eqref{vv}   imposes $q=0$. This leads to $g_1=g_2=g_3$ and therefore, with ignoring integrating constants, $X_1=X_2=X_3$. 
	Hence, the Bianchi-type $V$ model which was allowed to be anisotropic from the leading order $\beta$-function solution point of view, is constrained to be  isotropic in the solution of two-loop $\beta$-function equations. However, an anisotropy may appear because of the integrating constants of  \eqref{vx1}-\eqref{vx3}, but it can be removed by a redefinition or a shift of time.
	Generally, with $q=0$ the $g_i$ \eqref{gi} and $\rho$ \eqref{ro} are given by
	\begin{eqnarray}
	\begin{split}
	g_1&=g_2=g_3=\frac {  \sinh \left( p\tau
		\right) n}{4A \left( \cosh \left( n\tau \right)  \right) ^{5
		}{p}^{3}}\bigg( 4\,\sinh \left( p\tau \right) \sinh \left( 
	n\tau \right) n\cosh \left( p\tau \right)  \left( 4\,{n}^{2}+19\,{p}^{
		2} \right) p \left( \cosh \left( n\tau \right)  \right) ^{3}\\
	&+\left(  \left( -5\,{n}^{4}-58\,{n}^{2}{p}^{2}-33
	\,{p}^{4} \right)  \left( \cosh \left( p\tau \right)  \right) ^{2}+5\,
	{n}^{4}+10\,{n}^{2}{p}^{2}+33\,{p}^{4} \right)  \left( \cosh \left( n
	\tau \right)  \right) ^{4}\\
	&-2\,\sinh \left( p\tau \right) \sinh \left( n\tau \right) {n}^{3}
	\cosh \left( p\tau \right) p \left( 58\,f+61 \right) \cosh \left( 
	n\tau \right) -{n}^{4} \left( 58\,f+76 \right)  \left( \sinh
	\left( p\tau \right)  \right) ^{2}\\
	&+ \left( 
	\left(  \left( 58f+39 \right) {n}^{2}+{p}^{2} \left( 58f+
	131 \right)  \right)  \left( \cosh \left( p\tau \right)  \right) ^{2}-
	\left( 58f+39 \right) {n}^{2}+2{p}^{2} \left( f+40
	\right)  \right) {n}^{2} \left( \cosh \left( n\tau \right)  \right) ^
	{2} \bigg)
	\end{split}
	\end{eqnarray}
	\begin{eqnarray}
	\begin{split}
	\rho&=\frac {3\sinh \left( p\tau \right) n}{2A{p}^{3} \left( \cosh
		\left( n\tau \right)  \right) ^{5}} \bigg(  \left( 3\, \left( -{n}^{4
	}-10\,{n}^{2}{p}^{2}-5\,{p}^{4} \right)  \left( \cosh \left( p\tau
	\right)  \right) ^{2}+{n}^{4}+2\,{n}^{2}{p}^{2}+5\,{p}^{4} \right) 
	\left( \cosh \left( n\tau \right)  \right) ^{4}\\
	&+12\,n\sinh \left( p
	\tau \right) \sinh \left( n\tau \right) \cosh \left( p\tau \right) p
	\left( {n}^{2}+3\,{p}^{2} \right)  \left( \cosh \left( n\tau \right) 
	\right) ^{3}\\
	&+ \left(  \left(  \left( 50\,f+45 \right) {n}^{2}+{p}
	^{2} \left( 50\,f+89 \right)  \right)  \left( \cosh \left( p\tau
	\right)  \right) ^{2}+ \left( -50\,f-45 \right) {n}^{2}-2\,{p}^{2
	} \left( f+22 \right)  \right) {n}^{2} \left( \cosh \left( n\tau
	\right)  \right) ^{2}\\
	&-10\,\sinh \left( p\tau \right) \sinh \left( n
	\tau \right) {n}^{3}\cosh \left( p\tau \right)  \left( 10\,f+11
	\right) p\cosh \left( n\tau \right) - ( 10\,f+{ {188}/{
			15}} ) {n}^{4} \left( \sinh \left( p\tau \right)  \right) ^{2}
	\bigg) 
	\end{split}
	\end{eqnarray}
	A detailed solution with $p=n=1$ is given in appendix C in \eqref{v11}-\eqref{v12}.

	\subsection{Bianchi-type $VI_{-1}$  }
	In this Bianchi-type case, the (0,1) component of  $\beta$-function equations of \eqref{betaGR}
	imposes the following constraint 
	\begin{eqnarray}
	\begin{split}
	(\ln&{\frac{a_{3}}{a_{2}}})'+\alpha'[X_3'-X_2'-(\ln{\frac{a_{3}}{a_{2}}})'(\frac{3A^2}{4a^6}+\frac{1}{a_{1}^2})\\
	&+\frac{4}{3}a^{-6}e^{-2\phi_0}((\ln{\frac{a_{1}}{a_{2}}})'(\ln{a_{2}})''+(\ln{\frac{a_{3}}{a_{1}}})'(\ln{a_{3}})''-(\ln{\frac{a_{3}}{a_{2}}})'((\ln{\frac{a_{1}}{a_{2}a_{3}}})'\phi_0'-(\ln{a_{2}})'(\ln{a_{3}})'))]=0.
	\end{split}
	\end{eqnarray}
	Its zeroth order requires the condition $ a_{3}=a_{1}$ and imposing it  leads to vanishing of the first order constraint equation, automatically.
	Moreover, the Bianchi-type dependent $\hat{V}$ terms  are given by
	\begin{eqnarray}
	\hat{V}_1^{(1)}=-2a_2^4e^{2\phi_0}(1+2\alpha'(2X_1+\xi_4)),~~\hat{V}_2^{(1)}=\hat{V}_3^{(1)}=0,
	\end{eqnarray}
	\begin{eqnarray}
	\hat{V}_1^{(2)}&=&a_{1}^{-2}\big(-2\,(a_{2}a_{3}e^{\phi_0})^2-2(\ln a_{1})'^2-(\ln a_{2})'^2-(\ln a_{3})'^2+(5-f)A^2e^{2\phi_0}/2\big),\\
	\hat{V}_2^{(2)}&=&a_{1}^{-2}\big(2\,(a_{2}a_{3}e^{\phi_0})^2-2(\ln a_{1})'^2-(\ln a_{2})'^2+(\ln a_{2})'(\ln a_{1}{a_{3}})'+(1+f)A^2e^{2\phi_0}/2\big),\\
	\hat{V}_3^{(2)}&=&a_{1}^{-2}\big(2\,(a_{2}a_{3}e^{\phi_0})^2-2(\ln a_{1})'^2-(\ln a_{3})'^2+(\ln a_{3})'\ln (a_{1}{a_{2}})'+(1+f)A^2e^{2\phi_0}/2\big),\\
	\hat{V}_{0}^{(2)}&=&-a_{1}^{-2}\big(2(\ln a_{1} )'^2+(\ln a_{2} )'^2+(\ln a_{3} )'^2-(\ln a_{1} )'(\ln a_{2}a_{3} )'\big),\\
	\hat{V}_{\phi}&=&a_{1}^{-2}(A^2e^{2\phi_0}/4{-4(\ln a_{1})'^2-2(\ln a_{2})'^2-2(\ln a_{3})'^2+2(\ln a_{3})'(\ln a_{1}{a_{2}})'+2(\ln a_{1})'(\ln a_{2})'}).
	\end{eqnarray}			
	With these form of potentials and  $ a_{3}=a_{1}$ condition, the $\beta$-function equations \eqref{iifinal}  in zeroth order give  \cite{batakis2}
	\begin{eqnarray}
	{(\ln a_{1}^2 e^{\phi_{0}})}''-4a_{2}^4 e^{2\phi_{0}}=0, \label{145}\\
	{(\ln a_{2}^2 e^{\phi_{0}})}''=0\label{146},
	\end{eqnarray}
	and in the first order of $\alpha'$ give
	\begin{eqnarray}
	X_1''-4a_{2}^4 e^{2\phi_{0}}(2X_1+\xi_4)-\frac{1}{2}\,	{(\ln a_{1}^2 e^{\phi_{0}})}'\xi_4'+g_1=0,\label{147}\\
	X_2''-\frac{1}{2}\,	{(\ln a_{2}^2 e^{\phi_{0}})}'\xi_4'+g_2=0,\label{148}\\
	X_3''-	X_2''=0\label{149}.
	\end{eqnarray}
	The  solutions of \eqref{145} and \eqref{146} are given by \cite{ batakis2}
	\begin{eqnarray}
	a_{1}^2 e^{\phi_{0}}=q_1^2p_1e^{p_1\tau}e^{q_2^4e^{-p_2\tau}}	,\\
	a_{2}^2 e^{\phi_{0}}=a_{3}^2 e^{\phi_{0}}=q_2^2p_2e^{p_2\tau},
	\end{eqnarray}
	and    \eqref{147}-\eqref{149} admit the following solutions
	\begin{eqnarray}\label{vihx1}
	X_1=-8\,\iint (p_2^2q_2^2e^{2\,p_2\tau}\iint g_2\,d\tau\,d\tau)\,d\tau\,d\tau-\iint g_1\,d\tau\,d\tau+ (p_2q_2^2e^{2\,p_2\tau}+\frac{p_1}{2})\int\xi_4 \,d\tau,
	\end{eqnarray}
	\begin{eqnarray}
	X_2=X_3=-\iint g_2\,d\tau\,d\tau+\frac{p_2}{2}\int\xi_4 \,d\tau.
	\end{eqnarray}
	The $q_2$ constant in this Bianchi-type is given by
	\begin{eqnarray}
	q_2=(n^2-p_1p_2-p_2^2)/2.
	\end{eqnarray}
	For example with $p_1=p_2=n=1$ we have
	\begin{eqnarray}
	\begin{split}
	g_1=&\frac{12\,{{\rm e}^{-{{\rm e}^{2\,\tau}}-3\,\tau}}}{A\left( \cosh \left( \tau \right)  \right) ^{3}} \big( ( -\frac{1}{2}
	\cosh \left( \tau \right) {{\rm e}^{\tau}}+f+1 ) {{\rm e}^{2\,
			\tau}}+ ( -3\, \left( \cosh \left( \tau \right)  \right) ^{2}+\frac{1}{2}(f+1) {{\rm e}^{4\,\tau}}+\frac{3}{4} \left( \cosh \left( \tau
	\right)  \right) ^{2}{{\rm e}^{6\,\tau}}+{\frac {25\,f}{16}}\\
	&-\frac{3}{4}( \left( \cosh \left( 
	\tau \right)  \right) ^{2}-\cosh \left( \tau \right) \sinh
	\left( \tau \right)) +{\frac {67}{32}}+{\frac {25
		}{16
		} ( f+{\frac {11}{10}} ) }\,\tanh \left( \tau \right)  \left( 1+{{\rm e}^{2\,\tau}} \right)-{ { \left( \cosh \left( 
			\tau \right)  \right) ^{-2}} ( {\frac {25\,f}{32}}+{\frac {47}{48}
		} ) }
	\big).
	\end{split}
	\end{eqnarray}
	\begin{eqnarray}
	\begin{split}
	g_2=&\frac {{{\rm e}^{-{{\rm e}^{2\,\tau}}-3\,\tau}}}{4A \left( \cosh
		\left( \tau \right)  \right) ^{3}} \big(  ( -4\,\cosh \left( 
	\tau \right) {{\rm e}^{\tau}}+7\,f+4+8\,\tanh \left( \tau \right) 
	( f+{\frac {23}{16}} )  ) {{\rm e}^{2\,\tau}}+4\,
	( -7/2\, \left( \cosh \left( \tau \right)  \right) ^{2}\\
	&+\cosh
	\left( \tau \right) \sinh \left( \tau \right) +f+3/2 ) {{\rm e}
		^{4\,\tau}}+18\, \left( \cosh \left( \tau \right)  \right) ^{2}{
		{\rm e}^{6\,\tau}}-12\, \left( \cosh \left( \tau \right)  \right) ^{2}
	-12\,\cosh \left( \tau \right) \sinh \left( \tau \right) +{\frac {29\,
			f}{2}}\\
	&+{\frac {85}{4}}+{\frac {29}{2} ( f+{\frac {61}{58}} ) }\,\tanh \left( \tau \right) \cosh
	\left( \tau \right) -{\frac {
			1}{ 2\left( \cosh \left( \tau \right)  \right) ^{2}} ( {\frac {29
				\,f}{2}}+19 ) } \big) 
	.
	\end{split}
	\end{eqnarray}
	\begin{eqnarray}
	\begin{split}
	\rho=&\frac{12\,{{\rm e}^{-{{\rm e}^{2\,\tau}}-3\,\tau}}}{A\left( \cosh \left( \tau \right)  \right) ^{3}} \big( ( -\frac{1}{2}
	\cosh \left( \tau \right) {{\rm e}^{\tau}}+f+1 ) {{\rm e}^{2\,
			\tau}}+\frac{1}{2} ( -\frac{3}{2} \left( \cosh \left( \tau \right)  \right) ^
	{2}+f+1 ) {{\rm e}^{4\,\tau}}+\frac{3}{4} \left( \cosh \left( \tau
	\right)  \right) ^{2}{{\rm e}^{6\,\tau}}+{\frac {25\,f}{16}}\\
	&-\frac{3}{4} (\left( \cosh \left( 
	\tau \right)  \right) ^{2}+\cosh \left( \tau \right) \sinh
	\left( \tau \right) )+{\frac {67}{32}}+{\frac {25
			\,\tanh \left( \tau \right)  \left( 1+{{\rm e}^{2\,\tau}} \right) }{16
		} ( f+{\frac {11}{10}} ) }-{ { \left( \cosh \left( 
			\tau \right)  \right) ^{-2}} ( {\frac {25\,f}{32}}+{\frac {47}{48}
		} ) }
	\big).
	\end{split}
	\end{eqnarray}
	Unfortunately, there are not a clear results for the first integral of  \eqref{vihx1}.
	\subsection{Bianchi-type $VII_0$}
	Here, there is no constraint equation  and the Bianchi-type dependent $\hat{V}$ terms are given by
	\begin{eqnarray}
	\hat{V}_1^{(1)}=-\hat{V}_2^{(1)}=(a_1^4-a_2^4+2\alpha'(2a_1^4X_1-2a_2^4X_2+(a_1^4-a_2^4)\xi_4))e^{2\phi_0},\\
	\hat{V}_3^{(1)}=-(a_1^4+a_2^4-2a_1^2a_2^2+2\alpha'(2a_1^4X_1+2a_2^4X_2-2a_1^2a_2^2(X_1+X_2)+(a_1^4+a_2^4-2a_1^2a_2^2)\xi_4))e^{2\phi_0},
	\end{eqnarray}
	\begin{eqnarray}
	\begin{split}
	\hat{V}_1^{(2)}&=\frac{1}{2{{a_{3}}^{2}}}\big(-\frac{{a_{1}}^{2}}{{a_{2}}^{2}}( 2\,(\ln   a_{1} )' (\ln {
		\frac {a_{1}}{{a_{2}}^{2}{a_{3}}^{2}}} )' +(\ln a_{3}
	)' (\ln a_{2} )' + \sum (\ln a_{{ i}})'^{2} )- (\ln a_{3} )' (\ln {\frac {a_{2}}{a_{3}}} )'\\
	&-\frac{{a_{2}}^{2}}{{a_{1}}^{2}}
	(2\,(\ln a_{2} )' (\ln {
		\frac {a_{2}}{{a_{1}}^{2}{a_{3}}^{2}}} )'+(\ln a_{3} ) '
	(\ln a_{1} )' + \sum( (\ln a_{{ i}} ) '
	) ^{2} ) +e^{2\phi_0}(-2a_{2}^{4}+{
		\frac {{a_{1}}^{6}}{4{a_{2}}^{2}}}+{\frac {5{a_{2}}^
			{6}}{4{a_{1}}^{2}}} +{\frac {{a_{2}}^{2}{a_{1}}^{2}
		}{2}})\\
	&+A^2e^{2\phi_0}({ {3(f-1){a_{1}}^{2}}{a_{2}}^{-2}/{4}}-(f+1)/2-
	{ {(f-5){a_{2}}^{2}}{a_{1}}^{-2}/{4}}
	)\big)	,
	\end{split}
	\end{eqnarray}
	\begin{eqnarray}
	\begin{split}
	\hat{V}_2^{(2)}&=\frac{1}{2{{a_{3}}^{2}}}\big(-\frac{{a_{1}}^{2}}{{a_{2}}^{2}} ( (\ln a_{1} )' (\ln {
		\frac {{a_{1}}^{2}}{{a_{3}}^{3}{a_{2}}^{4}}} ) '+4\,\ln
	( a_{3} )'(\ln a_{2} ) '+ \sum (\ln a_{{ i}})'^{2} ) - (\ln a_{3}
	) '(\ln {\frac {a_{1}}{a_{3}}} ) '\\
	&-\frac{{a_{2}}^{2}}{{a_{1}}^{2}}( (\ln a_{2} )' (\ln
	{\frac {{a_{2}}^{2}}{{a_{1}}^{4}{a_{3}}^{3}}} )' +4\,
	(\ln a_{3} )' (\ln a_{1} ) '+\sum (\ln a_{{ i}})'^{2}  )+2e^{2\phi_0}({\frac {{a_{1}}^{2}{a_{2}}^{2}}{4}}-{a_{2}}^{4} +{\frac {5{a_{
					1}}^{6}}{8{a_{2}}^{2}}}+\frac {{a_{2}}^{6}}{{8}{a_{1
		}}^{2}})\\
	&	+A^2e^{2\phi_0}({ {3(f-1){a_{2}}^{2}}{a_{1}}^{-2}}/{4}-(f+1)/2-
	{ {(f-5){a_{1}}^{2}}{a_{2}}^{-2}}/{4}
	)\big),
	\end{split}
	\end{eqnarray}
	\begin{eqnarray}
	\begin{split}
	\hat{V}_3^{(2)}&=\frac{1}{2{a_{3}}^{2}}\big(-\frac{{a_{1}}^{2}}{{a_{2}}^{2}} ( (\ln a_{1} )' (\ln {
		\frac {{a_{1}}^{2}}{{a_{3}}^{4}{a_{2}}^{3}}} )' +4\,(\ln a_{3} )' (\ln a_{2} ) '+  
	\sum (\ln a_{{ i}})'^{2}  )+    (
	\ln a_{3} )'^{2}+(\ln a_{1} ) '
	(\ln a_{2} ) '  \\
	&-\frac{{a_{2}}^{2}}{{a_{1}}^{2}} ( (\ln a_{2} )' (\ln {\frac {{a_{2}
			}^{2}}{{a_{3}}^{4}{a_{1}}^{3}}} )' +4\,(\ln a_{3}
	)' (\ln a_{1} ) '+\sum (\ln a_{{ i}})'^{2}  )+e^{2\phi_0}( -
	{\frac {{a_{2}}^{2}{a_{1}}^{2}}{2}}+{\frac {5{a_{2}}^{6}}{4{a_{1}}^{2}}}-{\frac {{a_
				{1}}^{4}}{{a_{3}}^{2}}}\\
	&-4{a_{2}}^{4}+{\frac {5{a_{1}}^{6}}{4{a_{2}}^{2
			}{a_{3}}^{2}}}+\frac{(5-f)A^2}{4}({\frac {{a_{1}}^{2}}{{a_{2}}^{2}}}-2+
	{\frac {{a_{2}}^{2}}{{a_{1}}^{2}}}
	))\big),
	\end{split}
	\end{eqnarray}
	\begin{eqnarray}
	\begin{split}
	\hat{V}_{\phi}=&\frac{1}{16}a_{3}^{-2}e^{2\phi_0}\big(-12\,{a_{1}}^{4}+2\,{a_{1}}^{2}{a_{2}}^{2}-12\,{a_{2}}^{4}+{
		\frac {11{a_{1}}^{6}}{{a_{2}}^{2}}}+{\frac {11{a_{2}}^{6}}{
			{a_{1}}^{2}}}+A^2\big(\frac{{a_{1}}^{2}}{ a_{2}^2}+\frac{{a_{2}}^{2}}{ a_{1}^2}-2\big)	\big)\\
	&+\frac{a_{1}^2}{a_{2}^2a_{3}^2}\big(-\sum (\ln a_{i} ) '^2+
	\frac{1}{2}(7(\ln a_{1} ) '-5(\ln a_{2} ) ')(\ln a_{3} ) '-2(\ln a_{1} ) '^2+\frac{7}{2}(\ln a_{1} ) '(\ln a_{2} ) '\big)\\
	&+\frac{a_{2}^2}{a_{1}^2a_{3}^2}\big(-\sum (\ln a_{i} ) '^2+
	\frac{1}{2}(7(\ln a_{2} ) '-5(\ln a_{1} ) ')(\ln a_{3} ) '-2(\ln a_{2} ) '^2+\frac{7}{2}(\ln a_{2} ) '(\ln a_{1} ) '\big)\\
	&+a_{3}^{-2}\big(2(\ln a_{3} ) '^2-(\ln a_{1} a_{2} ')(\ln a_{3} ) '+(\ln a_{1} ) '(\ln a_{2} ) '\big),
	\end{split}
	\end{eqnarray}	
	With these potentials, at leading order of \eqref{iifinal},   no general solution has  been found
	unless by setting $a_{1}=a_{2}$ \cite{ batakis2}.  In this case, all the  $V$ terms will   vanish and the equations will be  reduced to the Bianchi-type $I$ set. Consequently, the solutions can be retrieved from there.
	%%%%%%%%%%%%%%%%%%%%%%%%%%%%%%%%%%%%%%%%%%%%%%%%%%%%%%%%%%%%%%%%%%%%%%%%%%%%%%%%%%%%%%%%%%%%%%%%%
	\subsection{Bianchi-type $VIII$}
	For space-time of $VIII$ type  the $\hat{V}$ terms are given by
	\begin{eqnarray}
	\hat{V}_1^{(1)}=\frac{1}{2}\big(a_1^4-(a_2^2+a_3^2)^2+2\alpha'(2a_{1}^4\,X_1-2a_{2}^4\,X_2-2a_{3}^4\,X_3-a_{2}^2a_{3}^2(X_2+X_3)+(a_1^4-(a_2^2+a_3^2)^2)\xi_4)\big)e^{2\phi_0},\\
	\hat{V}_2^{(1)}=\frac{1}{2}\big(a_2^4-(a_3^2+a_1^2)^2+2\alpha'(2a_{2}^4\,X_2-2a_{1}^4\,X_1-2a_{3}^4\,X_3-a_{2}^2a_{3}^2(X_1+X_3)+(a_2^4-(a_1^2+a_3^2)^2)\xi_4)\big)e^{2\phi_0},\\
	\hat{V}_3^{(1)}=\frac{1}{2}(a_3^4-(a_1^2-a_2^2)^2+2\alpha'(2a_{3}^4\,X_3-2a_{3}^4\,X_2-2a_{2}^4\,X_2+a_{1}^2a_{2}^2(X_1+X_2)+(a_3^4-(a_1^2+a_2^2)^2)\xi_4))e^{2\phi_0},
	\end{eqnarray}
	\begin{eqnarray}
	\begin{split}
	\hat{V}_1^{(2)}&=(-\sum(\ln a_{i})'-2(\ln a_{1})'(\ln\frac{a_{1}}{a_{2}^2a_{3}^2})'-(\ln a_{2})'(\ln a_{3})')\frac{a_{1}^2}{2a_{2}^2a_{3}^2}-\frac{(\ln a_{1}'^2+(\ln a_{2})'(\ln a_{3})')}{a_{1}^2}\\
	&+(-\sum(\ln a_{i})'+4(\ln a_{1})'(\ln\frac{a_{2}}{a_{3}})'-(\ln a_{2})'(\ln\frac{a_{2}^2}{a_{3}^3})')\frac{a_{2}^2}{2a_{1}^2a_{3}^2}-\frac{(\ln a_{2})'((\ln a_{2})'-(\ln a_{3})')}{a_{2}^2}\\
	&+(-\sum(\ln a_{i})'-4(\ln a_{1})'(\ln\frac{a_{2}}{a_{3}})'-(\ln a_{3})'(\ln\frac{a_{2}^2}{a_{3}^3})')\frac{a_{3}^2}{2a_{1}^2a_{2}^2}-\frac{(\ln a_{3})'((\ln a_{2})'-(\ln a_{3})')}{a_{2}^2}\\
	&+e^{2\phi}\bigg(-{\frac {{a_{2}}^{4}}{{a_{3}}^{2}}}+{\frac {{a_{3}}^{4}}{{
				a_{2}}^{2}}}+{\frac {{a_{2}}^{4}}{2{a_{1}}^{2}}}+
	{\frac {{a_{3}}^{4}}{2{a_{1}}^{2}}}+{\frac {{a_{1}}^{6
		}}{8{a_{3}}^{2}{a_{2}}^{2}}}+{\frac {5{a_{2}}^{6}}{8{{
					a3}}^{2}{a_{1}}^{2}}}+{\frac {5{a_{3}}^{6}}{8{{ a_{2}
			}}^{2}{a_{1}}^{2}}}+\frac{1}{2}{a_{1}}^{2}+{a_{3}}^{2}
	-{{  a_{2}
	}}^{2}-{\frac {{a_{3}}^{2}{a_{2}}^{2}}{4{a_{1}}^{2}}}\\
	&+
	{\frac {{a_{3}}^{2}{a_{1}}^{2}}{4{a_{2}}^{2}}}+{
		\frac {{a_{2}}^{2}{a_{1}}^{2}}{4{a_{3}}^{2}}}
	+\frac{A^2}{8}\big({\frac {3(f-1){a_{1}}^{2}}{{a_{2}}^{2}{a_{3}}^{2}}}+2(f+1)({\frac{1}{a_{2}^{2}}}-{\frac{1}{a_{3}^{2}}})+(5-f)({\frac{2}{a_{1}^{2}}}+
	{\frac {{a_{3}}^{2}}{{a_{2}}^{2}{a_{1}}^{2}}}+{\frac {{a_{2}}^{2}}{{a_{3}}^{2}{a_{1}}^{2}}})
	\big)\bigg),
	\end{split}
	\end{eqnarray}
	\begin{eqnarray}
	\hat{V}_2^{(2)}=	\hat{V}_1^{(2)}(1\leftrightarrow 2),
	\end{eqnarray}		
	\begin{eqnarray}
	\begin{split}
	\hat{V}_3^{(2)}=&(-\sum(\ln a_{l})'+4(\ln a_{2})'(\ln\frac{a_{3}}{a_{1}})'-2(\ln a_{3})'^2-3(\ln a_{1})'(\ln a_{3})')\frac{a_{3}^2}{2a_{1}^2a_{2}^2}+\frac{(\ln a_{3}')^2+(\ln a_{1})'(\ln a_{2})'}{a_{3}^2}\\
	&+\sum_{\substack{k\\(k\neq i\neq j)}}\bigg((-\sum(\ln  a_{l})'-(\ln a_{k})'(\ln\frac{a_{k}^2}{a_{j}^4a_{3}^3})'-4(\ln a_{j})'(\ln a_{3})')\frac{a_{k}^2}{2a_{j}^2a_{3}^2}-\frac{(\ln a_{k})'((\ln a_{k})'-(\ln a_{j})')}{a_{k}^2}\\
	&~~~~~~~~~~~~~~~~~~+e^{2\phi}\big({\frac {a_{3}^{2}a_{k}^{2}}{4a_{j}^{2}}}-{\frac {
			a_{2}^{2}a_{1}^{2}}{4a_{3}^{2}}}-{\frac {a_{k}^
			{4}}{2a_{3}^{2}}}
	+{\frac {a_{j}^{4}}{2a_{k}^{2}}}+{\frac {5{{
					a_{k}}}^{6}}{8a_{3}^{2}a_{j}^{2}}}+
	{\frac {a_{3}^{6}}{8a_{2}^{2}a_{1}^{2}}}-\frac{1}{2}a_{3}^{2}-a_{k}^{2}\\
	&~~~~~~~~~~~~~~~~~~+\frac{A^2}{8}\big(
	{\frac {3\,(f-1){a_{3}}^{2}}{{a_{2}}^{2}{a_{1}}^{2}}}
	+2(f+1){\frac{1}{a_{k}^{2}}}	
	+	(5-f)(	{\frac {{a_{k}}^{2}}{{a_{j}}^{2}{a_{3}}^{2}}}	-{\frac{2\,}{a_{3}^{2}}})
	\big)
	\big)\bigg),
	\end{split}
	\end{eqnarray}	
	\begin{eqnarray}
	\begin{split}
	\hat{V}_{0}^{(2)}=\sum_{i\neq j\neq k}\big(&(\sum(\ln a_{l})'+(\ln a_{i})'(\ln \frac{a_{i}^2}{a_{k}^4a_{j}^3})'+(\ln a_{j})'(\ln a_{k})')\frac{a_{i}^2}{2a_{j}^2a_{k}^2}\\
	&-((\ln a_{i})'-(\ln a_{j})')((\ln a_{i})'-(\ln a_{k})'){a_{i}^{-2}}\big),
	\end{split}
	\end{eqnarray}
	
	\begin{eqnarray}
	\begin{split}
	\hat{V}_{\phi}&=\sum_{\substack{i\\j\neq k\neq i}}\bigg(\frac{a_{i}^2}{2\,a_{j}^2\,a_{k}^2}(-2\sum (\ln a_{l} )'^2-4(\ln a_{i} )'^2+7\,(\ln a_{i}) '((\ln a_{j} )'+(\ln a_{k} )')-5(\ln a_{j}) '(\ln a_{k}) ')\\
	&+\frac{e^{2\phi_0}}{16}\big({11a_{i}^6}{a_{j}^{-2}\,a_{k}^{-2}}+{(2+A^2)a_{i}^2\,a_{j}^2}{a_{k}^{-2}}-12\sum{a_{i}^4}{a_{3}^{-2}}
	+A^2( {2}{a_{1}^{-2}}+ {2}{a_{2}^{-2}}- {2}{a_{3}^{-2}})\big)\\
	&	-({(\ln a_{2} )'(\ln a_{1}) '-3(\ln a_{3} )'^2-\sum(\ln a_{3}) '(\ln a_{l}) '}){a_{3} ^{-2}}\bigg)\\
	&	-\sum_{\substack{i\neq 3\\j\neq k\neq i}}\big(({(\ln a_{j}) '(\ln a_{3}) '+3(\ln a_{i} )'^2-\sum(\ln a_{i}) '(\ln a_{l}) '}){a_{i} ^{-2}}+12{a_{j}^4}{a_{i}^{-2}}e^{2\phi_0}\big).
	\end{split}
	\end{eqnarray}			
	With this form of potentials, the $\beta$-function equations \eqref{iifinal} recast in the following forms in the zeroth order \cite{batakis2}
	\begin{eqnarray}
	{(\ln a_{1}^2e^{\phi_{0}})}''+(a_{1}^4-(a_{2}^2+a_{3}^2)^2)e^{2\phi_0}=0,\label{191}\\
	{(\ln a_{2}^2 e^{\phi_{0}})}''+(a_{2}^4-(a_{3}^2+a_{1}^2)^2)e^{2\phi_0}=0,\label{192}\\
	{(\ln a_{3}^2 e^{\phi_{0}})}''+(a_{3}^4-(a_{1}^2-a_{2}^2)^2)e^{2\phi_0}=0,\label{193}
	\end{eqnarray}
	and in the first order of $\alpha'$
	\begin{eqnarray}\label{194}
	\begin{split}
	X_1''&+2(a_{1}^4\,X_1-a_{2}^4\,X_2-a_{3}^4\,X_3-a_{2}^2a_{3}^2(X_2+X_3))e^{2\phi_0}\\
	&~~~~~+(a_{1}^4-(a_{2}^2+a_{3}^2)^2)e^{2\phi_0}\xi_4
	-\frac{1}{2}\,	{(\ln a_{1}^2 e^{\phi_{0}})}'\xi_4'+g_1=0,
	\end{split}
	\end{eqnarray}
	\begin{eqnarray}\label{195}
	\begin{split}
	X_2''&+2(a_{2}^4\,X_2-a_{3}^4\,X_3-a_{1}^4\,X_1-a_{1}^2a_{3}^2(X_1+X_3))e^{2\phi_0}\\
	&~~~~~+(a_{2}^4-(a_{3}^2+a_{1}^2)^2)e^{2\phi_0}\xi_4
	-\frac{1}{2}\,	{(\ln a_{1}^2 e^{\phi_{0}})}'\xi_4'+g_2=0,
	\end{split}
	\end{eqnarray}
	\begin{eqnarray}\label{196}
	\begin{split}
	X_3''&+2(a_{3}^4\,X_3-a_{1}^4\,X_1-a_{2}^4\,X_2+a_{1}^2a_{2}^2(X_1+X_2))e^{2\phi_0}
	\\
	&~~~~~
	+(a_{3}^4-(a_{1}^2-a_{2}^2)^2)e^{2\phi_0}\xi_4
	-\frac{1}{2}\,	{(\ln a_{1}^2 e^{\phi_{0}})}'\xi_4'+g_3=0.
	\end{split}
	\end{eqnarray}
	General solutions of \eqref{191}-\eqref{193} are given by \cite{batakis2}
	\begin{eqnarray}\label{viii1}
	\begin{split}
	a_{1}^2 e^{\phi_{0}}&=a_{2}^2 e^{\phi_{0}}=\frac{p_1^2\cosh(p_2\tau)}{p_2\sinh^2(p_1\tau)},\\
	a_{3}^2 e^{\phi_{0}}&=\frac{p_2}{\cosh(p_2\tau)}.
	\end{split}
	\end{eqnarray}
	and the \eqref{194}-\eqref{196} admit the following general solutions
	\begin{eqnarray}\label{viiix1}
	\begin{split}
	X_1=X_2=&-X_3+{ Q_3}\,{\rm coth} (p_1\tau)+{Q_4}\, ( p_1\,\tau\,{\rm coth} (p_1\tau)-1 ) -p_1{\rm coth} (p_1\tau)\int \xi_4\, d\tau\\
	&+\frac{1}{2p_1}\bigg({\rm coth} (p_1\tau)\int ( p_1\,\tau\,{\rm coth} (p_1\tau)-1 ) (g_1+g_2)\,d\tau\\
	&~~~~~~~~~~-( p_1\,\tau\,{\rm coth} (p_1\tau)-1 )\int {\rm coth} (p_1\tau) (g_1+g_2)\,d\tau\bigg),
	\end{split}
	\end{eqnarray}
	\begin{eqnarray}\label{viiix3}
	\begin{split}
	X_3=&{ Q_5}\,{\rm \tanh} (p_2\tau)+{Q_6}\, ( p_2\,\tau\,{\rm tanh} (p_2\tau)-1 ) -p_2\,{\rm coth} (p_2\tau)\int \xi_4\, d\tau\\
	&+\frac{1}{p_1}\bigg({\rm tanh} (p_1\tau)\int ( p_1\,\tau\,{\rm tanh} (p_1\tau)-1 ) (g_1+g_2)\,d\tau\\
	&~~~~~~~~~~-( p_1\,\tau\,{\rm tanh} (p_2\tau-1 ))\int {\rm tanh} (p_2\tau) (g_1+g_2)\,d\tau\bigg).
	\end{split}
	\end{eqnarray}
	Again, the $g_i$ \eqref{gi} and $\rho$ \eqref{ro} of this Bianchi-type are obtained by using the \eqref{fioa} and \eqref{viii1}. Then, the appeared integral in  $\xi_4$ \eqref{xi4}, $\phi_1$ \eqref{fi1s} and the above $X_i$ could be calculated. In this Bianchi model the $q_2$ constant \eqref{q2} is given by
	\begin{eqnarray}
	q_2=(n^24-p_1^2+p_2^2)/2
	\end{eqnarray}
	and  for consistency in the solution of $\xi_4$ \eqref{xi4}, any set of parameterization   for $p_i$ and $n$ should satisfy $q_2\neq 0$.
	Because of the dense form of  $g_i$ and $\rho$ terms, here  we present  an example  with a choice of parameters $p_1=p_2=n=1$ as following
	\begin{eqnarray}
	\begin{split}
	g_1=&\frac { 1}{16A
		\left( \sinh \left( \tau \right)  \right) ^{2} \left( \cosh \left( 
		\tau \right)  \right) ^{6}}\big(\left( -7\,f-73 \right)  \left( \cosh \left( \tau
	\right)  \right) ^{6}+ \left( 85\,f+251 \right)  \left( \cosh
	\left( \tau \right)  \right) ^{4}\\
	&+ \left( -149\,f-313 \right) 
	\left( \cosh \left( \tau \right)  \right) ^{2}+71\,f+123\big)		.
	\end{split}
	\end{eqnarray}
	\begin{eqnarray}
	\begin{split}
	g_3=&\frac { 1}{16A
		\left( \sinh \left( \tau \right)  \right) ^{2} \left( \cosh \left( 
		\tau \right)  \right) ^{6}}\big(\left( 7\,f-159 \right)  \left( \cosh \left( \tau
	\right)  \right) ^{6}+ \left( 25\,f+535 \right)  \left( \cosh
	\left( \tau \right)  \right) ^{4}\\
	&+ \left( -71\,f-623 \right) 
	\left( \cosh \left( \tau \right)  \right) ^{2}+39\,f+235\big)
	.
	\end{split}
	\end{eqnarray}
	\begin{eqnarray}
	\begin{split}
	\rho=&\frac {1}{8A
		\left( \sinh \left( \tau \right)  \right) ^{2} \left( \cosh \left( 
		\tau \right)  \right) ^{6}}\big( \left( -7\,f-169 \right)  \left( \cosh \left( \tau
	\right)  \right) ^{6}+ \left( 163\,f+637 \right)  \left( \cosh
	\left( \tau \right)  \right) ^{4}\\
	&+ \left( -305\,f-797 \right) 
	\left( \cosh \left( \tau \right)  \right) ^{2}+149\,f+317\big)
	.
	\end{split}
	\end{eqnarray}
	A detailed form of metric and dilaton corrections for this choice of parameters is given in appendix $c$ in \eqref{viiix}-\eqref{53viii}.
	%%%%%%%%%%%%%%%%%%%%%%%%%%%%%%%%%%%%%%%%%%%%%%%%%%%%%%%%555
	\subsection{Bianchi-type $IX$}
	The  $\hat{V}$ terms of in Bianchi-type $IX$ case are given  as following
	\begin{eqnarray}
	\hat{V}_i^{(1)}=\frac{1}{2}(a_i^4-(a_j^2-a_k^2)^2+2\alpha'(a_{i}^4\,X_i-a_{j}^2\,X_j-a_{k}^2\,X_k+a_{k}^2a_{j}^2(X_k+X_j))+(a_i^4-(a_j^2-a_k^2)^2)\xi_4)e^{2\phi_0},
	\end{eqnarray}
	\begin{eqnarray}
	\begin{split}
	\hat{V}_i^{(2)}=\sum_{\substack{k\neq i\\(j\neq k)} }&\big(-\frac{a_{k}^2}{2a_{j}^2a_{i}^2}(\sum_{l}(\ln a_{l})'+(\ln a_{k})'(\ln \frac{a_{k}^2}{a_{i}^4a_{j}^3})'+4(\ln a_{j})'(\ln a_{i})')\\
	&-\frac{a_{i}^2}{2a_{k}^2a_{j}^2}(\sum_{l}(\ln a_{l})'-4(\ln a_{i})'(\ln\frac{a_{j}}{a_{k}})'+2(\ln a_{j})'^2+3(\ln a_{k})'(\ln a_{j})')\\
	&+{(\ln a_{k})'((\ln a_{k})'-(\ln a_{j})')}{a_{k}^{-2}}+{(\ln a_{j})'((\ln a_{j})'-(\ln a_{k})')}{a_{j}^{-2}}+{(\ln a_{i})'^2+(\ln a_{k})'(\ln a_{i})'}{a_{i}^{-2}}\\
	&+e^{2\phi}\big({
		\frac {a_{j}^{2}a_{i}^{2}}{4a_{k}^{2}}}-{\frac {a_{j}^{2}a_{k}^{2}}{4a_{i}^{2}}}-{\frac {a_{k}^
			{4}}{a_{j}^{2}}}-{\frac {a_{k}^{4}}{2a_{i}^{2}}}
	+{\frac {{{
					a_{i}}}^{6}}{8a_{j}^{2}a_{k}^{2}}}+{\frac {5a_{k}^{6}}{8a_{j}^{2}{{a_{i}}}^{2}}}
	+a_{k}^{2}-\frac{1}{2}a_{i}^{2}\\
	&+\frac{A^2}{8}\big((5-f)(\frac{a_{k}^2}{a_{i}^2\,a_{j}^2}-\frac{2}{a_{i}^2})+\frac{3(f-1)a_{i}^2}{a_{k}^2\,a_{j}^2}-2(f+1)\frac{1}{a_{k}^2}\big)\big)\big),
	\end{split}
	\end{eqnarray}	
	\begin{eqnarray}
	\begin{split}
	\hat{V}_{0}^{(2)}=\sum_{i\neq j\neq k}\big(&(\sum(\ln a_{l})'+(\ln a_{i})'(\ln \frac{a_{i}^2}{a_{k}^4a_{j}^3})'+(\ln a_{j})'(\ln a_{k})')\frac{a_{i}^2}{2a_{j}^2a_{k}^2}\\
	&-{((\ln a_{i})'-(\ln a_{j})')((\ln a_{i})'-(\ln a_{k})')}{a_{i}^{-2}}\big),
	\end{split}
	\end{eqnarray}
	\begin{eqnarray}
	\begin{split}
	\hat{V}_{\phi}&=\sum_{i}\sum_{j,k\neq i}\big(\frac{a_{i}^2}{2\,a_{j}^2\,a_{k}^2}(-2\sum _l(\ln a_{l} )'^2-4(\ln a_{i} )'^2+12\,(\ln a_{i}) '\sum_{l\neq i}(\ln a_{l}) '-5\sum_{l< j}(\ln a_{i}) '(\ln a_{l}) ')\\
	&+{\big(2\,(\ln a_{i} )'^2-2(\ln a_{i}) '\sum_{l\neq i}(\ln a_{l}) '+(\ln a_{k} )'(\ln a_{j}) '\big)}{a_{i}^{-2}}\\
	&+\frac{e^{2\phi_0}}{16}\big(({11a_{i}^6}{a_{j}^{-2}\,a_{k}^{-2}})+{2a_{i}^2\,a_{j}^2}{a_{k}^{-2}}+{A^2}\big({a_{i}^2}{a_{j}^{-2}\,a_{k}^{-2}}- {2}{a_{i}^{-2}}-{12\,a_{j}^2}{a_{i}^{-4}}-12 a_{i}^2\big)\big)
	\big)
	\end{split}
	\end{eqnarray}			
	With this form of potentials, the $\beta$-function equation of metric \eqref{iifinal} will be, respectively in the zeroth and first order of $\alpha'$, as following
	\begin{eqnarray}
	{(\ln a_{1}^2e^{\phi_{0}})}''+(a_i^4-(a_j^2-a_k^2)^2)e^{2\phi}=0,
	\end{eqnarray}
	\begin{eqnarray}
	\begin{split}
	X_i''+2(a_{i}^4\,X_i&-a_{j}^2\,X_j-a_{k}^2\,X_k+a_{k}^2a_{j}^2(X_k+X_j))e^{2\phi_0}\\
	&+(a_{i}^4-(a_{j}^2-a_{k}^2)^2)e^{2\phi_0}\xi_4
	-\frac{1}{2}\,	{(\ln a_{i}^2 e^{\phi_{0}})}'\xi_4'+g_i=0.
	\end{split}
	\end{eqnarray}
	General solutions admitted by the  zeroth order are given by \cite{batakis2}
	\begin{eqnarray}\label{ixi}
	\begin{split}
	a_{1}^2 e^{\phi_{0}}&=a_{3}^2 e^{\phi_{0}}=\frac{p_1^2\cosh(p_2\tau)}{p_2\cosh^2(p_1\tau)}\\
	a_{2}^2 e^{\phi_{0}}&=\frac{p_2}{\cosh(p_2\tau)}
	\end{split}
	\end{eqnarray}
	and solutions of 
	first order are given as following
	\begin{eqnarray}\label{ixx1}
	\begin{split}
	X_1=X_3=&-X_2+{ Q_3}\,{\rm \tanh} (p_2\tau)+{Q_4}\, ( p_2\,\tau\,{\rm tanh} (p_1\tau-1 ) -p_1{\rm tanh} (p_1\tau)\int \xi_4\, d\tau\\
	&+\frac{1}{p_1}\big({\rm tanh} (p_1\tau)\int ( p_1\,\tau\,{\rm tanh} (p_1\tau)-1 ) (g_1+g_2)\,d\tau\\
	&~~~~~~~~~~~~~-( p_1\,\tau\,{\rm tanh} (p_1\tau)-1 )\int {\rm tanh} (p_1\tau) (g_1+g_2)\,d\tau\big)+c_1,
	\end{split}
	\end{eqnarray}
	\begin{eqnarray}\label{ixx2}
	\begin{split}
	X_2&={ Q_5}\,{\rm \tanh} (p_2\tau)+{Q_6}\, ( p_2\,\tau\,{\rm tanh} (p_2\tau-1 ) -p_2{\rm tanh} (p_2\tau)\int \xi_4\, d\tau\\
	&+\frac{1}{p_2}\big({\rm tanh} (p_2\tau)\int ( p_2\,\tau\,{\rm tanh} (p_2\tau)-1 ) g_2\,d\tau-( p_2\,\tau\,{\rm tanh} (p_2\tau)-1 )\int {\rm tanh} (p_2\tau) g_2\,d\tau\big)+c_2.
	\end{split}
	\end{eqnarray}
	In this Bianchi-type the $q_2$ in \eqref{q2} is given by
	\begin{eqnarray} 
	q_2=(n^2-4p_1^2+p_2^2)/2. 
	\end{eqnarray}
	Now, using the \eqref{ixi} and \eqref{fioa} the $g_i$ \eqref{gi} and $\rho$ \eqref{ro} can be derived and then by performing the integrals of \eqref{fi1s}, \eqref{xi4}, \eqref{ixx1} and \eqref{ixx2} the corrections of dilaton and metric in this Bianchi-type are found. Here, as an example, we present a $g_i$ and $\rho$ with special choice of parameters as $n=1$ and $p_1=p_2=p$, where the $q_2\neq 0$ condition requires $p\neq \sqrt{3}/3$.
	\begin{eqnarray}
	\begin{split}
	g_1=&\frac {1}{{32A}\,{p}^{3}\cosh \left( p\tau \right)  \left( 
		\cosh \left( \tau \right)  \right) ^{5}} \big(   (  \left( -33\,{p}^{4
	}-58\,{p}^{2}-5 \right)  \left( \cosh \left( \tau \right)  \right) ^{4
	}\\
	&+ \left(  \left( 58\,f+131 \right) {p}^{2}+58\,f+39 \right) 
	\left( \cosh \left( \tau \right)  \right) ^{2}-58\,f-76 ) 
	\left( \cosh \left( p\tau \right)  \right) ^{4}\\
	&+76\, \left(  \left( {
		p}^{2}+{ {5}/{19}} \right)  \left( \cosh \left( \tau \right) 
	\right) ^{2}-{ {29\,f}/{19}}-{ {61}/{38}} \right) \sinh
	\left( p\tau \right) \sinh \left( \tau \right) \cosh \left( \tau
	\right) p \left( \cosh \left( p\tau \right)  \right) ^{3}\\
	&+16\,
	\left( \cosh \left( \tau \right)  \right) ^{2} \left(  \left( {p}^{2}
	+4 \right)  \left( \cosh \left( \tau \right)  \right) ^{2}-7/2\,f-
	{ {19}/{4}} \right) {p}^{2} \left( \cosh \left( p\tau \right) 
	\right) ^{2}\\
	&-32\,\sinh \left( \tau \right) \cosh \left( p\tau
	\right)  \left( \cosh \left( \tau \right)  \right) ^{3}\sinh \left( p
	\tau \right) {p}^{3}-64\, \left( \cosh \left( \tau \right)  \right) ^{
		4}{p}^{4}
	\big) 
	.
	\end{split}
	\end{eqnarray}
	\begin{eqnarray}
	\begin{split}
	g_2=	g_1+\frac {1}{\cosh \left( p\tau
		\right)  \left( \cosh \left( \tau \right)  \right) ^{3}}\big( &( 1- \left( {p}^{2}+1 \right)  \left( \cosh \left( \tau
	\right)  \right) ^{2} )  \left( \cosh \left( p\tau \right) 
	\right) ^{2}\\
	&+2\,\sinh \left( \tau \right) \cosh \left( p\tau \right) 
	\cosh \left( \tau \right) \sinh \left( p\tau \right) p+4\, \left( 
	\cosh \left( \tau \right)  \right) ^{2}{p}^{2}\big)
	\end{split}
	\end{eqnarray}
	\begin{eqnarray}
	\begin{split}
	\rho=&\frac {1}{{16A}\,{p}^{3}\cosh \left( p\tau \right)  \left( 
		\cosh \left( \tau \right)  \right) ^{5}} \big(  (  \left( -45\,{
		p}^{4}-90\,{p}^{2}-9 \right)  \left( \cosh \left( \tau \right) 
	\right) ^{4}\\
	&~~~+ \left(  \left( 150\,f+267 \right) {p}^{2}+150\,f_{1
	}+135 \right)  \left( \cosh \left( \tau \right)  \right) ^{2}-150\,f_{
		1}-188 )  \left( \cosh \left( p\tau \right)  \right) ^{4}\\
	&~~~+108\,
	(  \left( {p}^{2}+1/3 \right)  \left( \cosh \left( \tau \right) 
	\right) ^{2}-{ {25\,f}/{9}}-{ {55}/{18}} ) \sinh
	\left( \tau \right) \sinh \left( p\tau \right) \cosh \left( \tau
	\right) p \left( \cosh \left( p\tau \right)  \right) ^{3}\\
	&~~~+16\,
	\left( \cosh \left( \tau \right)  \right) ^{2} (  \left( {p}^{2}
	+11/2 \right)  \left( \cosh \left( \tau \right)  \right) ^{2}-9\,f
	-10 ) {p}^{2} \left( \cosh \left( p\tau \right)  \right) ^{2}\\
	&~~~-32
	\,\cosh \left( p\tau \right)  \left( \cosh \left( \tau \right) 
	\right) ^{3}\sinh \left( p\tau \right) \sinh \left( \tau \right) {p}^
	{3}-64\, \left( \cosh \left( \tau \right)  \right) ^{4}{p}^{4}
	\big) 
	.
	\end{split}
	\end{eqnarray}
	The string frame Ricci scalar of this $IX$ model with scale factors of \eqref{ixi} is given by $R=\frac{3}{2 \,A}$. Therefore, if the $A$ is assumed to be a very small constant, for example of order $\alpha'$, the model will be in the high curvature limit forever. Also, according to the given discussion in the last paragraph of section $3$, the kinetics of dilaton and $B$-field will be high as well and the $\alpha'$-correction had to be taken into account.	
	
	\subsubsection{Some Physical Properties of an  example in the Bianchi-type $IX$}			
	Avoiding of dense mathematical results,	 as an example we set $n=1$ and $p_1=p_2=1$ and present the resulted forms of $X_i$, $\phi_1$ and the lapse function $\xi_4$ in appendix $C$ in \eqref{ixx11}-\eqref{ixx4}. In this  subsection,  we are going to investigate the physical behavior of this example.	
	
	In this Bianchi model with the given scale factors \eqref{ixi}, transforming from $\tau$ to cosmic time $\bar{t}$ by \eqref{tautot} and finding the $\tau$ in terms of $\bar{t}$, as we have done for the Bianchi-type $I$  in \eqref{ti}, is not straightforward.  Hence, we rewrite the time derivatives in the physical quantities in terms of $\tau$ derivatives. With the following relation between the  cosmic time and $\tau$ 
	\begin{eqnarray}
	\begin{split}
	d\,\bar{t}=e^{\frac{\phi}{2}}\,\sqrt{g_{00}}\,d\,t=\bar{	a}^{-3}(1+\alpha'(\sum X_i+\xi_4))d\tau,
	\end{split}
	\end{eqnarray}
	up  to first order of $\alpha'$ we have
	\begin{eqnarray}\label{88}
	\begin{split}
	\dot{\bar{a}}_i=	\frac{d\, \bar{a}_i}{d\,\bar{t}}=\bar{	a}^{-3}(\bar{a}_i'(1-\alpha'(\sum X_j+\xi_4))+\alpha'(\bar{	a}_iX_i)'),
	\end{split}
	\end{eqnarray}
	\begin{eqnarray}
	\begin{split}
	\ddot{\bar{a}}_i=	\frac{d^2\, \bar{a}_i}{d\,\bar{t}^2}=\bar{	a}^{-6}\big(\bar{a}_i''-\bar{a}_i'\sum (\ln{\bar{a}_j})')(1&-2\alpha'(\sum X_j+\xi_4))\\
	&+\alpha'((\bar{a}_iX_i)''-(\bar{a}_iX_i)'\sum (\ln{\bar{a}_j})'-(\sum X_j'+\xi_4')\bar{a}_i')\big)
	.
	\end{split}
	\end{eqnarray}
	In this chosen parametrization,	the leading order solution of $IX$  \eqref{ixi}  is an isotropic model 	 with $\dot{\bar{a}}<0$ and $\ddot{\bar{a}}<0$. Hence, it describes a decelerated contracting universe with an initial singularity and no inflation \cite{batakis2}. 
	
	Considering the  two-loop  $\beta$-function equations solution, the  $\alpha'$-corrected scale factors \eqref{sf} with their correction given by   \eqref{ixx11} and \eqref{ixx22}, with setting $A = 1.5\,\alpha', Q_1 = 2, Q_3 = 10, Q_4 = -1.8,  Q_5 = 5, Q_6 = -1.2, c_1 = -0.1, c_2 = 1$ and $ \varphi_0 = 230$, give rise to $\dot{\bar{a}}_i>0$ and $\ddot{\bar{a}}_1>0$, $\ddot{\bar{a}}_3>0$ but the $\ddot{\bar{a}}_2$ changes sign such that $\ddot{\bar{a}}_2<0$ near $\tau=0$ and then turn to $\ddot{\bar{a}}_2>0$, such a way that $\ddot{\bar{	a}}>0$.\footnote{For choosing the constants, the  signature of  metric is considered to be adopted $(-1,1,1,1)$ after adding corrections. Also, a good cosmological behavior of the solutions has been considered  and this set of constants is not unique.} Consequently, the solution of two-loop $\beta$-functions in this case of $IX$ model describes an  expanding universe which is decelerated at first but then becomes accelerated and there is  no collapse to a singularity in the future. 
	
	The possibility of recollapse in the Bianchi-type $IX$ can be well understood by the weak energy condition. It is worth considering whether  this condition in our solution  is satisfied  or not. We need to analysis the equation \cite{wald}\footnote{Here, we consider the equation given in \cite{wald} with $\Lambda=0$ and assume that  the higher curvature terms contribute to the right hand of Einstein frame field equations in the energy momentum tensor.}
	\begin{eqnarray}
	\begin{split}
	k^2=-\frac{3}{2}R^{(3)}+24T_{\mu\nu}n^{\mu}n^{\nu}
	\end{split}
	\end{eqnarray}
	where $R^{(3)}$ is the scalar curvature of $3$-dimensional homogeneous hypersurface, $n^{\mu}$ is a time-like vector, the intrinsic curvature  $k=\sum \dot{\bar{a}}_i/\bar{a}_i$ and the weak energy condition requires $T_{\mu\nu}n^{\mu}n^{\nu}>0$.
	Unlike the other Bianchi models, the $R^{(3)}$ may become positive in Bianchi-type $IX$ and hence the $k$ can pass zero and it is possible for an expanding universe of this type to recollapse. But with a negative $R^{(3)}$, a change in the sign of $k$ is not allowed and an expanding model will expand forever. Here, in our example, the $R^{(3)}$ vanishes in zeroth order and is negative in the first order of $\alpha'$. Therefore, being no change in the sign of $\dot{\bar{a}}$   satisfies the weak energy condition.
	Furthermore, as we have discussed in the Bianchi-type $I$, the strong energy condition  requires the $\dot{\bar{k}}+{\bar{k}}^2<0$. 
	Here, the violation of this energy conditions, which is necessary for avoiding  the initial singularity \cite{Hawking}, occurs soon after $\tau=0$ by $\ddot{\bar{a}}>0$ in both $f=\pm 1$ schemes.
	
	In this example, an isotropic parameterization for the zeroth order metric has been chosen.	However,  the  $\alpha'$-corrected metric is anisotropic even if one ignores the constants of integration. However, the anisotropy measured by the mean anisotropy parameter $A_m=\frac{(\bar{H}_i-\bar{H})^2}{\bar{H}^2}$ increases at early times but then decreases  as time goes on, and the decrease rate is faster in $f=1$ scheme.

	\section{Conclusion}
	The  anisotropic homogeneous space-times have been emerged in the context  of string cosmology  during the search of  backgrounds for describing the early universe evolution. 
	In this work  the two-loop (order $\alpha'$) $\beta$-function equations with dilaton and $B$-field on the anisotropic homogeneous space-times, taking into account the two particular cases of RS   which are called  $R^2$ and Gauss-Bonnet schemes have been investigated.
	General forms of two-loop $\beta$-function equations  in terms of the  Hubble  coefficients, dilaton and $B$-field  are derived in   both  RS.
	Then, the solutions  in the all Bianchi-type space-times with zero  $\Lambda$  are computed  and then the metric, dilaton and axion field are found with perturbative $\alpha'$-corrections.

	The basic forms of  the first $\alpha'$ order  solutions  have been given in  certain integral terms in the all models.
	Because of mathematical complexity and dense results of the integrals, for obtaining an explicit answer for the first  $\alpha'$ order corrections,   fixed value for the arbitrary constants of zeroth order solutions  should be chosen. 
	Nevertheless, the integral forms are strongly  applicable in every chosen parameterizations of zeroth order as needed for any physical purposes.  A sample example of any Bianchi-type  with a specific parameter is presented.  
	
	Especially, the  behavior of the  $\alpha'$-corrected answers in the Bianchi  $I$ and $IX$ models have been investigated. In the Bianchi $I$ model  in an isotropic example, the decelerated expanding behavior of leading order solutions are modified in the two-loop solutions and the $\alpha'$-corrected answers describe a universe which starts decelerated expanding and then turns to an accelerated expanding phase and then with leaving the high curvature limit, reaches to the leading order solution behavior. The accelerated phase violates the strong energy condition and may  describe an inflationary phase. 
	In the Bianchi $IX$ model, as a special example,  an isotropic case of leading order solution is selected which describes a decelerated contraction. The $\alpha'$-corrected model of $IX$ (with choosing a set for arbitrary constants) describes an expanding universe which is  decelerated  in early times and then becomes accelerated violating the strong energy condition. Although an isotropic parameterization for the leading order  has been chosen, the first order correction	brings about an anisotropy that decreases as time passes. Also, the condition of the   recollapse  possibility in the Bianchi-type $IX$ model is investigated and showed that being no recollapse in the selected solution satisfies the weak energy condition. Beside it, considering this solution in an inflationary framework, no premature recollapse can  occur. In both selected Bianchi-type $I$ and $IX$,  the $\alpha'$-corrected solutions of two-loop $\beta$-functions in the both Gauss-Bonnet and $R^2$ schemes with $B$-field contribution are capable to well describe the accelerated expansion of the universe.
	
	The leading order solutions have had an initial singularity with no inflation \cite{batakis2}. We found that the $\alpha'$-corrected solutions of two-loop $\beta$-function equations in the selected Bianchi-type $I$ and $IX$ model in the  Bianchi-type $I$ and $IX$ model  avoid singularity in the violation of strong energy condition context and satisfies the weak energy condition.  It has been shown in \cite{nonsin}
	that including only the $\alpha'$-correction of Gauss-Bonnet multiplied by a dilaton field  to the leading order effective action does not violate  the  energy conditions  and hence it can not lead to a singularity-free solution. Therefore, the violation of strong energy condition in our selected solutions in Bianchi-type $I$ and $IX$ models may be interpreted as a result of contribution of the $B$-field. 
	
	Usually, in the recent studies  the solutions of the corrected action field equations with dilaton and modulus fields have been investigated where the field strength of the $B$-field has been taken to be zero for simplicity \cite{Yajima}. In this work, we have considered the two-loop $\beta$-functions with the contributions of dilaton and $B$-field (with nonzero but constant field strength tensor magnitude),  where in the presence of $B$-field a RS dependence appears.
	Noting the equivalence of the $\beta$-function and the equations of motion of the $\alpha'$-corrected effective action, the solutions of $\beta$-function equations, beside securing the conformal invariance up to two-loop,  are  higher derivative  quadratic gravitational action solutions in the Einstein frame. The dilaton and a $B$-field will  appear as an effective matter in the  field equations. Hence, from the Einstein frame point of view, our solutions are  solutions of
	quadratic curvature gravity with effective matter associated with the dilaton and $B$-field, where the $B$-field part is capable of violating the strong energy condition.

	In \cite{batakis2} the magnitude of the field strength tensor of $B$-field, $A$,  have relation with the anisotropy through the initial value equation. But here, a different solution for the zeroth order of dilaton \eqref{fioa} has been chosen which result in the elimination of $A$ from the initial value equation constraint \eqref{q2}. Then, taking advantageous of the arbitrariness, being in the high curvature and high kinetics limits as the necessary condition for including the $\alpha'$-corrections in the action is prescribed by the magnitude of $A$ such that it is required to be of order $\alpha'$.
	
	Moreover,	in the Bianchi-type $V$ model, it is observed that although the leading order solution allows an isotropic metric,  the two-loop solutions restrict the $\alpha'$-corrected metric to be isotropic in its zeroth and first order.
	The detailed behavior and cosmological implications of these Bianchi-type models would be investigated in another work.
	\section*{Acknowledgment}
	We would like to express our sincere gratitude to M. M. Sheikh-Jabbari for his
	useful comments.	This research	has been supported by Azarbaijan Shahid Madani university by a research fund No. 401.231.
	%%%%%%%%%%%%%%%%%%%%%%%%%%%%%%%%%%%%%%%%%%%%%%%%%%%%%%%
	
	\section*{Appendix A}
	In this appendix we present  some  definitions	and the classifications of real  $3$-dimensional Bianchi  groups and their associated vielbeins ${e_\alpha}^i(x)$ \cite{batakis2, mr}.
	
	\begin{center}
		\hspace{0.01mm}{{\bf Table 1}}:\hspace{1mm} Bianchi-types, their  Lie algebras and left invariant one forms.
		\\ \hspace{-1mm}\\
		\begin{tabular}{l l l l  l l p{15mm} }
			\hline\hline
			\vspace{-1mm}
			{\scriptsize ${\bf Algebra}$ }& {\scriptsize Non-zero commutation
				relations $[\sigma_i,\sigma_j]=f_{ij }^{~~k}\sigma_{k}$ }&{\scriptsize Class }&{\scriptsize $g^{-1} dg$}
			\\\hline
			
			\vspace{1mm}
			
			{\scriptsize ${I}$}& {\scriptsize $[\sigma_i,\sigma_j]=0$}&
			{\scriptsize $\cal{A}$} & {\scriptsize $dx^i \sigma_i$}\\

			\vspace{1mm}
			
			{\scriptsize ${II}$}& {\scriptsize$[\sigma_2,\sigma_3]=\sigma_1$}&
			{\scriptsize $\cal{A}$} &
			{\scriptsize $dx^1\sigma_1+dx^2(\sigma_2+x^3\sigma_1)+dx^3\sigma_3$} \\
			
			\vspace{-1mm}
			
			{\scriptsize ${III}$}& {\scriptsize$[\sigma_1,\sigma_3]=\sigma_3$}&
			{\scriptsize $\cal{B}$} &
			{\scriptsize $dx^1\sigma_1+dx^2(\sigma_2+x^3\sigma_1)+e^{x^1}dx^3(\sigma_3-x^2\sigma_1)$} \\
			
			\vspace{-1mm}
			
			{\scriptsize ${IV}$}&
			{\scriptsize$[\sigma_1,\sigma_2]=-\sigma_2+\sigma_3,[\sigma_1,\sigma_3]=-\sigma_3$}&{\scriptsize $\cal{B}$} & {\scriptsize $dx^1 [\sigma_1-x^2\sigma_2+(x^2-x^3)\sigma_3]+dx^2\sigma_2+dx^3\sigma_3$}\\

			\vspace{1mm}
			
			{\scriptsize ${V}$}&
			{\scriptsize$[\sigma_1,\sigma_2]=-\sigma_2,[\sigma_1,\sigma_3]=-\sigma_3$}& {\scriptsize
				$\cal{B}$} &
			{\scriptsize $dx^1(\sigma_1-x^2\sigma_2-x^3\sigma_3)+dx^2\sigma_2+dx^3\sigma_3$}\\
			
			\vspace{-1mm}
			
			{\scriptsize ${VI_0}$}&
			{\scriptsize$[\sigma_1,\sigma_3]=\sigma_2,[\sigma_2,\sigma_3]=\sigma_1$}& {\scriptsize
				$\cal{B}$} &
			{\scriptsize $dx^1(\sigma_1\cosh{x^3}+\sigma_2\sinh{x^3})+dx^2(\sigma_2\cosh{x^3}$}\\
			
			\vspace{1mm}
			
			& &  & {\scriptsize$+\sigma_1\sinh{x^3})+dx^3\sigma_3$}\\
			\vspace{1mm}

			{\scriptsize ${VI_{-1}}$}&
			{\scriptsize$[\sigma_1,\sigma_2]=-\sigma_2,[\sigma_1,\sigma_3]=\sigma_3$}& {\scriptsize
				$\cal{B}$} &
			{\scriptsize $dx^1(\sigma_1\cosh{x^3}+\sigma_2\sinh{x^3})+dx^2(\sigma_2\cosh{x^3}$}\\
			
			\vspace{1mm}
			
			& &  & {\scriptsize$+\sigma_1\sinh{x^3})+dx^3(\sigma_3-\cosh{x^3}(x^1\sigma_2+x^2\sigma_1)$}\\
			\vspace{1mm}
			
			& &  & {\scriptsize$+\sinh{x^3}(x^1\sigma_1+x^2\sigma_2))$}\\
			\vspace{1mm}
			
			{\scriptsize ${VII_0}$}&
			{\scriptsize$[\sigma_1,\sigma_3]=-\sigma_2,[\sigma_2,\sigma_3]=\sigma_1$}& {\scriptsize
				$\cal{A}$} &
			{\scriptsize $dx^1(\sigma_1\cos{x^3}-\sigma_2\sin{x^3})+dx^2(\sigma_2\cos{x^3}+\sigma_1\sin{x^3})$}\\
			
			\vspace{1mm}
			
			& &  & {\scriptsize$+dx^3\sigma_3$}\\
			
			\vspace{-1mm}
			
			{\scriptsize ${VII_a}$}&
			{\scriptsize$[\sigma_1,\sigma_2]=-a\sigma_2+\sigma_3,[\sigma_3,\sigma_1]=\sigma_2+a\sigma_3$}& {\scriptsize
				$\cal{B}$} &
			{\scriptsize $dx^1[\sigma_1-(ax^2+x^3)\sigma_2-(-ax^3+x^2)\sigma_3]+dx^2\sigma_2$}\\
			
			\vspace{1mm}
			
			& &  & {\scriptsize $+dx^3\sigma_3$}\\
			
			\vspace{-1mm}
			
			{\scriptsize ${VIII}$}&
			{\scriptsize$[\sigma_1,\sigma_3]=-\sigma_2,[\sigma_2,\sigma_3]=\sigma_1,[\sigma_1,\sigma_2]=-\sigma_3$}&
			{\scriptsize	$\cal{A}$} &
			{\scriptsize $dx^1(\sigma_1\cosh{x^2}\cos{x^3}-\sigma_2\cosh{x^2}\sin{x^3}-\sigma_3\sinh{x^2})$}\\
			
			\vspace{1mm}
			
			& &  &{\scriptsize$+dx^2(\sigma_2\cos{x^3}+\sigma_1\sin{x^3})+dx^3\sigma_3$}\\
			
			\vspace{-1mm}
			
			{\scriptsize ${IX}$}&
			{\scriptsize$[\sigma_1,\sigma_3]=-\sigma_2,[\sigma_2,\sigma_3]=\sigma_1,[\sigma_1,\sigma_2]=\sigma_3$}&
			{\scriptsize 	$\cal{A}$} &
			{\scriptsize $dx^1(\sigma_1\cos{x^2}\cos{x^3}-\sigma_2\cos{x^2}\sin{x^3}+\sigma_3\sin{x^2})$}\\
			
			\vspace{1mm}
		
		& &  &{\scriptsize$+dx^2(\sigma_2\cos{x^3}+\sigma_1\sin{x^3})+dx^3\sigma_3$}\\

			\hline\hline
		\end{tabular}
	\end{center}
	Where in the above table, the $\{\sigma_{i},~i=1,2,3\}$ are non-coordinate basis and their dual are $\{\sigma^{i},~i=1,2,3\}$. The relation between coordinate and non-coordinate basis is given by $\sigma^i={e_\alpha}^i(x) \,dx^{\alpha}$.
	Hence,  the non-zero components of the Riemann  and Ricci tensors of the metric \eqref{metric}   are expressed as \cite{nakahara}
	\begin{eqnarray}\label{rnon}
	\begin{split}
	R^i_{~jkl}&=\Gamma_{l j}^{m}\Gamma_{km}^{i}-\Gamma_{kj}^{m}\Gamma_{lm}^{i}-f_{kl}^{~~m}\Gamma_{mj}^{i},\\
	R^0_{~j0l}&=\dot{\Gamma_{lj}^{0}}+\Gamma_{l j}^{m}\Gamma_{0m}^{0}-\Gamma_{0j}^{m}\Gamma_{lm}^{0},\\
	R^0_{~jkl}&=\Gamma_{l j}^{m}\Gamma_{km}^{0}-\Gamma_{kj}^{m}\Gamma_{lm}^{0}-f_{kl}^{~~m}\Gamma_{mj}^{0},
	\end{split}
	\end{eqnarray}
	\begin{eqnarray}\label{ricnon}
	\begin{split}
	R_{ij}&=\dot{\Gamma_{j i}^{0}}
	+\Gamma_{j i}^{e}\Gamma_{d e}^{d}+\Gamma_{j i}^{0}\Gamma_{d 0}^{d}+\Gamma_{j i}^{0}\Gamma_{0 0}^{0}
	-\Gamma_{d i}^{e}\Gamma_{j e}^{d}-\Gamma_{d i}^{0}\Gamma_{j 0}^{d}-\Gamma_{0 i}^{e}\Gamma_{j e}^{0}
	-f_{d j}^{~~e}\Gamma_{e i}^{d},\\
	R_{0j}&=\Gamma_{j 0}^{e}\Gamma_{d e}^{d}-\Gamma_{d 0}^{e}\Gamma_{j e}^{d},\\
	R_{00}&=\dot{\Gamma_{d 0}^{d}}+\Gamma_{00}^{e}\Gamma_{d e}^{d}+\Gamma_{00}^{0}\Gamma_{d 0}^{d}-\Gamma_{d 0}^{e}\Gamma_{0 e}^{d},
	\end{split}
	\end{eqnarray}
	where the dot symbol stands for derivative with respect to $t$ and  the connection coefficients are given by
	\begin{eqnarray}\label{gnon}
	\begin{split}
	\Gamma_{ij}^{k}=-g^{d k}(f_{i d}^{~~e}g_{e j}+f_{j d}^{~~e}g_{e i})+f_{ij}^{~~k},\\
	\Gamma_{ij}^{0}=-\frac{1}{2}g^{00}\dot{g_{ij}},~~\Gamma_{00}^{0}=\frac{1}{2}g^{00}\dot{g_{00}},~~\Gamma_{i0}^{i}=\frac{1}{2}g^{ii}\dot{g_{ii}}.
	\end{split}
	\end{eqnarray}
	The Riemann  and Ricci tensors in coordinate basis are obtained by  multiplying the vielbeins $e_{\alpha}^{~i}$, for example $R_{\alpha\beta}=e_{\alpha}^{~i}e_{\beta}^{~j}R_{ij}$ and $R_{\alpha0}=e_{\alpha}^{~i}R_{i0}$. 	
	\section*{Appendix B}	
	In this appendix, for instructive purposes, we are going to  write the given calculation in the sections $3$ and $4$ with more details for an isotropic example, i.e. $a_1=a_2=a_3$, in the Bianchi-type $V$. In this case, using \eqref{rnon}-\eqref{ricnon} and the metric \eqref{metric} (along with \eqref{d}) we have
	\begin{eqnarray}\label{201}
	\begin{split}
	R^1_{~212}&=R^1_{~313}=R^2_{~323}=a_1^2H_1^2-1\\
	R^{1}_{~010}&=	R^{2}_{~020}=	R^{3}_{~030}=-H_1^2-\dot{H}_{1},\\
	R_{11}&=R_{22}=R_{33}=-2-g_{00}^{-1}a_1^2(\dot{H}_{1}+3\,H_1^2-H_1\dot{(\ln{g_{00}})}),\\
	R_{00}&=3(-\dot{H}_{1}-H_1^2+H_1\dot{(\ln{g_{00}})}),\\
	\Gamma_{ij}^{k}&=-1,~~
	\Gamma_{ij}^{0}=-g^{00}a_1^{2}H_1,~~\Gamma_{00}^{0}=\frac{1}{2}\dot{\ln{g_{00}}},~~\Gamma_{i0}^{i}=H_1,
	\end{split}
	\end{eqnarray}
	The constant terms in the Riemann and Ricci tensors and connection coefficients are related to the structure constants $f_{ij}^{~k}$, which differ in the Bianchi models and contribute in the Bianchi dependent terms, $V$, that have been introduced in the section $3$. The other terms in \eqref{201} are independent of the  structure constants and appear in all Bianchi-type,  taking part in  Bianchi independent terms $K$. In this isotropic case the $\beta$-functions of \eqref{betaGR} and \eqref{betafR} cast the following forms
	\begin{eqnarray}\label{2}
	\begin{split}
	\frac{1}{\alpha'}\bar{\beta}^{G~1}_{~1}=&	\dot{H_{1}}+3 H_1^2 + H_{1} \dot{\phi}-H_{1}\,\dot{(\ln{g_{00}})}+V_{1}^{(1)}-\frac{1}{2}A^2{{a_1}}^{-6}
	+\alpha'(	V_{1}^{(2)}+K_1),
	\end{split}
	\end{eqnarray}
	\begin{eqnarray}\label{4}
	\frac{1}{\alpha'}\bar{\beta}^{G~0}_{~0}=	3(\dot{H_1}+ H_1^2)+\ddot{\phi}-\dot{(\ln{g_{00}})}(\dot{\phi}+3\dot{H_1})-\alpha'(V_{0}^{(2)}+K_0),
	\end{eqnarray}
	\begin{eqnarray}\label{1}
	\begin{split}
	\frac{1}{\alpha'}	\tilde{\beta}^{\phi}=-2\ddot{\phi}-\dot{\phi}^2&-3\,(2\,\dot{\phi} H_1+V_{1}^{(1)}+{H_1}^2-2\,\dot{H_1}
	+3 H_1^2)\\
	&-2(\phi'+ H_1)\dot{(\ln{g_{00}})}+\frac{1}{2}A^2{a_1}^{-6}
	+\alpha'[K_{\phi}+V_{\phi}^{(2)}],
	\end{split}
	\end{eqnarray}
	where
	\begin{eqnarray}\label{205}
	\begin{split}
	V_{1}^{(1)}&=-2a_1^{-2}g_{00}\\
	V_{1}^{(2)}+K_1&=\dot{H_{1}}^2+2H_1\dot{H_{1}}+3H_1^4-4H_1^2a_1^{-2}+2a_1^{-4}+\frac{3}{8}A^2a_1^{-12}+\frac{1}{8}A^2(-H_1^2(92\,f+88)a_1^{-6}+4(f+7)a_1^{-8})\\
	V_{0}^{(2)}+K_0&=-3(H_1^2+\dot{H_1})^2+3A^2a_1^{-6}(\frac{1}{2}\dot{H_1}+(f-1)H_1^2)\\
	K_{\phi}&=
	-3(\dot{H_1}^2+2H_{1}^2\dot{H_1}+{H^4_{1}})-\frac{1}{4}A^2{a_1}^{-6}(3H_{1}^2-5A^2{a_1}^{-6})
	\\
	V_{\phi}&=\frac{1}{4} (-3 a_1^{-6} H_1^2+5 A^2a_1^{-12}+3 a_1^{-8}) A^2.
	\end{split}
	\end{eqnarray}
	Now, the equation of \eqref{1} with using \eqref{2} and \eqref{4} can be rewritten in the following form
	\begin{eqnarray}\label{000}
	\begin{split}
	-\ddot{\phi}-\dot{\phi}^2-3\,\dot{\phi} H_1-A^2{a_1}^{-6}
	+\alpha'[K_{\phi}+V_{\phi}^{(2)}+3V_{1}^{(2)}+3K_1-V_{0}^{(2)}-K_0]=0.
	\end{split}
	\end{eqnarray}
	By a time redefinition \eqref{tau} and using the following relations for any function $B$
	\begin{eqnarray}\label{207}
	\begin{split}
	B'=a_1^{3}e^{\phi}\dot{B},~~~
	B''=a_1^{6}e^{2\phi}(\ddot{B}+\dot{B}(3H_1+\dot{B})),
	\end{split}
	\end{eqnarray}
	the \eqref{000}	recast the following form
	\begin{eqnarray}\label{22}
	\begin{split}
	-{\phi}''-A^2e^{2\phi}
	+\alpha'[\hat{K}_{\phi}+\hat{V}_{\phi}^{(2)}+3\hat{V}_{1}^{(2)}+3\hat{K}_1-\hat{V}_{0}^{(2)}-\hat{K}_0],
	\end{split}
	\end{eqnarray}
	where we have used the hatted symbols for  the terms in $\tau$ coordinate with a $a_1^{6}e^{2\phi}$ factor which are obtained from \eqref{205}  by using \eqref{207}.
	Now, using the $\alpha'$ expansions
	of \eqref{18}-\eqref{17} and the series expansion of $e^{\phi}$ up to first order of $\alpha'$ as following
	\begin{eqnarray} \label{fise}
	e^{\phi}=e^{\phi_0}(1+\alpha' \phi_1),
	\end{eqnarray}
	the  \eqref{22} equation leads to the equations 
	\eqref{fio} and \eqref{fi1} in the zeroth and first order of $\alpha'$. Their solutions have been given in \eqref{fioa} and \eqref{fi1s}. For obtaining the $\xi_4$-containing term in the \eqref{fi1s} the following relation, which is obtained by integration by part, has been used
	\begin{eqnarray}\label{int1}
	\begin{split}
	\frac{1}{n}\bigg[\tanh(n\tau)&\int (n\tau \tanh(n\tau)-1)\big(\frac{2n^2\xi_4}{\cosh^2(n\tau)}+n\tanh(n\tau)\xi_4'\big)\,d\tau\\
	&-(n\tau \tanh(n\tau)-1)\int \tanh(n\tau)\big(\frac{2n^2\xi_4}{\cosh^2(n\tau)}+n\tanh(n\tau)\xi_4'\big)\,d\tau\big]=-n\tanh(n\tau)\int \xi_4d\tau.
	\end{split}
	\end{eqnarray}

	Then, adding the \eqref{22} (with a $-\frac{1}{2}$ coefficient) to the \eqref{2}, results in an equation of type \eqref{iifinal} which in the zeroth and first order of $\alpha'$ reads the following equations, respectively
	\begin{eqnarray}\label{211}
	{(\ln a_{1}^2 e^{\phi_{0}})}''-2a_{1}^4 e^{2\phi_{0}}=0,
	\end{eqnarray}
	\begin{eqnarray}\label{212}
	X_1''-2\,a_{1}^4 e^{2\phi_{0}}(2X_1+\xi_4)-\frac{1}{2}\,	{(\ln a_{1}^2 e^{\phi_{0}})}'\xi_4'+g_1=0,
	\end{eqnarray}
	with
	\begin{eqnarray}
	g_1=-\frac{1}{2}(\hat{K}_{\phi}+\hat{V}_{\phi}^{(2)}+\hat{V}_{1}^{(2)}+\hat{K}_1-\hat{V}_{0}^{(2)}-\hat{K}_0),
	\end{eqnarray}
	where we have used the following relation for rewriting the $H_i$ in the $\tau$ coordinate up to the first order of $\alpha'$
	\begin{eqnarray}\label{111}
	H_{i}=\dot{(\ln a_{i})}=a^{-3}e^{-\phi}[(\ln a_{i0})'+\frac{\alpha'\,X_i'}{1+\alpha'\,X_i}]\simeq a^{-3}e^{-\phi}[(\ln a_{i0})'+\alpha'\,X_i'].
	\end{eqnarray}
	The solutions of equations \eqref{211}-\eqref{212}  are given by
	\begin{eqnarray}
	a_{1}^2 e^{\phi_{0}}=\frac{p}{\sinh(p\tau)},
	\end{eqnarray}
	\begin{eqnarray}\label{216}
	\begin{split}
	X_1&={ Q_3}\,{\rm coth} (p\tau)+{Q_4}\, ( p\,\tau\,{\rm coth} (p\tau)-1 ) -\frac{1}{2}p\,{\rm coth} (p\tau)\int \xi_4\, d\tau\\
	&+\frac{1}{p}\bigg({\rm coth} (p\tau)\int ( p\,\tau\,{\rm coth} (p\tau)-1 ) (g_1)\,d\tau-( p\,\tau\,{\rm coth} (p\tau)-1 )\int {\rm coth} (p\tau) (g_1)\,d\tau\bigg).
	\end{split}
	\end{eqnarray}
	In finding the above answer the following relation has been used
	\begin{eqnarray}\label{int2}
	\begin{split}
	\frac{1}{p}\bigg[\coth(p\tau)&\int (p\tau \coth(p\tau)-1)\big(\frac{2p^2\xi_4}{\sinh^2(p\tau)}-p\coth(p\tau)\xi_4'\big)\,d\tau\\
	&-(p\tau \coth(p\tau)-1)\int \coth(p\tau)\big(\frac{2p^2\xi_4}{\sinh^2(p\tau)}-p\coth(p\tau)\xi_4'\big)\,d\tau\big]=p\coth(p\tau)\int \xi_4d\tau.
	\end{split}
	\end{eqnarray}
	
	Now, the time-time component of $\beta$-function equation \eqref{4} needs to be solved. In the new time coordinate $\tau$, using the \eqref{18}-\eqref{17}, \eqref{fise} and \eqref{111} it reads
	\begin{eqnarray}
	\begin{split}
	\phi_0''&-\phi_0'(3\,(\ln{a_1})'+\phi_0')+3\,(\ln{a_1})''-3\,(\ln{a_1})'(3(\ln{a_1})'+\phi_0')\\
	&+\alpha'[3\,X_1''-\frac{1}{2}\phi_1''+6(3(\ln{a_1})'+\phi_0')X_1'+\phi_0'\phi_1'+6(\ln{a_1})'X_1'+\hat{K}_0+\hat{V}_0]=0
	\end{split}
	\end{eqnarray}	
	Using the \eqref{fio}, \eqref{fi1}, \eqref{211}, \eqref{212} it can be rewritten as following
	\begin{eqnarray}
	\begin{split}
	\big[&\frac{3}{2}(\ln a_{1}^2e^{\phi_0})'^2-6\,a_1^4e^{2\phi_0}-\phi_0'^2-A^2e^{2\phi_0}\big](1+2\alpha'\xi_4)\\
	&+\alpha'\big[3(3(\ln a_{1}^2e^{2\phi_0})'-\phi_0') X_1'-\phi_0'\phi_1'+A^2e^{2\phi_0}\phi_1 -12\,a_1^4e^{2\phi_0}+3\,g_1+\frac{\rho}{2}+\hat{K}_{0}+\hat{V}_{0}^{(2)}\big]=0.
	\end{split}
	\end{eqnarray}	
	Then, following the given discussion in the first paraphrase of page $7$, solution of the above equation gives the $\xi_4$ and 
	$
	q_2=(n^2-3p^2)/2
	$
	.	The explicit answers with $p=n=1$ which are obtained after performing the integrals of $X_1$ and $\phi_1$ can be found in \eqref{v11}-\eqref{v12}.
	\section*{Appendix C}
	In this appendix some sample examples for discussed models in the section $4$ are presented.
	\subsection*{Bianchi-type I}	
	In Bianchi-type $I$ with choosing $3p_1=3p_2=3p_3= n=1$ the first $\alpha'$-corrections of metric and dilaton are found as following \footnote{The ${ Li_2}(\tau)$ is  Polylogarithm function.}
	\begin{eqnarray}\label{ix1}
	\begin{split}
	X_i=&\frac{1}{AL^3}\big[ \frac{1}{27}{ Li_2} \left( -{{\rm e}^{2\,\tau}
	} \right) -{\frac {\ln  \left( {{\rm e}^{2\,\tau}}+1 \right)  \left( 
			87\,{ f}-136 \right) }{216}}-{\frac {203\,{ f}+236}{144\,
			\left( {{\rm e}^{2\,\tau}}+1 \right) ^{2}}}+\frac{1}{24}{\frac {29\,{ f
			}+38}{ \left( {{\rm e}^{2\,\tau}}+1 \right) ^{3}}}\\
	&+{\frac {29\,{ f
			}-24}{72\,({{\rm e}^{2\,\tau}}+1)}}+{\frac {\tau\, \left( -1766+609\,{
				f}+128\,\ln  \left( 2 \right)  \right) }{1728}}+{\frac {2\,{\tau}
			^{2}}{27}} 							\big]
	+\frac{p_i}{2}
	\int\xi_4 \,d\tau+r_i\tau+c_i,
	\end{split}
	\end{eqnarray}
	\begin{eqnarray}\label{48}
	\begin{split}
	\phi_1&={ Q_1}\,\tanh (\tau ) +{ Q_2}\, (\tau\,
	\tanh (\tau) -1 )-\tanh(\tau)\int \xi_4 \,d\tau\\
	&+\frac {1}{2{L}^{3}A} \bigg[ \frac{1}{9}\tanh \left( \tau \right) 
	\big( -\frac { 1
	}{20 \left( {{\rm e}^{2\,\tau}}+1 \right) ^{5}}\big(\left(  \left( 5500\,f+5100 \right) \tau+
	1175\,f+6371 \right) {{\rm e}^{2\,\tau}}+ (  \left( -4000\,f_
	{1}-11680 \right) \tau\\
	&-7275\,f-3243 ) {{\rm e}^{4\,\tau}}+
	\left(  \left( 12000\,f+9560 \right) \tau-8775\,f-8523
	\right) {{\rm e}^{6\,\tau}}+ \left( 50\,f-880\,\tau+206 \right) {
		{\rm e}^{8\,\tau}}\\
	&+ \left( -100\,f+148 \right) \tau-375\,f+885\big)+4\,{\tau}^{2}+1/10\,
	\ln  \left( {{\rm e}^{2\,\tau}}+1 \right)  \left( -40\,\tau-77+25\,f_{
		1} \right) -2\,{ Li_2} \left( -{{\rm e}^{2\,\tau}} \right)\\
	& +{ {
			77\,\tau}/{5}}-5\,f\,\tau \big) -2\, \left( \tau\,\tanh \left( 
	\tau \right) -1 \right)  \big( -2/9\,\ln  \left( \cosh \left( \tau
	\right)  \right) +2/9\,\tau+{ { \left( 125\,f-1 \right) 
			\left( \tanh \left( \tau \right)  \right) ^{2}}/{96}}\\
	&+{\frac {\sinh
			\left( \tau \right) }{ \left( \cosh \left( \tau \right)  \right) ^{3}
		} \left( { {145\,f}/{72}}+{ {173}/{90}} \right) }+1/36\,
	\left( { {37}/{5}}-5\,f \right) \tanh \left( \tau \right) \\
	&-{\frac { \left( 39+25\,f \right)  \left( \sinh \left( \tau
			\right)  \right) ^{2}}{ 32\left( \cosh \left( \tau \right)  \right) ^{4
	}}}-{\frac {\sinh \left( \tau \right)  \left( 75\,f+94
			\right) }{ 40\left( \cosh \left( \tau \right)  \right) ^{5}}} \big) 
	\bigg]+\varphi_0
	,
	\end{split}
	\end{eqnarray}
	where the $r_i$, $c_i$ and $\varphi_0$ are constants of integrating and	The initial value equation \eqref{xi4} gives
	\begin{eqnarray}\label{53}
	\begin{split}
	\xi_4&=\frac {1}{5760\,{L}^{3}A} \big[ 960\, \left( 1-7/3\,\tanh
	\left( \tau \right)  \right) \ln  \left( \cosh \left( \tau \right) 
	\right) -28335\,f-960\,\tau+9260\\
	&+2240\,\tanh \left( \tau \right) 
	\left( \tau-{ {1959\,f}/{448}}-{ {5133}/{1120}} \right) +
	{ \left( \cosh \left( \tau \right)  \right) ^{-4}}\big( \left( 32940\,f+1960 \right)  \left( \cosh \left( \tau
	\right)  \right) ^{2}\\
	&+ \left( -28860\,f-20448 \right) \tanh
	\left( \tau \right)  \left( \cosh \left( \tau \right)  \right) ^{2}-
	4605\,f-11220+ \left( 34155\,f+43254 \right) \tanh \left( \tau
	\right) \big) \big]
	+{\sum r_i}
	.
	\end{split}
	\end{eqnarray}
	
	\subsection*{Bianchi-type II}	
	For simplicity, with the special choice of parameterization as  $p_1=p_2=p_3=n=1$, the $\alpha'$-corrections of metric and dilaton  are found as following
	\begin{eqnarray}\label{iix1}
	\begin{split}
	X_1=& Q_3\,\tanh(\tau)+Q_4\,(\tau \tanh(\tau)-1)-\frac{1}{2}\tanh(\tau)\int \xi_4 \,d\tau\\
	&+\frac {\tanh \left( \tau \right) }{{2A}} \big( \frac {-{
			{\rm e}^{-6\,\tau}}  }{46080\, \left( \cosh \left( \tau \right) 
		\right) ^{6}}\big(  \left(  \left( 4896\,f+28422 \right) 
	\tau-2088\,f+32859 \right) {{\rm e}^{2\,\tau}}+ (  ( -
	8280\,f-107100 ) \tau\\
	&-18864\,f-57258 ) {{\rm e}^{4
			\,\tau}}+ \left(  \left( 23760\,f+74700 \right) \tau-16692\,f-
	85004 \right) {{\rm e}^{6\,\tau}}+ \left( -810\,\tau+252\,f-2031
	\right) {{\rm e}^{10\,\tau}}\\
	&+ \left( -17820\,\tau+1386\,f-1248
	\right) {{\rm e}^{8\,\tau}}+ \left( -504\,f+1632 \right) \tau-
	1050\,f+4330 \big)+{\frac {9\,{\tau}^{2}}{8}}\\
	&+{\frac {  \left( -135\,\tau+2\,f-257 \right) }{
			120}}\ln  \left( 1+{
		{\rm e}^{2\,\tau}} \right)-{\frac {9\,{\it Li_2} \left( -{{\rm e}^{2\,\tau}} \right) 
		}{16}}-{\frac {11\,\tau\, \left( -24+f \right) }{30}} \big)\\
	& -
	\frac {\tau\,\tanh \left( \tau \right) -1}{1920\,{A}} \big( -
	1080\,\ln  \left( \cosh \left( \tau \right)  \right) +750\,f+1080
	\,\tau-1525-176\, ( {\frac {21\,f}{11}}-{\frac {68}{11}}
	) \tanh \left( \tau \right) \\
	&+\frac {1}{ \left( \cosh \left( 
		\tau \right)  \right) ^{6}} \big(  \left( -1320\,f-2970 \right) 
	\left( \cosh \left( \tau \right)  \right) ^{4}+1232\,\sinh \left( 
	\tau \right)  ( {\frac {72\,f}{77}}+{\frac {43}{14}} ) 
	\left( \cosh \left( \tau \right)  \right) ^{3}\\
	&+ ( 1350\,f+
	9195 )  \left( \cosh \left( \tau \right)  \right) ^{2}-1056\,
	\sinh \left( \tau \right)  ( {\frac {17\,f}{22}}+{\frac {62}{
			11}} ) \cosh \left( \tau \right) -780\,f-4700 \big) 
	\big) 	
	\end{split}
	\end{eqnarray}
	\begin{eqnarray}
	\begin{split}
	X_2=&-X_1 +\frac{1}{2}\int \xi_4 \,d\tau
	-\frac {1}{{30 A}} \big( -{\frac {45\,f-92}{1+{{\rm e}^{2\,\tau}}}}-{\frac {90\,f+
			191}{4 \left( 1+{{\rm e}^{2\,\tau}} \right) ^{2}}}-{\frac {3\,(55\,f
			+179)}{ \left( 1+{{\rm e}^{2\,\tau}} \right) ^{4}}}+{\frac {195\,f+
			604}{ \left( 1+{{\rm e}^{2\,\tau}} \right) ^{3}}}\\
	&+\frac{1}{8}(-{ {105\,{Li_2 
			} \left( -{{\rm e}^{2\,\tau}} \right) }}-2{ {105\,{\tau}^
			{2}}}+\ln  \left( 1+{{\rm e}^{2\,\tau}} \right)  \left( 360\,f
	-631 \right) -\tau\, \left( -1546+210\,\ln  \left( 2 \right) 
	+375\,f \right))
	\big) 
	\end{split}
	\end{eqnarray}
	\begin{eqnarray}
	\begin{split}
	\phi_1=& Q_1\,\tanh(\tau)+Q_2\,(\tau \tanh(\tau)-1)+\frac {\tanh \left( \tau \right) }{2A} \big( \frac {{{\rm e}^{-
				6\,\tau}}  }{23040\, \left( \cosh \left( \tau \right) 
		\right) ^{6}}\big(  (  \left( 21600\,\tau+3240 \right) f+40518
	\,\tau\\
	&+35451 ) {{\rm e}^{2\,\tau}}+ \left(  \left( -40680\,\tau-
	58320 \right) f-130140\,\tau-88362 \right) {{\rm e}^{4\,\tau}}+
	(  \left( 79920\,\tau-61420 \right) f+117900\,\tau\\
	&-118636
	) {{\rm e}^{6\,\tau}}+ \left( -810\,\tau+420\,f-1839
	\right) {{\rm e}^{10\,\tau}}+ \left( -17820\,\tau+2310\,f-192
	\right) {{\rm e}^{8\,\tau}}+ \left( -840\,\tau-1750 \right) f\\
	&+
	1248\,\tau+3530 \big)+{\frac {
			\left( 70\,f-135\,\tau-239 \right) }{60}}\ln  \left( 1+{{\rm e}^{2\,\tau}} \right) +\frac{9}{4}{\tau}^{2}-{\frac 
		{9\,{Li_2} \left( -{{\rm e}^{2\,\tau}} \right) }{8}}-\frac{1}{30}
	\tau\, \left( 70\,f-239 \right)  \big) \\
	&-\frac {\tau\,\tanh
		\left( \tau \right) -1}{960\,A} \big( -1080\,\ln  \left( \cosh
	\left( \tau \right)  \right) +1080\,\tau+1810\,f-605+ \left( -560
	\,f+832 \right) \tanh \left( \tau \right) \\
	&+\left( \cosh \left( \tau \right) 
	\right) ^{-6}\big( \left( -4440\,
	f-5370 \right)  \left( \cosh \left( \tau \right)  \right) ^{4}+
	\left( 4160\,f+6056 \right) \sinh \left( \tau \right)  \left( 
	\cosh \left( \tau \right)  \right) ^{3}\\
	&+ \left( 5610\,f+12315
	\right)  \left( \cosh \left( \tau \right)  \right) ^{2}+ \left( -3600
	\,f-7968 \right) \sinh \left( \tau \right) \cosh \left( \tau
	\right) -2980\,f-6340\big) \big)-\tanh(\tau)\int \xi_4 \,d\tau, 
	\end{split}
	\end{eqnarray}
	The initial value equation \eqref{xi4} gives
	\begin{eqnarray}\label{53ii}
	\begin{split}
	\xi_4=&	Q_{4}( 2+{\frac {\sinh \left( \tau \right) \tau}{ \left( \cosh
			\left( \tau \right)  \right) ^{3}}}- \left( \cosh \left( \tau
	\right)  \right) ^{-2} ) +{\frac {{ Q_3}\,\sinh \left( 
			\tau \right) }{ \left( \cosh \left( \tau \right)  \right) ^{3}}}
	-
	\,{\frac {{{\rm e}^{-2\,\tau}} }{16 A
			\, \left( \cosh \left( \tau \right)  \right) ^{6}}}\big( 4\,\sinh \left( \tau \right) 
	\cosh \left( \tau \right)  ( 16\, \left( \cosh \left( \tau
	\right)  \right) ^{2}\\
	&-16\,f-37 ) +71\, \left( \cosh \left( 
	\tau \right)  \right) ^{4}+ \left( -80\,f-165 \right)  \left( 
	\cosh \left( \tau \right)  \right) ^{2}+64\,f+148 \big) +{\frac {\ln 
			\left( \cosh \left( \tau \right)  \right) }{2A}}\\
	&+\frac {1}{480\,A
	} \big( 240\,\ln  \left( \cosh \left( \tau \right)  \right) -240
	\,\tau-20\,f-2295+ \left( -32\,f+1948 \right) \tanh \left( 
	\tau \right) \\
	&+\left( \cosh \left( \tau
	\right)  \right) ^{-6}\big(2070\, \left( \cosh \left( \tau \right) 
	\right) ^{4}+ \left( -16\,f-1156 \right) \sinh \left( \tau
	\right)  \left( \cosh \left( \tau \right)  \right) ^{3}+ \left( 60\,f
	_{1}-4815 \right)  \left( \cosh \left( \tau \right)  \right) ^{2}\\
	&+
	\left( -72\,f+2568 \right) \sinh \left( \tau \right) \cosh
	\left( \tau \right) -40\,f+5040\big) \big) .
	\end{split}
	\end{eqnarray}
	
	\subsection*{Bianchi-type III}
	With choosing a parameterization of  $p_1=p_2=p_3=n=1$ and calculating the integrals of \eqref{fi1s}, \eqref{94} and \eqref{95}, we obtain
	\begin{eqnarray}\label{iiix1}
	\begin{split}
	X_1&=X_3=-X_2+{ Q_3}\,{\rm coth} (\tau)+{Q_4}\, ( \tau\,{\rm coth} (\tau)-1 ) +(\frac{1}{2}-{\rm coth} (\tau))\int \xi_4\, d\tau\\
	&+\frac {{\rm coth} \left(\tau\right)}{A} \big( {\frac {11\,{
				{\rm e}^{-2\,\tau}}}{16} ( 3\,f+{\frac {247}{22}}+2\,f\,
		\tau+{\frac {45\,\tau}{11}} ) }+\frac{1}{8} ( -{\frac {35}{4}}-7
	\,\tau ) {{\rm e}^{-4\,\tau}}-\frac {1}{ 24\left( {{\rm e}^{2\,\tau}}+1 \right) ^{4}}( (  \left( 398
	\,\tau+63 \right) f\\
	&-130\,\tau-145 ) {{\rm e}^{2\,\tau}}+
	\left(  \left( 360\,\tau-347 \right) f-192\,\tau-461 \right) {
		{\rm e}^{4\,\tau}}+ \left(  \left( 30\,\tau-55 \right) f-186\,\tau
	+5 \right) {{\rm e}^{6\,\tau}}+ \left( 68\,\tau+7 \right) f\\
	&-124\,
	\tau-135)+\frac{f+25}{16} \left( -{ 
		Li_2} \left( -{{\rm e}^{2\,\tau}} \right) +2\,{\tau}^{2} \right) 
	+\frac{1}{12}\tau\, \left( 103\,f_
	{1}+232 \right) -\frac{1}{24}\ln  \left( {{\rm e}^{2\,\tau}}+1 \right) 
	( 3\,f\,\tau+103\,f\\
	&+75\,\tau+232 )  \big) -\frac{1}{48}
	\,\frac { 1 }{A \left( \cosh \left( \tau
		\right)  \right) ^{3}}\left( \tau\,{\rm coth} \left(\tau\right)-1 \right) 
	\big( -6\, \left( \cosh \left( \tau \right)  \right) ^{3} \left( f_{
		1}+25 \right) \ln  \left( \cosh \left( \tau \right)  \right) \\
	&-336\,
	\left( \cosh \left( \tau \right)  \right) ^{7}+336\,\sinh \left( \tau
	\right)  \left( \cosh \left( \tau \right)  \right) ^{6}+ \left( 132\,
	f+606 \right)  \left( \cosh \left( \tau \right)  \right) ^{5}+
	\left( -132\,f-438 \right) \sinh \left( \tau \right)  \left( 
	\cosh \left( \tau \right)  \right) ^{4}\\
	&+ \left(  \left( 6\,\tau+39
	\right) f+150\,\tau+45 \right)  \left( \cosh \left( \tau \right) 
	\right) ^{3}+ \left( 68\,f-124 \right) \sinh \left( \tau \right) 
	\left( \cosh \left( \tau \right)  \right) ^{2}\\
	&+ \left( -39\,f-45
	\right) \cosh \left( \tau \right) + \left( 58\,f+76 \right) \sinh
	\left( \tau \right)  \big)
	,
	\end{split}
	\end{eqnarray}
	\begin{eqnarray}
	\begin{split}
	X_2&=-\frac{1}{2}\int \xi_4\, d\tau+\frac {1}{24\,{ A}} \big( -\ln  \left( {{\rm e}^{2\,\tau}}+1 \right)  \left( 28\,f+89
	\right) -1/8\, \left( -348\,f-1560 \right) {\tau}^{2}+1/8\,\tau\,
	( 348\,f\,\ln  \left( 2 \right)\\
	& +1560\,\ln  \left( 2
	\right) +161\,f+952 ) -{\frac {27\,{{\rm e}^{-2\,\tau}}f_{1
		}}{4}}+3/4\,{Li_2} \left( -{{\rm e}^{2\,\tau}} \right)  \left( 
	29\,f+130 \right) -{\frac {41\,f+100}{ 2\left( {{\rm e}^{2
					\,\tau}}+1 \right) ^{2}}}\\
	&+{\frac {29\,f+38}{ \left( {{\rm e}^{2\,
					\tau}}+1 \right) ^{3}}}+{\frac {15\,{{\rm e}^{-4\,\tau}}}{8}}+{
		\frac {2\,(5\,f+43)}{{{\rm e}^{2\,\tau}}+1}}-{\frac {279\,{{\rm e}^{-2
					\,\tau}}}{8}}
	\big) ,
	\end{split}
	\end{eqnarray}
	\begin{eqnarray}
	\begin{split}
	\phi_1=&{ Q_1}\,\tanh (\tau )+Q_2(\tau\tanh(\tau)-1)-\tanh(\tau)\int \xi_4 \,d\tau +\frac {\tanh \left( \tau \right) }{A} \big({\frac {
			\left( -105-84\,\tau \right) {{\rm e}^{-4\,\tau}}}{68}}\\
	&+ 1/32\, \left( 92\,f
	\,\tau+138\,f+454\,\tau+513 \right) {{\rm e}^{-2\,\tau}}+\frac {
		1}{120\,
		\left( {{\rm e}^{2\,\tau}}+1 \right) ^{5}}(\left(  \left( 880\,\tau-1885 \right) f+9400\,\tau-2377 \right) {
		{\rm e}^{2\,\tau}}\\
	&+ \left(  \left( 6680\,\tau+1125 \right) f+20540
	\,\tau-459 \right) {{\rm e}^{4\,\tau}}+ \left(  \left( -420\,\tau+2685
	\right) f+8940\,\tau+1041 \right) {{\rm e}^{6\,\tau}}+ ( 
	\left( 690\,\tau+80 \right) f\\
	&+3840\,\tau-712 ) {{\rm e}^{8
			\,\tau}}+ \left( 290\,\tau-405 \right) f+2984\,\tau-165)-{\frac {5\,  \left( 13\,f+55 \right) }{16
	}}{ Li_2}
	\left( -{{\rm e}^{2\,\tau}} \right)\\
	&- \left(  \left( 
	975\,\tau+140 \right) f+4125\,\tau-142 \right)\ln  \left( {{\rm e}^{2\,\tau}}+1 \right)/120+5/8\,{\tau}
	^{2} \left( 13\,f+55 \right) +1/30\,\tau\, \left( 70\,f-71
	\right)  \big) 
	\\
	&-\frac {\tau\,\tanh \left( \tau \right) -1}{480\,A
	} \big( -3900\, ( f+{\frac {55}{13}} ) \ln  \left( 
	\cosh \left( \tau \right)  \right) -5040\, \left( \cosh \left( \tau
	\right)  \right) ^{4}+5040\,\sinh \left( \tau \right)  \left( \cosh
	\left( \tau \right)  \right) ^{3}\\
	&+ \left( 2760\,f+18660 \right) 
	\left( \cosh \left( \tau \right)  \right) ^{2}+ \left( -2760\,f-
	16140 \right) \sinh \left( \tau \right) \cosh \left( \tau \right) +
	\left( 3900\,\tau-3195 \right) f\\
	&+16500\,\tau-8325+ \left( -580\,f
	_{1}-5968 \right) \tanh \left( \tau \right) +\frac { 1}{ \left( \cosh
		\left( \tau \right)  \right) ^{5}}(\left( 60\,f
	-840 \right)  \left( \cosh \left( \tau \right)  \right) ^{3}\\
	&+ \left( 
	375\,f+585 \right) \cosh \left( \tau \right) + \left(  \left( 340
	\,f+1696 \right)  \left( \cosh \left( \tau \right)  \right) ^{2}-
	900\,f-1128 \right) \sinh \left( \tau \right) )\big) 
	,
	\end{split}
	\end{eqnarray}
	\begin{eqnarray}\label{53iii}
	\begin{split}
	\xi_4&=-2\,{ Q_4}-\frac {1}{960\,A} \bigg( 1920\, ( f+{\frac 
		{45}{32}} ) \ln  \left( \cosh \left( \tau \right)  \right) +
	\left( 3600\,f+13560 \right)  \left( \cosh \left( \tau \right) 
	\right) ^{2}-10800\, \left( \cosh \left( \tau \right)  \right) ^{4}\\
	&+
	\frac {1}{\sinh \left( \tau \right)  \left( \cosh \left( \tau \right) 
		\right) ^{5}} \big(  ( 8640\, \left( \cosh \left( \tau \right) 
	\right) ^{9}-8640\,\sinh \left( \tau \right)  \left( \cosh \left( 
	\tau \right)  \right) ^{8}+ \left( -1440\,f-19260 \right)  \left( 
	\cosh \left( \tau \right)  \right) ^{7}\\
	&+ \left( 1440\,\sinh \left( 
	\tau \right) f+14940\,\sinh \left( \tau \right)  \right)  \left( 
	\cosh \left( \tau \right)  \right) ^{6}+ \left( 2400\,f+10740
	\right)  \left( \cosh \left( \tau \right)  \right) ^{5}\\
	&+ \left( -1920
	\,f-9060 \right) \sinh \left( \tau \right)  \left( \cosh \left( 
	\tau \right)  \right) ^{4}+ \left( -480\,f-180 \right)  \left( 
	\cosh \left( \tau \right)  \right) ^{3}+ \left( 1200\,f+3780
	\right) \sinh \left( \tau \right)  \left( \cosh \left( \tau \right) 
	\right) ^{2}\\
	&+ \left( -480\,f+60 \right) \cosh \left( \tau
	\right) + \left( -720\,f-60 \right) \sinh \left( \tau \right) 
	) {{\rm e}^{-\tau}}+10800\, \left( \cosh \left( \tau \right) 
	\right) ^{10}\\
	&+ \left( -3600\,f-18960 \right)  \left( \cosh
	\left( \tau \right)  \right) ^{8}+ \left( 5780\,f+5608 \right) 
	\left( \cosh \left( \tau \right)  \right) ^{6}-1920\,\sinh \left( 
	\tau \right)  (  ( -{\frac {65}{32}}+\tau ) f\\
	&+{
		\frac {45\,\tau}{32}}-{\frac {549}{128}} )  \left( \cosh \left( 
	\tau \right)  \right) ^{5}+ \left( 260\,f+4036 \right)  \left( 
	\cosh \left( \tau \right)  \right) ^{4}+ \left( -2280\,f-2820
	\right) \sinh \left( \tau \right)  \left( \cosh \left( \tau \right) 
	\right) ^{3}\\
	&+ \left( -1540\,f-356 \right)  \left( \cosh \left( 
	\tau \right)  \right) ^{2}+ \left( 60\,f-15 \right) \sinh \left( 
	\tau \right) \cosh \left( \tau \right) -900\,f-1128 \big) 
	\bigg) \nonumber
	.
	\end{split}
	\end{eqnarray}
	\subsection*{Bianchi-type V}
	As an example In this Bianchi-type we set $p=1,~n=1,~q=0$ and find the following results for $X_i$ \eqref{vx1}, dilaton \eqref{fi1s} and $\xi_4$ \eqref{xi4}
	\begin{eqnarray}\label{v11}
	\begin{split}
	X_i&={ Q_3}\,{\rm coth} (\tau)+{Q_4}\, ( \tau\,{\rm coth} (\tau)-1 ) -\frac{1}{2}\,{\rm coth} (\tau)\int \xi_4\, d\tau+\frac {1}{3{A}} \bigg( {\rm coth} \left(\tau\right) [ 
	220\,\tau\,f-134\,\tau\\
	&-\ln  \left( {
		{\rm e}^{2\,\tau}}+1 \right)  \left( 110\,f-67 \right)+{\frac {2}{ \left( {{\rm e}^{2\,\tau}}+1 \right) ^{4}}}\big(\left(  \left( -446\,f+49 \right) \tau+52\,f-143 \right) {
		{\rm e}^{2\,\tau}}\\
	&+ \left(  \left( -330\,f+201 \right) \tau+452\,f
	_{1}+170 \right) {{\rm e}^{4\,\tau}}+ \left(  \left( 6\,f+219
	\right) \tau+52\,f-143 \right) {{\rm e}^{6\,\tau}}+ \left( -110\,
	f+67 \right) \tau
	\big)
	] \\
	&-{
		\frac { \sinh
			\left( \tau \right) }{ \left( \cosh \left( \tau \right)  \right) ^{3
	}}} \left( \tau\,{\rm coth} \left(\tau\right)-1 \right)   \left( 110\, \left( \cosh \left( \tau \right) 
	\right) ^{2}f-67\, \left( \cosh \left( \tau \right)  \right) ^{2}
	+58\,f+76 \right)\bigg)
	,
	\end{split}
	\end{eqnarray}
	\begin{eqnarray}
	\begin{split}
	\phi_1=&{ Q_1}\,\tanh (\tau )+Q_2(\tau\tanh(\tau)-1)-\tanh(\tau)\int \xi_4 \,d\tau +\\
	&+ \frac {1}{10A} \bigg( -\frac { \left( \sinh \left( \tau \right)  \right) ^{3} }{ \left( \cosh \left( \tau
		\right)  \right) ^{5}}\left( \tau\,\tanh \left( \tau
	\right) -1 \right)  
	\left(  \left( \cosh \left( \tau \right)  \right) ^{2} \left( -83+90
	\,f \right) +150\,f+188 \right)\\
	&+\tanh ( \tau )  \big( 180\,\tau\,
	f-166\,\tau+\frac {2		}{ \left( {{\rm e}^{2\,\tau}}+1 \right) ^{5}}( \left(  \left( -480\,f-210 \right) 
	\tau+30\,f-354 \right) {{\rm e}^{2\,\tau}}+ (  \left( 540\,f_
	{1}+1460 \right) \tau\\
	&+1290\,f+442 ) {{\rm e}^{4\,\tau}}+
	\left(  \left( -1440\,f-630 \right) \tau+1290\,f+442 \right) 
	{{\rm e}^{6\,\tau}}+ \left(  \left( 30\,f+625 \right) \tau+30\,f_{
		1}-354 \right) {{\rm e}^{8\,\tau}}\\
	&+ \left( -90\,f+83 \right) \tau)-\ln  \left( {{\rm e}^{2
			\,\tau}}+1 \right)  \left( -83+90\,f \right)  \big)  \bigg) 
	.
	\end{split}
	\end{eqnarray}
	The initial value equation \eqref{xi4} gives
	\begin{eqnarray}\label{v12}
	\begin{split}
	\xi_4&=6\,{ Q_4}-\frac {113\,\ln  ( {{\rm e}^{2\,\tau}}+1 ) }{4{{ A}
			\,\cosh ( \tau )  ( \sinh ( \tau ) 
			) ^{3}}}-\frac { 1}{320\,{A}\, \left( \cosh \left( \tau
		\right)  \right) ^{5} \left( \sinh \left( \tau \right)  \right) ^{3}}\bigg(\left( 460\,f-252 \right)  \left( \cosh \left( \tau
	\right)  \right) ^{8}\\
	&+ \left( -170\,f-197 \right)  \left( \cosh
	\left( \tau \right)  \right) ^{6}- \left( 550\,f-335 \right) 
	\sinh \left( \tau \right) \tau\, \left( \cosh \left( \tau \right) 
	\right) ^{3}+ \left( -630\,f-1172 \right)  \left( \cosh \left( 
	\tau \right)  \right) ^{2}\\
	&- \left( 290\,f+380 \right) \tau\,\sinh
	\left( \tau \right) \cosh \left( \tau \right) +270\,f+10436+
	\left(  \left( 1100\,\tau+70 \right) f-670\,\tau+1423 \right) {
		{\rm e}^{4\,\tau}}\\
	&+ \left(  \left( 4460\,\tau+800 \right) f-490\,
	\tau+4262 \right) {{\rm e}^{2\,\tau}}+ + \left(  \left( -60\,\tau+800 \right) f-2190\,
	\tau+4262 \right) {{\rm e}^{-2\,\tau}}\\
	&+ \left( 70\,f+1423 \right) 
	{{\rm e}^{-4\,\tau}}\left( 3300\,\tau+4940 \right) 
	f-2010\,\tau\bigg)
	.
	\end{split}
	\end{eqnarray}
	
	\subsection*{VIII}
	A sample of answers with choosing a simple parameterization of $p_1=p_2=A=1$  is found as following
	\begin{eqnarray}\label{viiix}
	\begin{split}
	X_1=&X_2={ Q_3}\,{\rm coth} (\tau)+{Q_4}\, ( \tau\,{\rm coth} (\tau)-1 ) -{\rm coth} (\tau)\int \xi_4\, d\tau\\
	&+\frac {1}{A} \bigg( {\rm coth} \left(\tau\right) \big( \frac {
		1 }{
		7680\, \left( \sinh \left( \tau \right)  \right) ^{2} \left( \cosh
		\left( \tau \right)  \right) ^{5}}(\left( 210\,f+2370 \right) \tau\,\cosh \left( 5\,\tau \right) +
	\left( -4050\,f-3570 \right) \tau\,\cosh \left( 3\,\tau \right)\\
	& +
	\left( 3840\,f+6960 \right) \tau\,\cosh \left( \tau \right)+\frac {{{\rm e}^{-7\,\tau}}
	}{2 }(  \left( 3405\,f+7185 \right) {{\rm e}^{10\,\tau}}+ \left( 
	-105\,f-1005 \right) {{\rm e}^{12\,\tau}}\\
	&+ \left( -3\,f+2721
	\right) {{\rm e}^{2\,\tau}}+ \left( -3441\,f-6613 \right) {
		{\rm e}^{4\,\tau}}+ \left( 10230\,f+13710 \right) {{\rm e}^{6\,
			\tau}}+ \left( -10050\,f-16570 \right) {{\rm e}^{8\,\tau}}\\
	&-36\,f_{
		1}+572 )) -3/8\,({ Li_2}
	\left( -{{\rm e}^{2\,\tau}} \right) -{ Li_2} \left( {{\rm e}^{
			2\,\tau}} \right))  \big) - \left( \tau\,{\rm coth} \left(\tau\right)
	-1 \right)  \big( 3/4\,\ln  \left( \tanh \left( \tau \right) 
	\right) \\
	&+\frac { \left( 14\,f+146 \right)  \left( \cosh
		\left( \tau \right)  \right) ^{6}+12\, \left( \cosh \left( \tau
		\right)  \right) ^{4}+ \left( -85\,f-257 \right)  \left( \cosh
		\left( \tau \right)  \right) ^{2}+71\,f+123}{32 \left( \sinh
		\left( \tau \right)  \right) ^{2} \left( \cosh \left( \tau \right) 
		\right) ^{4}} \big)  \bigg)
	,
	\end{split}
	\end{eqnarray}
	\begin{eqnarray}
	\begin{split}
	\phi_1=&{ Q_1}\,{\rm tanh} (\tau)+{Q_2}\, ( \tau\,{\rm tanh} (\tau)-1 ) -{\rm tanh} (\tau)\int \xi_4\, d\tau\\
	&+\frac {1}{{2A}} \bigg( \tanh \left( \tau \right)  \big( {
		\frac { 1}{2304\, \left( \cosh \left( \tau \right)  \right) ^{6}\sinh
			\left( \tau \right) }}(\left( 126\,f+2826 \right) \tau\,\sinh \left( 5\,\tau
	\right) + \left( -5238\,f-8802 \right) \tau\,\sinh \left( 3\,\tau
	\right) \\
	&- \left( -8940\,f-18804 \right) \tau\,\sinh \left( \tau
	\right) )+{\frac {{{\rm e}^{-7\,\tau}} }{4608\,
			\left( \cosh \left( \tau \right)  \right) ^{6}\sinh \left( \tau
			\right) }}  (  \left( 
	3960\,f+6012 \right) {{\rm e}^{10\,\tau}}+ \left( -63\,f-1629
	\right) {{\rm e}^{12\,\tau}}\\
	&+ \left( -808\,f-4564 \right) {
		{\rm e}^{2\,\tau}}+ \left( 2619\,f+729 \right) {{\rm e}^{4\,\tau}}
	+ \left( -3152\,f-8360 \right) {{\rm e}^{6\,\tau}}+ \left( -2407\,
	f-5425 \right) {{\rm e}^{8\,\tau}}\\
	&-149\,f-587 )\big) - \left( \tau {\tanh} \left( \tau \right) 
	-1 \right)  \big( -3\,\ln  \left(  { \tanh} \left( \tau \right) 
	\right) +\frac {1}{24
		\left( \cosh \left( \tau \right)  \right) ^{6}}( \left( 21\,f+471 \right)  \left( \cosh
	\left( \tau \right)  \right) ^{4}\\
	&+ \left( -234\,f-720 \right) 
	\left( \cosh \left( \tau \right)  \right) ^{2}+149\,f+317) \big)
	\bigg) 
	,
	\end{split}
	\end{eqnarray}
	\begin{eqnarray}
	\begin{split}
	X_3=&{ Q_5}\,{\rm \tanh} (\tau)+{Q_6}\, ( \tau\,{\rm tanh} (\tau-1 ) -{\rm coth} (\tau)\int \xi_4\, d\tau\\
	&+\frac {1}{2A} \bigg( \tanh \left( \tau \right)  \big( \frac {1 }{9216\, \left( 
		\cosh \left( \tau \right)  \right) ^{6}\sinh \left( \tau \right) }\big(2
	\,\tau\, (  \left( 126\,f+1098 \right) \sinh \left( 5\,\tau
	\right) + \left( -2430\,f-3546 \right) \sinh \left( 3\,\tau
	\right) \\
	&+ \left( 4260f\,\tau+7164 \right) \sinh \left( \tau
	\right)  ) +{{\rm e}^{-7\tau}} (  \left( 1854f+
	1638 \right) {{\rm e}^{10\,\tau}}+ \left( -63f-765 \right) {
		{\rm e}^{12\,\tau}}+ \left( -418f-2730 \right) {{\rm e}^{2\,\tau
	}}\\
	&+ \left( 1215\,f-1899 \right) {{\rm e}^{4\,\tau}}+ \left( -1436
	\,f-5820 \right) {{\rm e}^{6\,\tau}}+ \left( -1081\,f-3855
	\right) {{\rm e}^{8\,\tau}}-71\,f-393 )\big)\\
	&+3/
	4\,({ Li_2} \left( -{{\rm e}^{2\,\tau}} \right) -{ Li_2}
	\left( {{\rm e}^{2\,\tau}} \right))  \big) - \left( \tau\,\tanh
	\left( \tau \right) -1 \right)  \big( -3/2\,\ln  \left( \tanh
	\left( \tau \right)  \right) +\frac { 1}{ 48\left( \cosh \left( \tau \right)  \right) ^{6}}(( 21\,f\\
	&+183
	)  \left( \cosh \left( \tau \right)  \right) ^{4}+ \left( -117
	\,f-285 \right)  \left( \cosh \left( \tau \right)  \right) ^{2}+71
	\,f+123)
	\big)  \bigg),
	\end{split}
	\end{eqnarray}
	\begin{eqnarray}\label{53viii}
	\begin{split}
	\xi_4=&Q_{4}-Q_{6}-{\frac {3\,\ln  \left( \tanh \left( \tau \right)  \right) 
		}{A}}-\frac { 1}{96\, \left( \sinh \left( \tau \right)  \right) ^{2} \left( 
		\cosh \left( \tau \right)  \right) ^{6}A}\big(\left( 84\,f+876 \right)  \left( \cosh \left( \tau
	\right)  \right) ^{8}+418\,f_{1
	}\\
	&+ \left( -84\,f-1494 \right)  \left( \cosh
	\left( \tau \right)  \right) ^{6}+ \left( 165\,f+2841 \right) 
	\left( \cosh \left( \tau \right)  \right) ^{4}+ \left( -583\,f-
	3503 \right)  \left( \cosh \left( \tau \right)  \right) ^{2}+1352\big).
	\end{split}
	\end{eqnarray}
	\subsection*{IX}
	By setting $n=1$ and $p_1=p_2=1$,   the forms of $X_i$ \eqref{ixx1} and \eqref{ixx2}, $\phi_1$ \eqref{fi1s} and the lapse function $\xi_4$ \eqref{xi4} are found as following
	\begin{eqnarray}\label{ixx11}
	\begin{split}
	X_1=X_3=( Q_3+Q_5)\,&{\rm \tanh} (n\tau)+{(Q_4+Q_6)}\, (\tau\,{\rm tanh} (\tau)-1 ) -{\rm tanh} (\tau)\int \xi_4\, d\tau\\
	&-\frac {\tanh \left( \tau \right)  \left( f-31 \right) }{31A}
	\bigg( {\frac {2\,\tau-{{\rm e}^{\tau}}-1}{ \left( \cosh \left( \tau
			\right)  \right) ^{2}}}+ \left( \tau\,\tanh \left( \tau \right) -1
	\right) \tanh \left( \tau \right)  \bigg) +c_1-c_2
	,
	\end{split}
	\end{eqnarray}
	\begin{eqnarray}\label{ixx22}
	\begin{split}
	X_2={ Q_5}\,{\rm \tanh} (\tau)&+{Q_6}\, (\tau\,{\rm tanh} (\tau)-1 ) -{\rm tanh} (\tau)\int \xi_4\, d\tau\\
	&-\frac {\tanh \left( \tau \right)  \left( f+17 \right) }{31A}
	\bigg( {\frac {2\,\tau-{{\rm e}^{\tau}}-1}{ \left( \cosh \left( \tau
			\right)  \right) ^{2}}}+ \left( \tau\,\tanh \left( \tau \right) -1
	\right) \tanh \left( \tau \right)  \bigg) +c_2
	,
	\end{split}
	\end{eqnarray}
	\begin{eqnarray}
	\begin{split}
	\phi_1=&{ Q_1}\,\tanh (\tau )+Q_2(\tau\,{\rm tanh} (\tau)-1 ) -\tanh(\tau)\int \xi_4 \,d\tau\\
	&-\frac {\tanh \left( \tau \right)  \left( 3f-25 \right) }{31A}
	\bigg( {\frac {2\,\tau-{{\rm e}^{\tau}}-1}{ \left( \cosh \left( \tau
			\right)  \right) ^{2}}}+ \left( \tau\,\tanh \left( \tau \right) -1
	\right) \tanh \left( \tau \right)  \bigg)
	,
	\end{split}
	\end{eqnarray}
	\begin{eqnarray}\label{ixx4}
	\begin{split}
	\xi_4=4\,{ Q_4}-2\,{ Q_6}+\frac {1}{A \left( \cosh \left( \tau \right) 
		\right) ^{2}} \big( &1-3/16\,f+ \left( \tanh \left( \tau \right) 
	\right) ^{2} (  \left( f-5/3 \right)  \left( \cosh \left( 
	\tau \right)  \right) ^{2}\\
	&+ \left( -f+{ {25}/{3}} \right) 
	\left( 1+\tau\,\tanh \left( \tau \right)  \right)  )  \big) 
	.
	\end{split}
	\end{eqnarray}
	%%%%%%%%%%%%%%%%%%%%%%%%%%%%%%%%%%%%%%%%%%%%%%%%%%%%%%%%%%%%%%%%%%%%%%%%%%%%%%%%%%%%%%%%%%%%%%%%%%
	
\end{document}